\documentclass[prx, reprint, aps, notitlepage,longbibliography,floatfix,nobalancelastpage,tightenlines,nofootinbib]{revtex4-2}

\usepackage{cancel}
\usepackage[utf8]{inputenc}
\usepackage{float}
\usepackage{graphicx}  
\usepackage{physics}
\usepackage{esint}
\usepackage{amssymb}  
\usepackage{mathtools}
\usepackage{amsmath}
\usepackage{amsthm}
\usepackage{mathrsfs}
\usepackage{soul}
\usepackage[colorlinks=true,linkcolor=blue,citecolor=cyan,urlcolor=cyan,plainpages=false,pdfpagelabels]{hyperref}
\usepackage{color,xcolor,colortbl}
\usepackage{bm}
\usepackage[most]{tcolorbox}
\usepackage{enumerate}

\newcommand{\e}{\text{e}}
\newcommand{\im}{\text{i}}

\numberwithin{footnote}{section}
\numberwithin{equation}{section}

\begin{document}


\title{Entangled sensor-networks for dark-matter searches}

\author{Anthony J. Brady$^{1}$}
\email{ajbrady4123@email.arizona.edu}

\author{Christina Gao$^{2,3,4}$} 
\email{yanggao@fnal.gov}

\author{Roni Harnik$^{2}$} 
\email{roni@fnal.gov}

\author{Zhen Liu$^{5}$} 
\email{zliuphys@umn.edu}

\author{Zheshen Zhang$^{6,7}$}%
\email{zsz@email.arizona.edu}

\author{Quntao Zhuang$^{1,7}$}%
\email{zhuangquntao@email.arizona.edu}
\affiliation{
$^1$Department of Electrical and Computer Engineering, University of Arizona, Tucson, Arizona 85721, USA
}
\affiliation{
$^2$Theoretical Physics Division, Fermi National Accelerator Laboratory, P.O. Box 500, Batavia, Illinois 60510, USA}
\affiliation{$^3$Department of Physics, University of Illinois at Urbana-Champaign, Urbana, IL 61801, USA}
\affiliation{$^4$Illinois Center for Advanced Studies of the Universe, University of Illinois at Urbana-Champaign, Urbana, IL 61801, USA}
\affiliation{
$^5$School of Physics and Astronomy, University of Minnesota, Minneapolis, MN 55455, USA
}
\affiliation{
$^6$Department of Materials Science and Engineering, University of Arizona, Tucson, Arizona
85721, USA}
\affiliation{
$^7$J. C. Wyant College of Optical Sciences, University of Arizona, Tucson, Arizona 85721, USA}

\date{\today}

\begin{abstract}

The hypothetical axion particle (of unknown mass) is a leading candidate for dark matter (DM). Many experiments search for axions with microwave cavities, where an axion may convert into a cavity photon, leading to a feeble excess in the output power of the cavity. Recent work [Nature {\bf 590}, 238 (2021)] has demonstrated that injecting squeezed vacuum into the cavity can substantially accelerate the axion search. Here, we go beyond and provide a theoretical framework to leverage the benefits of quantum squeezing in a network setting consisting of many sensor-cavities. By forming a local sensor network, the signals among the cavities can be combined coherently to boost the axion search. Furthermore, injecting multipartite entanglement across the cavities---generated by splitting a squeezed vacuum---enables a global noise reduction. We explore the performance advantage of such a local, entangled sensor-network, which enjoys both coherence between the axion signals and entanglement between the sensors. 
Our analyses are pertinent to next-generation DM-axion searches aiming to leverage a network of sensors and quantum resources in an optimal way. Finally, we assess the possibility of using a more exotic quantum state, the Gottesman-Kitaev-Preskill (GKP) state. Despite a constant-factor improvement in the scan-time relative to a single-mode squeezed-state in the ideal case, the advantage of employing a GKP state disappears when a practical measurement scheme is considered. 

\end{abstract}

\maketitle

\section{Introduction}

The nature of Dark Matter (DM) poses an unsolved mystery in physics. 
Axions are a well motivated DM hypothesis. Originally proposed to address the strong CP problem~\cite{PhysRevLett.38.1440,PhysRevD.16.1791,PhysRevLett.40.223,PhysRevLett.40.279}, light pesudoscalars are also common in top-down constructions of high energy physics~\cite{Svrcek:2006yi, Arvanitaki:2009fg}. Such light bosonic fields may be produced in the early Universe in large occupation numbers and pose a low average momentum in late times. The axion can hence be described today as a coherent state---a classical non-relativistic wave, oscillating at a frequency set by the axion mass.

The energy density in the axion field is a good candidate to serve as the dark matter~\cite{Preskill:1982cy, Abbott:1982af, Dine:1982ah}.
The axion dark matter hypothesis can be tested experimentally in cavity based  searches~\cite{Sikivie:1983ip} in which axions can convert to photons in a quiet cavity mode in the presence of a background magnetic field,~\footnote{Axions are part of a broader class of wave-like DM models. Another DM model in this class are dark photons~\cite{Holdom:1985ag} which also can serve a dark matter candidate~(see e.g.~\cite{McDermott:2019lch}) and searched for in electromagnetic cavities, though without the need for a magnetic field. Our results also generalize to dark photon searches, as well as other wave-like candidates. However, we refer to axions throughout the paper for brevity of presentation.} with searches actively ongoing (e.g. ~ADMX~\cite{ADMX:2019uok,ADMX:2021nhd}, HAYSTAC~\cite{HAYSTAC:2018rwy}, ORGAN~\cite{McAllister:2017lkb}, and CAPP~\cite{CAPP:2020utb}).
The resonant frequency of the cavity mode must match the axion mass $m_a$, an unknown parameter, and should thus be scanned as part of the search. The size of the cavity $L_\mathrm{cav}$ is parametrically set by the inverse axion mass $m_a$, $L_\mathrm{cav}\sim m_a^{-1}$.
A challenge of axion cavity searches lies in detecting a small displacement of the electromagnetic field from the initial state, thus distinguishing the faint signal from the noise. Enhancing the sensitivity to small signals allows one to enhance the rate at which the cavity frequency is scanned, leading to a more effective search. Quantum resources and quantum measurement-techniques can play a vital role here. 

Quantum information science brings an unprecedented capability in ultra-precise sensing~\cite{giovannetti2004,giovannetti2006,giovannetti2011advances,degen2017,pirandola2018advances,lawrie2019quantum}. Non-classical phenomena such as entanglement and squeezing have been utilized to improve the measurement precision of various application scenarios including: bio-sensing~\cite{taylor2013biological}, radar target detection~\cite{tan2008quantum,zhang2015entanglement,zhuang2022ultimate}, the detection of gravitational waves~\cite{LIGO,abadie2011gravitational,aasi2013enhanced,tse2019quantum} and the search for dark matter~\cite{girvin2016axdm,malnou2019,backes2021,dixit2021,wurtz2021state_swap,Wang2021,Harnik+PRXQ2021}. Indeed, a recent experiment~\cite{backes2021} has utilized squeezing in microwave cavity sensors to speed up the dark matter search by a factor of two despite loss and noise. However, quantum sensing has much more to offer: entangling multiple sensors is known to fundamentally change how the precision scales with the number of sensors, from the standard quantum limit (imposed by the law-of-large-numbers) to Heisenberg scaling attained via multipartite entanglement~\cite{giovannetti2004}. Moreover, entangled sensor-networks enable Distributed Quantum Sensing (DQS), thus permitting an enhanced extraction of global properties of the network~\cite{zhuang2018DQSCV,proctor2017multi,ge2017distributed,xia2020demonstration,guo2020distributed,zhang2021dqs}.

\begin{figure}[t]
    \centering
    \includegraphics[width=\linewidth]{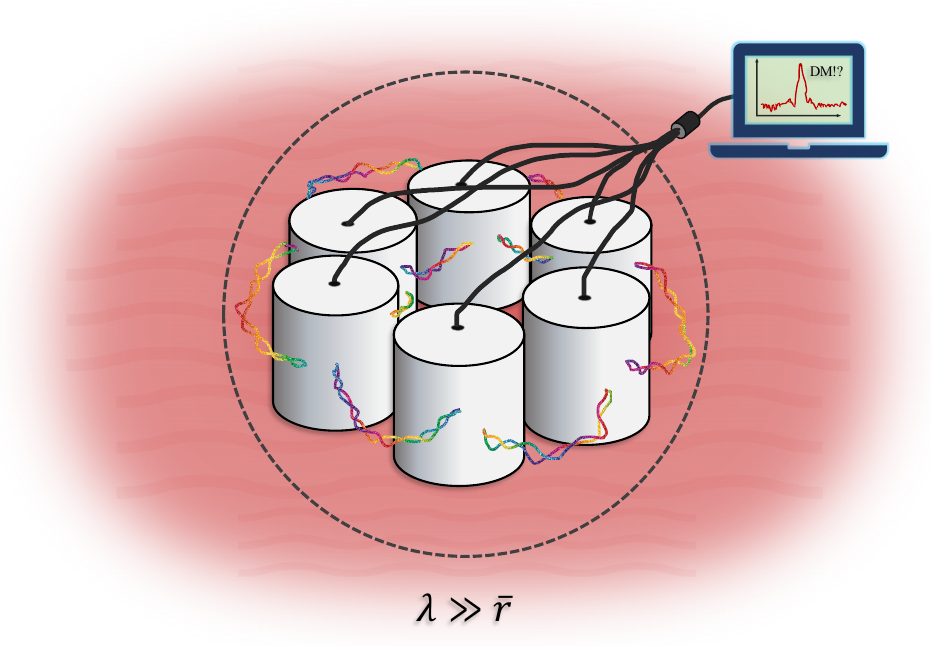}
    \caption{Illustration of entangled sensor-cavities within a network-volume of radius $\bar{r}$, taken to be much smaller than the axion wavelength $\lambda$. Thanks to this hierarchy of scales, the entire network experiences a coherent signal which can be combined at the amplitude level. The cylinders represent the cavities and the colored lines represent their entanglement. The black solid lines represent joint processing of the measurement to search for dark matter.}
    \label{fig:cavity_network_pic}
\end{figure}

In this work, we propose a design of entangled sensor-networks for axion dark matter searches. A local network of sensors can benefit the search thanks to a feature of the axion dark matter signal that lies in its non-relativistic nature. The virial velocity of dark matter in our galaxy is of order $v\sim 10^{-3}$ in units of the speed of light. The momentum of axion particles $m_a v$, which sets the gradient in the axion field, is parametrically suppressed compared to the axion mass. In any moment in time the axion field will change over a spatial length of order $\lambda\sim 10^3 m_a^{-1}$, which is the signal coherence length. Since the size of each cavity-based sensor is of order $m_a^{-1}$, a local network of sensors, sketched in Figure~\ref{fig:cavity_network_pic}, will be subject to a coherent axion signal common to all sensors within the network. 
By forming a local sensor network, the signals from different cavities can be coherently combined to boost the scan rate. Furthermore, injecting multipartite entanglement across the cavities---generated by splitting a squeezed vacuum---enables a global noise reduction, leading to a further enhancement in the scan rate. 

Our paper is organized as the following. Sec.~\ref{sec:preliminary} contains the preliminary knowledge to prepare the presentation of our main results (Sec.~\ref{sec:multi_sensor}) on entangled sensor-networks. In Sec.~\ref{sec:axion}, we provide some background on dark matter and axions. In Sec.~\ref{sec:models_search}, we introduce a quantum model of microwave cavities for dark matter search. In Sec.~\ref{sec:dicke_scan_rate}, we revisit the Dicke radiometer equation and give the figure of merit of a dark matter search system---the scan rate. In Sec.~\ref{sec:squeezing_DM_search}, we discuss how squeezing can boost the scan rate, as proposed in refs.~\cite{girvin2016axdm,malnou2019,dixit2021}. In Sec.~\ref{sec:multi_sensor}, we introduce our quantum network scheme, beginning with the ideal identical sensor case in Sec.~\ref{sec:entangled_sensor_identical} and generalizing to non-identical case in Sec.~\ref{sec:entangled_sensor_non_identical}.
To fully explore quantum sensing techniques, in Appendix~\ref{appendix:gkp_app} we consider more exotic quantum states, such as the Gottesman-Kitaev-Preskill (GKP) state~\cite{gkp2001}, which is shown to be valuable in sensing both quadratures of displacements~\cite{duivenvoorden2017single,zhuang2020DQSgkp}. More detailed tutorials and analyses are provided in the appendices for further reference.

\section{Preliminary}
\label{sec:preliminary}

\subsection{Axion Dark Matter}
\label{sec:axion}

Amongst the many orders of magnitudes in mass that a dark matter candidate can reside in, the ultralight regime
(with DM masses 
between
$10^{-21}~{\rm~eV}
\sim {\rm~keV}$), with the wave-like aspect of dark matter being more prominent, is  interesting in many respects. In particular,  quantum-noise-limited devices can play a critical role here in probing dark matter. In this regime, bosonic dark matter---such as axions, scalar DM, dark photon etc.---are all possible dark matter candidates. The optimal experimental setup in the search will depend 
on how they couple to standard model particles. 



Even within axion DM searches, one can consider various couplings, such as its coupling to spin or to electromagnetic fields. In a cavity-like setup, it is natural to exploit its coupling to photons via the following interaction,
\begin{equation}
    \frac a {f_a} F^{\mu\nu} \tilde F_{\mu\nu}\longrightarrow \frac a {f_a} \vec E\cdot \vec B\,.
\end{equation}
Here $a$ represents the dynamical axion dark matter field, $f_a$ is the axion decay constant, and we have expressed the general covariant interaction in terms of local quantities in the laboratory, the electric and magnetic fields, $\vec E$ and $\vec B$, respectively. This implies, for instance, in a background magnetic field $\vec B$, an axion DM will induce an electric field $\vec E$ within the cavity~\cite{Sikivie:1983ip}. We will consider this standard approach in this work, though the principle of leveraging a quantum network to benefit the scan rate is applicable more generally to wave-like dark matter.


Many of the features that we discuss apply to any wavelike dark matter, but we refer to the axion for concreteness.
The local DM density in our region of the Milky Way is approximately,
\begin{equation}
    \rho_a\sim 0.3{\rm-}0.4~{\rm~GeV/cm^3}.
\end{equation}
The local number density of ultralight DM with mass $m_a$ is then
\begin{equation}
    \frac {\rho_a} {m_a} \sim 10^{15}\left(\frac {{\rm \mu eV}} {m_a}\right) {\rm cm^{-3}}\sim 10^{15} \left(\frac {\lambda} {\rm km}\right) {\rm cm^{-3}}.
    \label{eq:n_DM}
\end{equation}
Here $\lambda$ is the De Broglie wave length of the cold dark matter, $\sim (m_a v)^{-1}$ with a typical virial velocity of order~$v\sim 200$~km/sec\footnote{Since the dark matter halo is virialized~\cite{binney2011book,eilers2019velcurve}, the dark matter has a velocity distribution, usually assumed to be Maxwell–Boltzmann, which is peaked around this velocity. Here we use natural units. In SI units, $\lambda=h/m_a v\sim 0.8$km, where $h$ is the Planck's constant.}. 
Due to the large number density of the axion field, locally, within a coherence time, the axion DM behaves like a classical wave,
\begin{equation}
\label{eq:axion-field}
    \langle a \rangle(m_a,\vec k_0, t)=
    \sqrt{\frac{2\rho_a}{m^2_a}}
     \cos(\omega_0 t+\vec k_0\cdot \vec x+\phi_a).
\end{equation}
 Here 
$
\sqrt{\frac{2\rho_a}{m^2_a}}$
 is the classical wave amplitude, $\phi_a$ is the phase, $\vec k_0$ is the wave vector with $|\vec k_0|=2\pi/\lambda$, and $\omega_0\equiv\sqrt{m_a^2+|\vec k_0|^2}\sim m_a$ is the axion central frequency. In reality, all parameters drift continuously with time, however we work in a discrete approximation. Hence, within a coherence time, the phase $\phi_a$ is fixed; while above the coherence time, $\phi_a$ is completely random in $[0,2\pi)$. Similarly, after a coherence time, the direction of $\vec k_0$ is expected to change by order one, its magnitude by order $10^{-3} m_a$, and thus, $\omega_0$ will change by order $10^{-6} m_a$. Therefore, the axion can effectively be thought of as a coherent background source of frequency $m_a$ and bandwidth of order $\Delta_a\sim m_a/Q_a$, where $Q_a\sim10^6$ is the ``axion quality factor''.

The axion De Broglie wavelength also sets a coherence length for the axion field, above which spatial variations of the axion field become appreciable. Because the earth moves through the DM halo with a similar velocity, the time the lab spends within a coherence length is $(m_a v^2)^{-1}$, which is the coherence length of the axion signal. 

In gist, the classical field in Equation~(\ref{eq:axion-field}) is a good approximation within a coherence time. A more accurate description is a superposition of nearly coherent waves with a frequency spread of order $10^{-6}$ around the central frequency of $m_a$. It is thus sometimes said that the axion is a coherent oscillator with a quality factor $Q_a\sim 10^6$.
As a result of, the phase $\phi_a$ in the approximate description above drifts slowly and is randomized roughly every $10^6$ periods or so. 

The prevailing way to search for axion dark matter is to place an electromagnetic cavity with a high quality factor $Q_c$ in a static magnetic field $\vec B$. Cavity modes in which $\vec E\cdot \vec B\ne 0$ will be excited by the coherent oscillating axion background. The power of the axion induced signal in the cavity is 
\begin{equation}
    \mathcal{P}_\mathrm{cav} \sim
    g_{a\gamma}^2\frac{\rho_{a}}{m_a}B^2V\eta\, \mathrm{min}\left[Q_c, Q_a\right]
    \label{P_cav}
\end{equation}
where $\eta$ is an order 0.1-1 form factor and $V$ is the volume of the cavity (see Table~\ref{tab:dictionary} for a summary of physical parameters). It is useful for our analysis to write the typical (unitless) axion-induced displacement of the cavity field
\begin{equation}
     \delta E \sim\sqrt{\frac{\mathcal{P}_{\rm cav}}{m_a\Delta_a}}\,.\label{eq:nu_physical}
\end{equation}
The goal of the axion search is to sense this displacement over the noise, be it thermal or quantum in nature. With this effective description of the axion dynamics and its interaction in a cavity, we now move on to writing a quantum model for the detector to account for the relevant effects.

\begin{table}[t]
    \renewcommand{\arraystretch}{1.6}
    \centering
    \begin{tabular}{c  c}
    \hline\hline
    Physical Parameters & Description  \\ \hline
    $Q_c$ & Intrinsic cavity quality factor \\
    $V$ & Cavity volume \\
    $B$ & Magnetic field \\
    $g_{a\gamma}$ & Axion-photon coupling \\
    $m_a$ & Axion mass \\
    $\rho_a$ & DM energy density\\
    $\Delta_a\sim\frac{m_a}{Q_a}$ & Axion bandwidth, $Q_a\sim 10^6$\\
    \hline\hline
    \end{tabular}
    \caption{Description of physical parameters.}
    \label{tab:dictionary}
\end{table}

\begin{figure}[t]
    \centering
    \includegraphics[width=\linewidth]{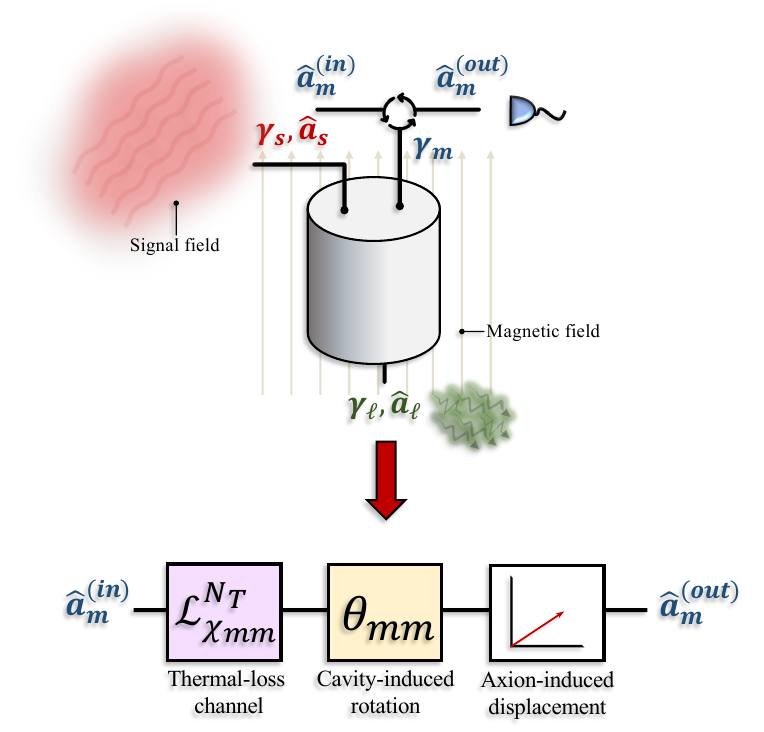}
    \caption{Cavity model. (Top) Illustration of the various couplings to the cavity: $\gamma_m$ denotes the measurement-port coupling parameter (with the input and output measurement ports being accessible and controllable) for the $\hat{a}_m$ mode; $\gamma_\ell$ denotes intrinsic cavity loss via the $\hat{a}_\ell$ mode; and $\gamma_s$ determines the coupling strength between the cavity mode and the axion-field mode $\hat{a}_s$ (note $\gamma_s\ll\gamma_m,\gamma_\ell$; cf. \cite{girvin2016axdm}). (Bottom) An equivalent single-mode Gaussian quantum-channel description which faithfully describes the input-output transformation of the measurement-port mode, $\hat{a}_m$. Channel decomposes into a thermal-loss channel with attenuation parameter $\abs{\chi_{mm}}^2$ and thermal-noise parameter $N_T=1+2\bar{n}_T$, followed by a cavity-induced rotation by an angle $\theta_{mm}$ and a random, axion-induced displacement.
    }
    \label{fig:cavity_channel}
\end{figure}

\subsection{Quantum model of a cavity}
\label{sec:models_search}

We utilize the workhorse cavity-setup described in refs.~\cite{girvin2016axdm,malnou2019} to model the coupling of a hypothetical axion field to the electromagnetic mode of a (microwave) cavity (see Fig.~\ref{fig:cavity_channel}) and eventually develop a quantum channel model for the input-output transformations of the cavity. Mathematical details justifying this model and channel decomposition are found below and in Appendix~\ref{appendix:cavity_model}. We emphasize that the model presented below works for detection of any fluctuating classical field which linearly drives the quantum field of a microwave cavity. Thus, throughout the document, we use the terminology ``signal field" and ``axion-field" interchangeably.

We model axion-to-photon conversion in a microwave cavity by treating the axion-field as a coherent (within the DM coherence time) linear-drive, which drives a damped microwave cavity-mode at a feeble rate $\gamma_s$ (a signal conversion rate, with units of inverse seconds; its connection to physics parameters are discussed in Appendix~\ref{app:physical_property}).
A transmission line running into and emanating from the ``measurement-port" of the cavity allows access to the cavity field. We assume full control of the measurement-port in the sense that: 1) the coupling rate between the cavity and the measurement-port, $\gamma_m$, is tunable; 2) we can design and inject fields into the cavity through the measurement-port input; and 3) we can measure the quadrature variables of the measurement-port output with homodyne or heterodyne detection. As shown in Fig. \ref{fig:cavity_channel}, formally, the cavity model consists of three input-output modes---the measurement-port $\hat{a}_m$, the loss-port $\hat{a}_\ell$ describing damping of the microwave cavity, and the axion-field (linear drive) $\hat{a}_s$, respectively---with only the measurement-port $\hat{a}_m$ being experimentally accessible. Here we have described each mode by its annihilation operator.

\begin{table*}[t]
\renewcommand{\arraystretch}{2}
    \centering
    \begin{tabular}{c c c}
    \hline\hline
      Model Parameters & Description & Connection to axion model \\ \hline
      $\omega$ & Cavity detuning (resonance at $\omega=0$) & Frequency variable $=\omega_c + \omega$   \\
      $\gamma_\ell$ & Internal dissipation rate of cavity & $\gamma_\ell\equiv \frac{\omega_c}{Q_c}$ \\
      $\gamma_m$ & Measurement-port coupling rate & $\gamma_m \equiv \beta\frac{\omega_c}{Q_c}$, with coupling $\beta$ \\
      $\gamma_s$ & Signal coupling rate; $\gamma_s\ll\gamma_\ell,\gamma_m$ & 
      $\gamma_s \equiv \frac{g_{a\gamma}^2 B^2\eta}{4\Delta_a}$\\
      $\bm\mu_s$ & Signal amplitude & $\bm\mu_s\equiv \frac{\sqrt{\rho_a V}}{m_a}$\\
      $T$ & System temperature & ---\\
      \hline
      $\gamma$ & Loaded cavity linewidth& $\gamma \approx\gamma_m + \gamma_\ell=\frac{\omega_c}{Q_l}$
      \\
      $\bm\chi_{ij}$ & Complex coefficients of susceptibility matrix \ \ \ \ & Equations \eqref{eq:chimm_expanded} and \eqref{eq:chima_expanded}\\
      $\theta_{ij}$ & Complex angles of susceptibility matrix & Equations \eqref{eq:theta_mm} and \eqref{eq:theta_ma}\\
      $\bar{n}_T$ & Bosonic thermal-occupation at temp. $T$ & Bose-Einstein distribution\\
      $N_T$ & Additive noise parameter; $N_T=1+2\bar{n}_T$ & Vacuum plus thermal noise\\
      $n_s$ & Occupation number of signal & $n_s=\frac{\rho_a V}{m_a}$ number of axions in cavity \\
    \hline\hline
    \end{tabular}
    \caption{Cavity-model parameters are in the top half of the table. Several useful parameters that are derived in terms of the model parameters are shown in the bottom half.
    Their connections to a few physical parameters can be found in Table.~\ref{tab:dictionary}.
    }\label{tab:notation}
\end{table*}

The input-output relations for the system of modes $(\hat{a}_m,\hat{a}_\ell,\hat{a}_s)$ can be found in the spectral domain, in the rotating reference frame of the cavity (rotating at cavity resonance frequency $\omega_c$), resulting in the linear relation (see Appendix~\ref{appendix:cavity_model} and ref.~\cite{girvin2016axdm})
\begin{equation}
    \hat{a}_k^{(\text{out})}(\omega)=\sum_{j\in\{m,s,\ell\}}\bm{\chi}_{kj}(\omega)\hat{a}_j^{(\text{in})}(\omega),\label{eq:in_out}
\end{equation}
where $\omega$ is the cavity detuning, with $[\hat{a}_k(\omega),\hat{a}_j^\dagger(\omega^\prime)]=2\pi\delta_{kj}\delta(\omega-\omega^\prime)$ being the commutation relations for both the input and output fields. Hence, $\hat{a}_k(\omega)$ has units of $\sqrt{\text{quanta/Hz}}$. The linear susceptibility matrix, $\bm{\chi}$, has been defined with matrix elements
\begin{equation}
    \bm{\chi}_{kj}(\omega)=\delta_{kj}-\frac{\sqrt{\gamma_k\gamma_j}}{\gamma/2+\im\omega},\label{eq:chi}
\end{equation}
where $\gamma=\sum_{j\in\{m,s,\ell\}}\gamma_j$ is the total coupling rate. It can be shown that $\sum_{j}\bm{\chi}_{ij}^*\bm{\chi}_{jk}=\delta_{ik}$, and thus $\bm{\chi}$ is a unitary matrix which, in the quadrature basis ($Q$'s and $P$'s; see below), corresponds to a symplectic orthogonal transformation (see Appendix~\ref{appendix:cavity_model} for explicit details). The corresponding unitary dynamics is therefore Gaussian~\cite{weedbrook2012gaussian}, which will allow us to reduce the full, unitary dynamics of the three-mode system to a single-mode quantum channel of the measurement-port mode, as we describe below. The forthcoming analyses are with respect to a single detuned-frequency $\omega$.

To simplify the signal-to-noise ratio (SNR) calculations, we model the continuous spectrum as a set of discrete frequency modes with bin size inverse to the total observation time~\footnote{See the supplemental materials of Ref.~\cite{zhuang2022ultimate} for an example of discretization.} and define the quadrature operators for the measurement port, 
\begin{equation}
     \hat{Q}_m\equiv\frac{1}{\sqrt{2}}\left(\hat{a}_m+\hat{a}_m^\dagger\right)\qq{and} \hat{P}_m\equiv\frac{\im}{\sqrt{2}}\left(\hat{a}_m^\dagger-\hat{a}_m\right),
\end{equation}
so that the canonical commutation relations $[\hat{Q}_m,\hat{P}_m]=\im\hat{\mathbb{I}}$. Here it is understood that this definition holds for both ``in" and ``out" modes. We now define the vector of quadrature operators $\hat{\bm{R}}_m\equiv(\hat{Q}_m,\hat{P}_m)^\top$. From which we define the mean vector and the covariance matrix for the measurement port (i.e. the vector of first moments and matrix of second moments, respectively),
\begin{align}
    \bm{\mu}_m&\equiv\ev{\hat{\bm{R}}_m}\\
    \left(\bm{\sigma}_m\right)_{ij}&\equiv\ev{\left\{\left(\hat{\bm{R}}_m-\bm{\mu}_m\right)_i,\left(\hat{\bm{R}}_m-\bm{\mu}_m\right)_j\right\}},
\end{align}
where the index $i$ corresponds to the $i$th entry of the vectors, the expectation value is taken with respect to some quantum state of the mode $\hat{a}_m$, and $\{\cdot,\cdot\}$ is the symmetric, anti-commutator.

We work in the Heisenberg picture and thus describe the input-output dynamics of the mode $\hat{a}_m$ by first specifying the input moments $(\bm{\mu}_m^{\rm (in)},\bm{\sigma}_m^{\rm (in)})$---which is equivalent to specifying the input quantum state to the measurement port, assuming an initial Gaussian quantum state---and then determine the Gaussian quantum channel $\mathcal{G}:(\bm{\mu}_m^{\rm (in)},\bm{\sigma}_m^{\rm (in)})\rightarrow(\bm{\mu}_m^{\rm (out)},\bm{\sigma}_m^{\rm (out)})$ where $(\bm{\mu}_m^{\rm (out)},\bm{\sigma}_m^{\rm (out)})$ are the moments of the measurement-port exiting the cavity. Although the full transformations implied by Eq.~\eqref{eq:in_out} are unitary, the reduced transformation induced by the channel $\mathcal{G}$ (which can be derived from the full unitary transformation) is non-unitary, due to our restriction to a single mode. Nonetheless, $\mathcal{G}$ serves as a valid quantum operation.
Indeed, as described in detail in Appendix~\ref{appendix:cavity_model}, we can decompose the Gaussian quantum channel $\mathcal{G}$ into a concatenation of three standard channels: 
\begin{enumerate}
\item A thermal-loss channel  $\mathcal{L}_{\bm{\chi}_{mm}}^{N_T}$, with attenuation parameter $\abs{\bm{\chi}_{mm}}^2$ and noise spectral-density parameter $N_T=1+2\bar{n}_T\geq1$, where $\bar{n}_T$ is the bosonic, thermal occupation number for a harmonic oscillator in equilibrium at temperature $T$ and oscillating at the detuned frequency $\omega_c+\omega$. Physically, the attenuation parameter $\abs{\bm{\chi}_{mm}}^2$ describes the ability of quanta injected into the cavity to efficiently transfer to the output. The noise parameter $N_T$ describes the noise added in this transfer process; such a process always adds at least a unit of vacuum noise (hence, $N_T\geq1$). For instance, near the operating conditions (see, for instance, \cite{malnou2019, backes2021}) $\omega_c\approx2\pi\times7\,{\rm GHz}$ and $T\approx35\,{\rm mK}$, $\bar{n}_T\approx1.1\times10^{-4}$; the noise is thus vacuum-dominated.

\item A cavity induced phase-rotation $\Phi_{\theta_{mm}}$, where $\theta_{mm}\equiv\arg(\bm{\chi}_{mm})$ is the relative angle between the input and output fields of the measurement port. This defines a natural reference frame for the measurement-port fields, which we can go to by applying the complementary phase shift $\Phi_{-\theta_{mm}}$ to input fields prior to the cavity interaction. See Eqs.~\eqref{eq:theta_ma} and \eqref{eq:theta_mm} and Appendix~\ref{appendix:cavity_model} for details regarding the cavity-induced angle $\theta_{mm}$ (and $\theta_{ms}$ defined below).

\item 
An axion-induced displacement, $\mathcal{D}_{\bm\nu}$, where $\bm{\nu}=\abs{\bm{\chi}_{ms}}\bm{O}(\theta_{ms})\bm{\mu}_s$ is the signal-amplitude to be read out from the measurement port of the cavity. \footnote{A typical readout displacement, $\nu$, is related to the cavity displacement, $\delta E$ of Eq.~\eqref{eq:nu_physical}, via $\nu/\delta E\approx\sqrt{\frac{\gamma_m/\gamma_\ell}{(1+\gamma_m/\gamma_\ell)^2}}$, where the approximation assumes $Q_a\gg Q_c$.} Here, $\abs{\bm{\chi}_{ms}}$ is the absolute value of the $(m,s)$ element of the susceptibility matrix $\bm\chi$, $\bm{O}(\theta_{ms})$ is a $2\times2$ rotation matrix by the angle $\theta_{ms}\equiv \arg(\bm{\chi}_{ms})$, and $\bm{\mu}_s$ is the spectral-amplitude of the axion field in phase-space (frequency dependence has been dropped for brevity).   
\end{enumerate}
In gist, the quantum channel mapping the input to the output is $\mathcal{G}=\mathcal{D}_{\bm\nu}\circ\Phi_{\theta_{mm}}\circ\mathcal{L}_{\bm{\chi}_{mm}}^{N_T}$, where `$\circ$' means the concatenation of quantum channels. 
In deriving this result, we have assumed the axion field to be a classical, coherent field. However, it is easy to generalize this to a classical ensemble (associated with the stochastic evolution of the axion field in phase space), by specifying a phase-space probability density function (PDF), $p(\bm\mu_s)$, for the axion field.

Utilizing the Gaussian formalism (see Appendices~\ref{appendix:gaussian} and \ref{appendix:cavity_model}), we find a general expression for the measurement-port output moments in terms of the input moments and the channel parameters,
\begin{equation}
 \bm{\mu}_m^{\rm (out)}=\abs{\bm{\chi}_{mm}}\bm{O}(\theta_{mm})\bm{\mu}_m^{\rm (in)} + \abs{\bm{\chi}_{ms}}\bm{O}(\theta_{ms})\bm{\mu}_s\label{eq:cavity_out_mu}
 \end{equation}
and
\begin{multline}
\bm{\sigma}_m^{\rm (out)}=\abs{\bm{\chi}_{mm}}^2\bm{O}(\theta_{mm})\bm{\sigma}_m^{\rm (in)}\bm{O}^\top(\theta_{mm})\\+N_T\left(1-\abs{\bm{\chi}_{mm}}^2\right)\mathbb{I}_2.\label{eq:cavity_out_sigma}
\end{multline}
For later reference, we write the relevant susceptibility coefficients in terms of the original coupling rates,
\begin{align} 
    \abs{\bm{\chi}_{mm}}^2&=1-\frac{\gamma_m\gamma_\ell+\gamma_m\gamma_s}{(\gamma/2)^2+\omega^2}\approx\frac{(\gamma_m-\gamma_\ell)^2/4+\omega^2}{(\gamma/2)^2+\omega^2},\label{eq:chimm_expanded}\\
    \abs{\bm{\chi}_{ms}}^2&=\frac{\gamma_m\gamma_s}{(\gamma/2)^2+\omega^2},\label{eq:chima_expanded}
\end{align}
where we have expanded $\gamma^2=(\sum_{j\in\{m,s,\ell\}}\gamma_j)^2$ and used the well-justified approximation $\gamma_s\ll\gamma_m,\gamma_\ell$. To be clear, the approximations are at $\order{\gamma_s^2}$. A plot of the susceptibility coefficients and the mixing-angles are shown in Fig.~\ref{fig:susc_plot} of Appendix~\ref{appendix:cavity_model}. We include a table of cavity-model parameters that we use throughout the manuscript in Table~\ref{tab:notation}.


\subsection{Revisiting the Dicke radiometer equation and the scan rate}
\label{sec:dicke_scan_rate}

In this section we derive the SNR for an axion search (known as the Dicke radiometer equation~\cite{dicke1946radiometer}) and introduce the standard figure of merit for the search, the so-called scan-rate. For this purpose, we assume that the input to the cavity consists of only thermal fluctuations, which reduce to vacuum fluctuations at zero temperature. We extend this to other quantum inputs in later sections. After describing the detection methods, we formally introduce the scan-rate, which is the rate at which one can tune the cavity-resonance frequency in search for an axion-signal in frequency space. The scan-rate is an important figure of merit because the mass of axion is unknown, spanning a range covering at least three decades (or more), and thus scanning over such a large range as swiftly as possible is desirable.  

We compute the SNR of the power-spectral-density (PSD) in a homodyne detection scheme, when the measurement-port input consists of thermal/vacuum fluctuations only, i.e. $\bm{\mu}_m^{\rm(in)}=\bm0$ and $\bm\sigma_m^{\rm(in)}=N_T\mathbb{I}_2$. For completeness, we present the heterodyne result in Appendix~\ref{app:heterodyne}. 

\subsubsection{Within the axion coherence time}
We begin our analyses for detection within the coherence time of the axion-field, taking the axion-signal as unknown but coherent.
A homodyne detection scheme enables measurement of a single quadrature of an electromagnetic field, e.g., the $Q$-quadrature, at the fundamental quantum noise level. An optical homodyne measurement consists of mixing the signal mode with a strong local oscillator of the same frequency (i.e., a high-amplitude laser of known phase) at a balanced beam-splitter and measuring the difference in the intensities at each output-port of the beam-splitter. In the context of axion search in the microwave domain, a homodyne measurement differs from that in the optical domain. In a typical configuration for microwave homodyne measurement, a high-gain phase-sensitive amplifier first amplifies a selected quadrature without introducing additional noise, followed by a phase-insensitive amplifier that further boosts both quadratures. The amplified signal is then multiplied with a microwave local oscillator on a frequency mixer. A low-pass filter then rejects the high-frequency components of the frequency mixer's output, leaving the measured quadrature on the baseband signal.

If we assume that mixing with the local-oscillator only adds about $N_T$ amount of noise to the signal (which is approximately vacuum-dominant; $N_T\approx1+\order{10^{-4}}$), we can use Eqs.~\eqref{eq:cavity_out_mu} and~\eqref{eq:cavity_out_sigma} directly to find an expression for the homodyne SNR of the signal power containing the axion signature,
\begin{equation}
    \begin{split}
   {\rm SNR}^{\rm (hom)}\equiv\frac{\ev*{\hat{Q}_m}^2}{{\rm Var}(\hat{Q}_m)}&=\frac{\abs{\bm{\chi}_{ms}}^2\abs{\bm{\mu}_s}^2\cos^2(\phi_s + \theta_{ms})}{N_T}\\
   &=\frac{\gamma_m\gamma_s\abs{\bm{\mu}_s}^2\cos^2(\phi_s + \theta_{ms})}{N_T\left((\gamma/2)^2+\omega^2\right)},
    \end{split}\label{eq:homo_psd}
\end{equation}
where the quadrature variance ${\rm Var}(\hat{Q}_m)=N_T$, $\abs{\bm{\mu}_s}$ is the amplitude of the axion field, $\phi_s$ is the phase of the axion field with respect to the $Q$-quadrature (in polar coordinates), and $\theta_{ms}$ is a cavity-induced rotation angle. Here $\phi_s=\phi_a + \vec k_0 \cdot \vec x$, where $\vec x$ is the location of the measurement device. Eq.~\eqref{eq:homo_psd} represents the expected ``single-shot" SNR, which we expect to hold within an axion coherence time.

\subsubsection{Long integration times}
We now consider observation times $T_O\gg 1/\Delta_a$, where $\Delta_a$ is the axion bandwidth, and derive the SNR of the PSD at the cavity-detuned frequency $\omega$. We shall assume that there are (Shannon-Nyquist sampling) $M=2\Delta_a T_O\gg1$ independent and identically distributed samples of the power within a total time $T_O$, from which we may acquire a significant SNR via the law of large numbers.

Formally, denote the the measured (normalized) power spectral-density along the $Q$-quadrature as $\mathcal{P}$, which takes on a random value for each detection interval. As typical for DM-axion searches, we shall assume that, over many detection intervals, the classical axion-field undergoes a random walk about the origin in phase-space. The phase-space PDF $p(\bm\mu_s)$ is then a uniform, bi-variate Gaussian distribution with zero mean and uniform variance $\bm\sigma_s=2n_s\mathbb{I}_2$ (i.e. $n_s=\mathbb{E}[{\bm\mu_s^\top\bm\mu_s}]/2$ is the occupation-number spectral-density of the axion field; see Appendix~\ref{app:physical_property} for connection to physical parameters). Under this assumption, it is easy to see that the sample-averaged power, taken over $2T_O\Delta_a$ samples, is $\mathbb{E}[\mathcal{P}]=\abs{\bm{\chi}_{ms}}^2n_s$, which one can ascertain by inspection of Eq.~\eqref{eq:homo_psd}. Furthermore, due to the underlying Gaussian statistics of both the homodyne detection results and the phase-space PDF, we have that the power variance of an \textit{individual, random sample} is
\begin{equation}
\begin{split}
{\rm Var}(\mathcal{P})&=2\left({\rm Var}(\hat{Q}_{m})+\abs{\bm{\chi}_{ms}}^2n_s\right)^2\\
&\approx 2{\rm Var}^2(\hat{Q}_{m}),
\end{split}
\end{equation}
where ${\rm Var}(\hat{Q}_{m})$ can be taken directly from Eq.~\eqref{eq:homo_psd}. The first term is due to the variance in measuring $Q$ (including thermal/vacuum fluctuations) while the second term is the variance of the power due to the underlying phase-space PDF $p(\bm\mu_s)$. The factor of 2 out front is due to the fact that the (uni-variate) PDF for $Q$ is Gaussian and thus ${\rm Var}(\mathcal{P})\sim {\rm Var}(Q^2)= 2{\rm Var}^2(Q)$. In the approximation, we have omitted the signal's contribution to the variance due to its relative smallness. 


Combining these results with the assumption of Nyquist-Shannon sampling $M=2\Delta_aT_O$, we find an expression for the SNR about the detuned frequency $\omega$,
\begin{equation}
\overline{{\rm SNR}}^{\rm(hom)}\approx\underbrace{\frac{\gamma_m\gamma_s n_s}{N_T\left((\gamma/2)^2+\omega^2\right)}}_{\equiv\alpha_{\rm QL}(\omega)}\sqrt{\Delta_aT_O},\label{eq:alpha}
\end{equation}
where we have defined the quantum-limited (QL) visibility $\alpha_{\rm QL}(\omega)$, which refers to the intrinsic limit set by the vacuum fluctuations of the modes. The SNR is peaked on resonance and is given as,\footnote{See ref.~\cite{kim2020revisit} for a definition of the total signal-power (e.g., over a band of frequencies), which includes the finite bandwidth of the axion. The main results of that reference can likewise be computed using the quantum cavity-model above by specifying the lineshape of $n_s(\omega)$ and calculating the total output signal power.}
\begin{equation}
\begin{split}
    \overline{{\rm SNR}}^{\rm(hom)}_{\omega=0}&=\frac{4\gamma_m\gamma_s n_s}{N_T\gamma^2}\sqrt{\Delta_aT_O}\\
    &=\frac{\gamma_m/\gamma_\ell}{(\gamma_m/\gamma_\ell+1)^2}\frac{4\gamma_sn_s}{N_T \gamma_\ell}\sqrt{\Delta_aT_O}.
\end{split}
\end{equation}
This is just the Dicke radiometer equation~\cite{dicke1946radiometer}, which has peak sensitivity at critical coupling, $\gamma_m/\gamma_\ell=1$---a known result; see e.g. \cite{krauss1985,kim2020revisit}. [It is common, in the literature, to define the ratio $\beta\equiv\gamma_m/\gamma_\ell$, but we bypass this convention in the main text to avoid adding extra notation.]

Using similar logic as above, one can show that heterodyne has the same average SNR performance at the quantum limit. In other words, heterodyne and homodyne detection perform equally well when sampling a random signal (at least in this setting, where phase-insensitive amplifier noise is negligible; see Appendix~\ref{app:heterodyne}). We shall henceforth drop the superscript labeling the detection scheme and restrict ourselves primarily to homodyne measurements. Only homodyne measurement benefits from quantum squeezing and allows us to surpass the quantum limit.


\subsubsection{Introducing the scan rate}
We now review the (spectral) scan-rate. Our presentation follows a similar line of argument as that provided in Appendix A of ref.~\cite{malnou2019}, but we include it here for completeness.

Since the DM-axion's mass is unknown over a large frequency range, a more relevant quantity than the SNR at a given cavity-resonance setting is the frequency-integrated SNR, ${\rm SNR}_{\rm I}$, where the subscript ``I" indicates integration over many resonance-frequencies. Given that one spends a time $T_O$ at each resonance-frequency and takes infinitesimal steps $\varepsilon$ from one resonance-frequency to the next, the SNRs at each resonance-frequency step add in quadrature such that, upon using Eq.~\eqref{eq:alpha}, the SNR around a single (detuned) frequency $\omega$ is
\begin{equation}
\begin{split}
    {\rm SNR}_{\rm I}^{\,2}=\frac{\Delta_aT_O}{\varepsilon}\sum_{k=-n}^n\alpha^2(\omega; \omega_c+k\varepsilon)\varepsilon,\label{eq:scan_rate_disc}
\end{split}
\end{equation}
where $n$ is the number of tuning steps and $\alpha(\omega;\omega')$ is the visibility function for an arbitrary input state (not necessarily the vacuum) when the resonant frequency is $\omega'$. In other words, for a fixed frequency $\omega$---that we arbitrarily measure with respect to some central resonance-frequency, $\omega_c$, at $k=0$---we square and sum the independent contributions from each cavity resonance-frequency setting.

Ideally, we can tune the resonance frequency continuously, such that the ratio $\varepsilon/2\pi T_O$ converges to some (optimal) non-zero value, assuming some desired target SNR, $\zeta_{\rm SNR}$. The reason we expect this convergence is that, as the tuning-step becomes smaller and smaller (requiring also that $n$ becomes larger), we need to spend less and less time, $T_O$, at a given resonance-frequency, since contributions from the many tuning-steps, which are infinitesimally far away, will contribute significantly to the SNR. This leads to a natural definition of the spectral scan-rate, $\mathcal{R}$, via $\varepsilon/2\pi T_O\rightarrow\mathcal{R}\equiv\dv*{\nu_c}{t}$ in a continuum limit (to be defined precisely below), where $\dd\nu_c=\dd\omega_c/2\pi$ denotes an infinitesimal change in the resonance-frequency.\footnote{This definition of the scan-rate is \textit{a priori} dependent on the chosen frequency $\omega$, as a different amount of time is generally required to reach a target SNR, $\zeta_{\rm SNR}$, depending on the value of $\omega$. However, for the cavity setup, this frequency dependence drops out altogether, as we show explicitly in the main text.} 

We now impose a continuum limit on Eq.~\eqref{eq:scan_rate_disc} by taking $\varepsilon\rightarrow0$ and $n\rightarrow\infty$ whilst keeping the product $n\varepsilon$ (practically on the order of a few bandwidths) and the ratio $\mathcal{R}\equiv \varepsilon/2\pi T_O$ fixed. Doing so, we have that Eq.~\eqref{eq:scan_rate_disc} becomes,
\begin{equation}
    {\rm SNR}_{\rm I}^{\,2}=\frac{\Delta_a}{2\pi \mathcal{R}}\int_{-\Omega_{\rm max}}^{\Omega_{\rm max}}\alpha^2(\omega;\omega_c+\Omega)\dd\Omega.\label{eq:scan_intermediate}
\end{equation}
We now make a crucial observation: namely, $\alpha(\omega;\omega_c+\Omega)=\alpha(\omega-\Omega)$, where on the right-hand-side we omitted the notation about the fixed resonant frequency $\omega_c$. In other words, changing the resonance frequency, $\omega_c$, by an amount $\Omega$ is equivalent to fixing $\omega_c$ and evaluating the SNR at the detuned frequency $\omega-\Omega$. 

If we now implement a change of variables $\Omega^\prime\equiv\omega-\Omega$ in Eq.~\eqref{eq:scan_intermediate} and make the simplifying assumption $\Omega_{\rm max}\rightarrow\infty$, the $\omega$ dependence in the integral above drops out entirely. This substitution then reduces Eq.~\eqref{eq:scan_intermediate} to the average SNR of the grand spectrum. Therefore, assuming a target value for the SNR of the grand spectrum, $\zeta_{\rm SNR}$, we can solve for the scan-rate, $\mathcal{R}$, required to achieve the target, 
\begin{equation}
\begin{split}
    \mathcal{R}&=\frac{\Delta_a}{2\pi \zeta_{\rm SNR}^2}\int_{-\infty}^\infty\alpha^2(\Omega^\prime)\dd\Omega^\prime,\\
\end{split}\label{eq:scan_rate}
\end{equation}
which is a frequency-independent result. This result is fairly intuitive. For example, consider increasing the target SNR, $\zeta_{\rm SNR}$, while keeping all else fixed. This necessarily reduces the scan-rate, as more observation time at each resonance-frequency is required in order to reach the target.

We emphasize that Eq.~\eqref{eq:scan_rate} applies for general input quantum-states (upon substituting a proper visibility, $\alpha(\omega)$, which is dependent on the quantum state of the modes). For vacuum input, we take the quantum-limited visibility, $\alpha_{\rm QL}(\omega)$, from Eq.~\eqref{eq:alpha} to obtain the quantum-limited scan-rate,
\begin{equation}
\begin{split}
    \mathcal{R}_{\rm QL}
    &\approx \frac{2\Delta_an_s^2\gamma_s^2}{\zeta_{\rm SNR}^2N_T^2\gamma_\ell}\frac{\left(\frac{\gamma_m}{\gamma_\ell}\right)^2}{\left(\frac{\gamma_m}{\gamma_\ell}+1\right)^3},
\end{split}\label{eq:rate_vac}
\end{equation}
where we have assumed that $N_T\approx{\rm constant}$ over the integration range and have used $\gamma\approx\gamma_m+\gamma_\ell$ (ignoring the $\gamma_s$ contribution). The optimal value for the scan-rate $\mathcal{R}^\star_{\rm QL}$ is found at the over-coupled parameter setting $\gamma_m/\gamma_\ell=2$, a known result in the $\Delta_a\ll\gamma_\ell$ regime~\cite{krauss1985,Chaudhuri:2018rqn,kim2020revisit,chaudhuri2021optimal}. This parameter setting differs from critical-coupling, $\gamma_m/\gamma_\ell=1$, where the SNR at zero detuning is maximal. The reason behind this difference is due to the trade-off between bandwidth and sensitivity when searching for an axion signal. Since the frequency-range over which we expect to find an axion-signal is large, we require that our cavity be fairly sensitive over a large bandwidth, however increasing the bandwidth comes at the price of decreasing the peak-sensitivity of the cavity (see Fig.~\ref{fig:susc_plot} in Appendix~\ref{appendix:cavity_model} for a visualization of this trade-off). The scan-rate is a good figure of merit which takes both the bandwidth and sensitivity into account.



\subsection{Squeezing-enhanced dark matter search}
\label{sec:squeezing_DM_search}

The final part of the preliminary section is devoted to summarizing a squeezing-enhanced scan introduced in ref.~\cite{girvin2016axdm} and developed and implemented in refs.~\cite{malnou2019,backes2021}. We provide a qualitative description as to how squeezing can help a DM search and then, for completeness, quote the results for the squeezing-enhanced scan-rate, $\mathcal{R}_{\rm sq}$, while leaving detailed derivations to Appendix~\ref{app:squeezing}.

{Ideally, we want good sensitivity over a large bandwidth in order to quickly scan frequency space for a DM signal. For a quantum limited setup (referring to a classical setup where the noise is dominated by vacuum fluctuations of the microwave fields), there is a sensitivity-bandwidth tradeoff, which ultimately limits a DM search performance. [This is quantified by the optimal quantum limited scan-rate, $\mathcal{R}_{\rm QL}^\star$; see discussion surrounding Eq.~\eqref{eq:rate_vac}.] For a squeezing-enhanced setup, it turns out that squeezing \emph{cannot} change the peak sensitivity of the microwave cavity receiver, which is set by the on-resonance sensitivity in the quantum-limited regime (see below and Appendix~\ref{app:squeezing} for more details). However, a key point is that squeezing \emph{can} increase the effective bandwidth of the cavity receiver \emph{without} sacrificing sensitivity. In succinct terms, squeezing allows for a more effective DM search by bypassing the bandwidth-sensitivity tradeoff set by quantum-limited setups~\cite{malnou2019,backes2021}. A mathematical analysis of such enhancement is just below.}

\begin{figure}[t]
    \centering
    \includegraphics[width=\linewidth]{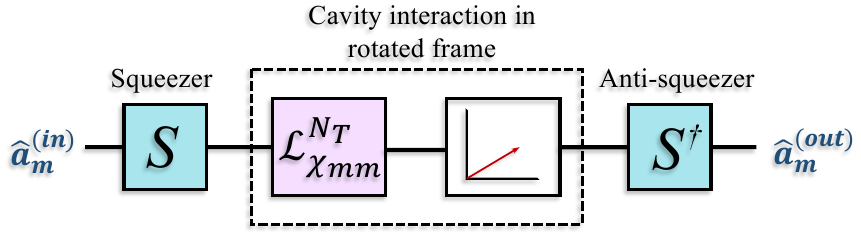}
    \caption{Squeezing the measurement port prior to the cavity interaction reduces input fluctuations in one quadrature. Anti-squeezing after the cavity interaction amplifies the axion-induced displacement in one quadrature (relative to, e.g., noise added after the cavity interaction by phase-insensitive amplification).}
    \label{fig:squeeze_cavity_channel}
\end{figure}

The squeezing setup introduced in ref. \cite{malnou2019} is shown in Fig.~\ref{fig:squeeze_cavity_channel}. A squeezed-vacuum is prepared and injected into the cavity, reducing input noise fluctuations along the squeezed quadrature. For instance, a (noisy) squeezed-vacuum, with squeezing along the $Q$ quadrature, has a covariance matrix $\bm\sigma_{m}^{(\rm in)}=N_T{\rm diag}(1/G, G)$, where $G$ is the gain of the squeezer, $N_T$ represents the initial noise fluctuations (including vacuum and thermal fluctuations), and `diag' denotes a diagonal matrix specified by the diagonal elements. Post-cavity interaction, an anti-squeezer amplifies the output signal along the initially squeezed quadrature. This is beneficial considering the potential noise added in signal-processing. As discussed in refs.~\cite{girvin2016axdm,malnou2019}, squeezing does not increase the peak-sensitivity on resonance but instead enhances the off-resonance sensitivity---resulting in an effectively increased bandwith of the cavity receiver which, in turn, yields an accelerated scan-rate proportional to the amount of squeezing (see below). 

In Appendix~\ref{app:squeezing}, we derive the SNR (and thus the visibility function) for squeezed-vacuum input. Using the general Eq.~\eqref{eq:scan_rate}, we then compute the  squeezing-enhanced scan-rate, $\mathcal{R}_{\rm sq}$. A natural figure of merit is the ratio of the squeezing-enhanced scan-rate, $\mathcal{R}_{\rm sq}$, to the quantum-limited scan-rate, $\mathcal{R}^\star_{\rm QL}$ [see discussion surrounding Eq.~\eqref{eq:rate_vac} for precise definition of $\mathcal{R}^\star_{\rm QL}$]. We compute this ratio to be,
\begin{equation}
    \frac{\mathcal{R}_{\rm sq}}{\mathcal{R}^\star_{\rm QL}}=\frac{27\sqrt{G}\left(\frac{\gamma_m}{\gamma_\ell}\right)^2}{32\left(\frac{\left(\frac{\gamma_m}{\gamma_\ell} -1\right)^2}{4G}+\frac{\gamma_m}{\gamma_\ell}\right)^{3/2}},\label{eq:scan_rate_sq}
\end{equation}
in agreement with ref.~\cite{malnou2019}. This quantity has a maximum corresponding to an optimal coupling of $\gamma_m$, which is around $\gamma_m/\gamma_\ell\approx 2G$. In the limit of $G\gg1$ and at optimal coupling, $\gamma_m/\gamma_\ell\approx 2G$, the optimal squeezing-enhanced scan-rate, $\mathcal{R}_{\rm sq}^\star$, approaches $\mathcal{R}_{\rm sq}^\star/\mathcal{R}_{\rm QL}^\star\approx 0.7G$. Thus the scan-rate (approximately) scales linearly with the squeezing gain.

Before moving forward, let us make a few final comments. First, if measurement noise (originating from, e.g., phase-insensitive amplification prior to detection) is not too large, then we can omit the anti-squeezer in the process (shown in Fig.~\ref{fig:squeeze_cavity_channel} as $\bm S^\dagger$) without loss of generality, as including such does not alter the performance of the setup. From here on, unless otherwise stated, we shall assume amplifier noise in the detector setup is negligible and thus omit the anti-squeezer. As a matter of practice though, some form of amplification prior to measurement is typically necessary to transform a very weak or quantum-limited signal into something that is classically detectable. The anti-squeezer (or more generally, a phase-sensitive amplifier) is thus a practical necessity for homodyne detection. Furthermore, a phase-\textit{insensitive} amplifier cannot be used as a substitute for the phase-\textit{sensitive} amplifier in this scenario, as the former generally degrades the performance of homodyne detection, even in the presence of squeezing. For more discussion on the effects phase-insensitive amplification to homodyne detection, see Appendix~\ref{app:heterodyne}.

Second, a more practical (though formally equivalent) implementation than the single-mode squeezed-vacuum setup considered here is to use two-mode squeezing generated by a Josephson parametric amplifier (JPA) (see, for instance, refs.~\cite{malnou2018jpa,backes2021}). Using two-mode squeezing can naturally resolve the cavity-induced phase $\theta_{mm}$. Therefore, frequency-dependent squeezing is unnecessary in the microwave cavity setup to achieve a quantum advantage. For completeness, we analyze the two-mode squeezing setup in Appendix~\ref{app:squeezing} and show its equivalence in performance to the single-mode setup discussed here. 



\section{Entangled sensor-networks for dark matter search}\label{sec:multi_sensor}


We extend the single-sensor model to a sensor-network consisting of $M$ cavities positioned at spatially distinct locations, each of which couples to the same background axion field. In general, the cavities can be in close proximity or well-separated, but we primarily focus on a local sensor network, as depicted in Fig.~\ref{fig:cavity_network_pic}. %
This choice allows us to neglect
 the position-dependent phase $\vec k_0\cdot \vec x$ and hence maintain coherence among the sensors.



Define the set of $M$ measurement-port input modes $\{\hat{a}^{\rm(in)}_{m_i}\}_{i=1}^M$, where the subscript $i$ refers to the $i$th sensor-cavity. The measurement-port output modes are likewise defined, i.e., $\{\hat{a}^{\rm(out)}_{m_i}\}_{i=1}^M$. Then, within the axion coherence time, the quantum channel $\mathcal{G}_M$ mapping the set of input modes to the set of output modes is given by a tensor product of the individual cavity channels, 
\begin{equation}
    \mathcal{G}_M=\bigotimes_{i=1}^M \mathcal{D}_{\bm\nu_i}\circ\Phi_{\theta_{m_im_i}}\circ\mathcal{L}_{\bm{\chi}_{m_im_i}}^{N_{T_i}}.
\end{equation}
The subscript $i$ here not only labels the individual cavities but also signifies the fact that each cavity may have different operating conditions, loss rates, resonance frequencies, etc.

\begin{figure}[t]
    \centering
    \includegraphics[width=\linewidth]{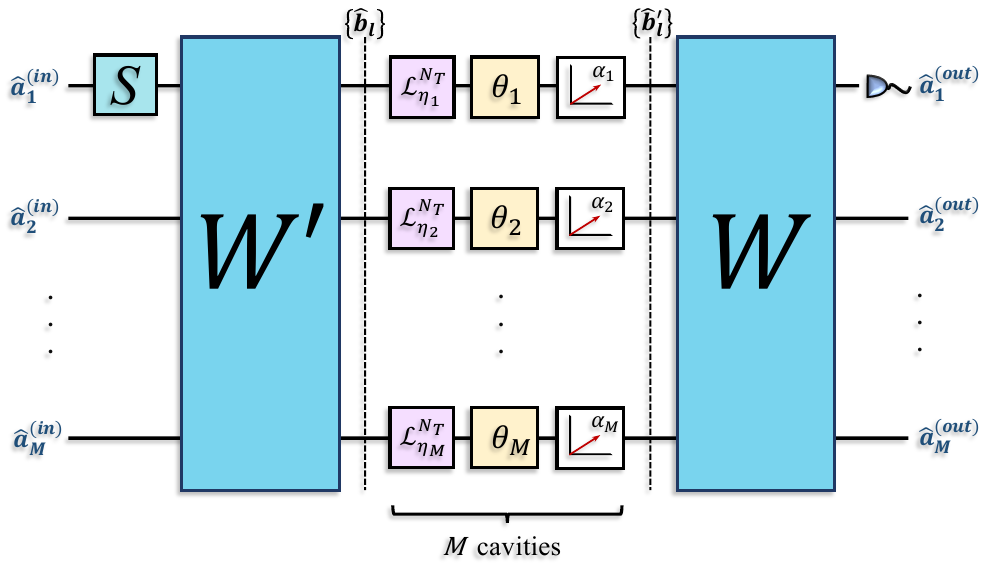}
    \caption{DQS setup. A single-mode squeezed vacuum is distributed to an array of $M$ cavities which are coupled via passive linear networks, $\bm{W}^\prime$ and $\bm{W}$. The network utilizes classical correlations between the axion-induced displacements at each cavity to coherently combine the signal fields into the primary output mode, $\hat{a}_1^{\rm (out)}$, and generate a larger signal amplitude. The linear networks are optimized to maximize the signal and minimize the noise in the primary mode. We have relabeled parameters in the diagram for brevity: $\eta_k\equiv\abs{\bm\chi_{m_km_k}}^2$; $\theta_k\equiv\theta_{m_km_k}$; and $\alpha_k$ is the complex amplitude of the output signal (i.e., the axion-induced displacement) of the $k$th cavity.}
    \label{fig:correlated_rand_walk}
\end{figure}

We have thus reduced axion detection to a model of displacement sensing with a sensor network, a topic generally studied in the field of DQS~\cite{zhuang2018DQSCV,zhang2021dqs}, where continuous-variable multi-partite entanglement---generated by passing a single-mode squeezed-vacuum through a linear network---plays an crucial role. Utilizing techniques from Distributed Quantum Sensing (DQS)~\cite{zhuang2018DQSCV,zhang2021dqs}, we consider the situation depicted in Fig.~\ref{fig:correlated_rand_walk}, where a squeezed-vacuum state is distributed to $M$ local sensor-cavities, which are coupled via passive, linear networks, $\bm W^\prime$ and $\bm{W}$.\footnote{In this section, we ignore the anti-squeezer at the end of the protocol for the sake of brevity, as including such does not change the main results. } The linear network, $\bm W$, can be replaced by local homodyne measurements and post-processing to achieve the same performance. 

We often refer to the setup depicted in Fig.~\ref{fig:correlated_rand_walk} as a DQS setup, since we generate quantum entanglement by sending the squeezed vacuum through the linear network, $\bm W^\prime$. Later in the paper, we shall also refer to a Quantum Limited, Distributed Classical Sensing (QL-DCS) setup, where the input radiation consists of only vacuum fluctuations but joint post-processing of the output signals is allowed.

Given the DQS setup of Fig.~\ref{fig:correlated_rand_walk}, our goal is to maximize the scan-rate of Eq.~\eqref{eq:scan_rate} by assuming control of the passive linear networks, $\bm W^\prime$ and $\bm W$, and by taking advantage of input quantum resources as well as the classical correlations of the axion signals between individual cavities. In doing so (details provided below), we arrive at our main result:

    \emph{\bf Main result---} In the ideal scenario of a susceptibility-matched DQS network consisting of $M$ identical microwave cavities, the scan-rate of the network scales as $M^2\mathcal{R}_{\rm sq}$, where $\mathcal{R}_{\rm sq}$ is the scan-rate of a single squeezed cavity with the same level of squeezing. This scaling is achieved by using one squeezer and one homodyne detector---all situated on the ``primary" mode, $\hat{a}_1$---as well as balanced $M$-mode linear networks, $\bm W$ and $\bm W^\prime=\bm W^{-1}$.

Thus, in the ideal case, we achieve a performance enhancement of $M^2$ by operating the network coherently and making use of the classical correlations of the axion field (i.e., via joint post-processing). Moreover, we obtain a simultaneous boost to the scan-rate (relative to the quantum limited case) via multi-partite entanglement shared between the cavities (generated from a single squeezed vacuum). We also extend these results to a susceptibility-mismatched array of non-identical cavities and discuss subtleties that arise therein. Main derivations and discussions leading to these results are just below.

For simplicity, we shall assume each cavity operates at roughly the same temperature $T$, such that $N_{T_i}\approx N_T$, and for brevity, we introduce the temporary notation: $\eta_k\equiv\abs{\bm\chi_{m_km_k}}^2$ and  $\alpha_k$ is defined as the complex amplitude of the output signal such that $\abs{\alpha_k}\equiv\abs{\bm\chi_{m_ks_k}}\abs{{\bm \mu}_{s_k}}$ and $\arg(\alpha_k)\equiv\phi_{s_k}+\theta_{m_ks_k}$. We assume the phases $\theta_{m_km_k}$ can be practically resolved by utilizing the two-mode squeezing setup discussed in Appendix~\ref{app:resonance_frequency_why}; we thus ignore the phases $\theta_{m_km_k}$ from hereon. Furthermore, we shall assume that the radius of the network volume is much smaller than the wavelength of the axion field, implying that the axion field at all sensors are homogeneous, i.e., the amplitude $\abs{{\bm \mu}_{s_k}}=\mu_{s}$ and phase $\phi_{s_k}=\phi_{s}$, $\forall\,k$. We therefore ignore spatial variations of the axion field across the network.\footnote{In cases there the relative phase between difference cavities $\vec k_0\cdot \vec x$ cannot be ignored, new protocols shall be investigated for efficient scan.}

Our derivations follow by stepping through the circuit illustrated in Fig.~\ref{fig:correlated_rand_walk}, with focus on the primary mode $\hat{a}_1$, which we assume is in a (noisy) squeezed-vacuum state and squeezed along the $\Re(\hat{a}_1^{\rm(in)})$ quadrature. The other input modes are quiet and, thus, only populated by uncorrelated thermal/vacuum fluctuations. 

Consider the set of modes $\{\hat{b}_l\}$ just after the linear network $\bm W^\prime$. The transformation from the input modes $\{\hat{a}_k^{\rm(in)}\}$ to these intermediary modes is dictated by the coefficients of the network (the ``weights"), $\{w_{lk}^\prime\}$, which obey the orthogonality relation $\sum_{k}w_{mk}^\prime w_{kn}^{\prime\,*}=\delta_{mn}$. In terms of the input modes, the intermediary mode $\hat{b}_l$ can be written as,
\begin{equation}
    \hat{b}_l=\sum_{k=1}^M w^\prime_{lk}\hat{a}_k^{\rm(in)}=w^\prime_{l1}\hat{a}_1^{\rm(in)}+\sqrt{1-\abs{w^\prime_{l1}}^2}\hat{e}_l^{\rm(in)},\label{eq:b_modes}
\end{equation}
where we have singled-out the primary mode $\hat{a}_1^{\rm (in)}$ and defined the input ``environmental mode" $\hat{e}_l^{\rm(in)}=\sum_{k=2}^Mw^\prime_{lk}/\sqrt{1-\abs{w^\prime_{l1}}^2}\hat{a}_k^{\rm(in)}$, which is populated by the uncorrelated thermal/vacuum fluctuations of the remaining input modes, $\{\hat{a}_k^{\rm(in)}\}_{k=2}^M$. 

Now consider the second set of intermediary modes $\{\hat{b}_l^\prime\}$ just after the cavity interaction but before the linear network $\bm W$. Defining the environmental modes $\{\hat{e}_l\}$, which are introduced from cavity transmission loss, the intermediary mode $\hat{b}_l^\prime$ can be written as,
\begin{equation}
    \hat{b}_l^\prime = \sqrt{\eta_l}\hat{b}_l+\sqrt{1-\eta_l}\hat{e}_l+\alpha_l,\label{eq:b_prime_modes}
\end{equation}
where $\hat{b}_l$ is taken from Eq.~\eqref{eq:b_modes} and $\alpha_l$ denotes the axion-induced displacement at the $l$th cavity. Next, we introduce the weights $\{w_{jl}\}$, which likewise obey an orthogonality relation $\sum_{k}w_{mk} w_{kn}^*=\delta_{mn}$. We then obtain a formal relation for the output modes, ${\hat{a}_j^{\rm(out)}=\sum_{l=1}^Mw_{jl}\hat{b}_l^\prime}$. Combining this relation with Eqs.~\eqref{eq:b_modes} and \eqref{eq:b_prime_modes} and choosing $j=1$, we find a generic expression for the output primary-mode of the network,
\begin{widetext}
\begin{equation}
    \hat{a}_1^{\rm(out)}=\sum_{l=1}^M\bigg(\underbrace{\left(w_{1l}w_{l1}^\prime\sqrt{\eta_l}\right)\hat{a}_1^{\rm(in)}}_{\text{Squeezed noise}}+\underbrace{w_{1l}\sqrt{1-\abs{w_{l1}^\prime}^2}\sqrt{\eta_l}\hat{e}_l^{\rm(in)}+w_{1l}\sqrt{1-\eta_l}\hat{e}_l}_{\text{Thermal/vacuum noise}}+\underbrace{w_{1l}\alpha_l}_{\text{Signal}}\bigg),\label{eq:a1_out}
\end{equation}
\end{widetext}
where we have identified the parts which contribute to the signal and to the noise. Performing a homodyne measurement along the real quadrature of $\hat{a}_1^{\rm(out)}$ then gives an estimate of the signal power and the noise power. Hence, understanding and manipulating this mode relation is of primary significance to our forthcoming analyses.

From here, the objective is to optimize the weights, $w$ and $w^\prime$, in order to maximize the scan-rate given by homodyne measurement. As the exact optimization is challenging, we take a two-step approach by first maximizing the signal and then minimizing the noise in the output. A heuristic solution to the optimization for identical sensors is just below. A more formal derivation of the optimization strategy (applied to the general case of non-identical sensors) is given in Appendix~\ref{app:opt_weights}. In the identical sensor case, the two-step optimization solution is the exact solution due to symmetry; while in the non-identical sensors case, we numerically show that the solution from the two-step optimization is close to the exact optimal.

\subsection{Identical sensors}
\label{sec:entangled_sensor_identical}

For identical sensor-cavities, $\eta_l=\eta$ and $\alpha_l=\alpha$ $\forall\,l$; thus, we have from Eq.~\eqref{eq:a1_out} 
\begin{multline}
    \hat{a}_1^{\rm(out)}= \sqrt{\eta}\left(\sum_{l=1}^M w_{1l}w^\prime_{l1}\right)\hat{a}_1^{\rm(in)}+\alpha\left(\sum_{l=1}^M w_{1l}\right)\\+\,\text{thermal noise}.\label{eq:a1_ideal}
\end{multline}
The signal amplitude along the homodyne-measurement direction is given by
\begin{equation}
    \ev{\Re\left(\hat{a}_1^{\rm(out)}\right)}= \abs{\alpha}\sum_{l=1}^M\abs{w_{1l}}\cos\big(\arg w_{1l}+\arg\alpha\big).
\end{equation}
Obviously, the amplitude is maximized for ${\arg w_{1l}=-\arg\alpha}$, which aligns the quadrature measurement along the direction of the axion-field displacement. However, $\arg\alpha=\phi_s+\theta_{ms}$, where $\phi_s$ is the randomly fluctuating (and presumably unknown) phase of the axion field. Hence, no choice of $\arg w_{1l}$---other than the sensor-independent choice $\arg w_{1l}=\arg w_1\,\forall\,l$---is beneficial for the output signal if $\phi_s$ is unknown. In this sense, an arbitrary identical phase can be chosen for the weights $\arg w_{1l}$. 

On the other hand, the magnitude of the weights $\abs{w_{1l}}$ should take on a specific value. Since each cavity is displaced by an equivalent amount $\alpha$, no cavity is preferred, and thus it is necessary to choose uniform weights, $\abs{w_{1l}}=1/\sqrt{M}$. Assuming uniform weights, taking $\arg w_{1l}=0\,\forall\,l$, using the definitions $\abs{\alpha_l}\equiv\abs{\bm\chi_{m_la_l}}\mu_{s}$ and $\arg\alpha\equiv\phi_s$, and averaging over the coherence time of the axion-field leads to an expression for the average signal power,
\begin{equation}
\begin{split}
    &\mathbb{E}\left[\ev{\Re\left(\hat{a}_1^{\rm(out)}\right)}^2\right]
    \\
    &\quad=\mathbb{E}\left[\left(\abs{\alpha}\sum_{l=1}^M\abs{w_{1l}}\cos\big(\arg\alpha\big)\right)^2\right]\\ 
    &\quad=\mathbb{E}\left[\left(\abs{\bm\chi_{ms}}\mu_s\sum_{l=1}^M \cos\left(\phi_s+\theta_{ms}\right)/\sqrt{M}\right)^2\right]\\
    &\quad=\left(\sum_{l=1}^M1/\sqrt{M}\right)^2\abs{\bm\chi_{ms}}^2\mathbb{E}\left[\mu^2_a\cos^2\left(\phi_s+\theta_{ms}\right)\right]\\
    &\quad=M\abs{\bm{\chi}_{ms}}^2n_s,\label{eq:signal_dqs}
\end{split}
\end{equation}
where we have used the fact that $\mathbb{E}[\mu_s^2\cos^2(\phi_s+\theta_{ms})]=n_s$, for any angle $\theta_{ms}$ that is independent of the random variable $\phi_s$. This result is just $M$ times the signal power of a single cavity, which is intrinsically derived from the classical correlations of the axion-field displacements at the various sensors. [Such scaling is not permissible with independently operated sensors, which do not take advantage of the classical correlations of the field; see Fig.~\ref{fig:scanrate_dqs} for how this affects the scan-rate.]

To minimize the noise power, we have to optimally utilize the squeezing injected into the $\hat{a}_1^{\rm(in)}$ mode. In other words, we must ensure that (1) all of the squeezing is along the direction of the homodyne measurements, $\Re(\hat{a}_1^{\rm(out)})$, and (2) the squeezing is distributed properly to all the sensor cavities. Now we previously assumed that the phases $\theta_{m_km_k}$ can be practically resolved via the two-mode squeezing setup laid out in Appendix~\ref{app:resonance_frequency_why}. We thus justifiably ignored these phases in our analysis. This (partially) ensures no excess noise from anti-squeezing will appear in the measured quadrature, however we must also choose $\arg w_{l1}^\prime=-\arg w_{1l}$ when dividing the input and combining the signals to further avoid anti-squeezing noise. These observations follow by inspection of first term Eq.~\eqref{eq:a1_ideal}. To accomplish (2), we choose uniform weights, $\abs{w_{l1}^\prime}=1/\sqrt{M}$, for the linear network $\bm W^\prime$, since all cavities perform equally well and should thus get an equal share of squeezing. Observe that, with these choices, $\bm W^\prime = \bm W^{-1}$. Considering that the $\hat{a}_1^{\rm(in)}$ mode is in a (noisy) squeezed-vacuum state and environmental modes are filled with thermal/vacuum fluctuations, the noise power becomes
\begin{equation}
\begin{split}
{\rm Var}\left(\Re\left(\hat{a}_1^{\rm(out)}\right)\right)&=N_T\left(\eta/G + 1-\eta\right)\\
&=N_T\left(\abs{\bm\chi_{mm}}^2/G +1-\abs{\bm\chi_{mm}}^2\right),\label{eq:noise_dqs}
\end{split}
\end{equation}
which, remarkably, is just the squeezed noise power of a single cavity.

Combining Eqs.~\eqref{eq:signal_dqs} and \eqref{eq:noise_dqs}, and substituting the explicit expressions \eqref{eq:chima_expanded} and \eqref{eq:chimm_expanded} for $\bm\chi_{ms}$ and $\bm\chi_{mm}$, we obtain the SNR after integrating over many axion coherence-times,
\begin{equation}
\begin{split}
    \overline{\rm SNR}_{M;\,{\rm ideal}}&=\frac{\mathbb{E}\left[\ev{\Re\left(\hat{a}_1^{\rm(out)}\right)}^2\right]}{{\rm Var}\left(\Re\left(\hat{a}_1^{\rm(out)}\right)\right)}\sqrt{\Delta_aT_O}\\
    &=\frac{M\abs{\bm{\chi}_{ms}}^2n_s}{N_T\left(\abs{\bm\chi_{mm}}^2/G +1-\abs{\bm\chi_{mm}}^2\right)}\sqrt{\Delta_aT_O}\\
    &=\frac{M\gamma_m\gamma_s n_s}{N_T\left(\frac{(\gamma/2)^2 + \omega^2-\gamma_m\gamma_\ell)}{G}+\gamma_m\gamma_\ell\right)}\sqrt{\Delta_aT_O},\label{eq:dqs_snr}
\end{split}
\end{equation}
which is $M$ times the SNR of a single cavity when a (noisy) squeezed-vacuum state is injected into it; see Eq.~\eqref{eq:sq_snr}. Therefore, given the squeezing-enhanced scan-rate for a single cavity, $\mathcal{R}_{\rm sq}$ from Eq.~\eqref{eq:scan_rate_sq} and the definition of scan rate in Eq.~\ref{eq:scan_rate}, an $M$-cavity scan-rate of $M^2\mathcal{R}_{\rm sq}$ is achievable, as claimed. 

We want to emphasize that such a performance is achieved by utilizing a {\em single} squeezed-vacuum input, which is split into equal copies to entangle the $M$ sensors. Another approach to achieve the same performance with separable sensors would require {\em $M$ copies} of squeezed vacuum together with coherent post-processing of the signals. In this sense, our proposed distributed sensing scheme reduces the number of squeezers from $M$ to one, at the cost of requiring tunable, passive, linear couplings (e.g., $\bm W^\prime$) between the input/output microwave fields.

\subsection{Non-identical sensors with the same resonance frequencies}
\label{sec:entangled_sensor_non_identical}

Ideally, we want the scenario described just above---a coherent quantum sensor-network consisting of many identical copies of a single, spectacular cavity. However, practical realizations are often far from this ideality and discrepancies between sensors naturally arise. Any such differences---originating from, e.g., differing intrinsic quality-factors, axion-photon conversion rates, etc.---generally cause a susceptibility-mismatch between the cavities, leading to varying signal and noise outputs across the network. A further detriment is the fact that the output signal-amplitudes will have relative phases due to the susceptibility-mismatch, quantified by differences in the cavity-induced angles $\{\theta_{m_ks_k}\}$, causing non-optimal interference of the amplitudes when attempting to combine the signals.

Here, we discuss this more general case of non-identical sensors and systematically analyze optimization of the linear networks, $\bm W$ and $\bm W^\prime$, which maximize the signal and minimize the noise in the output power. 
In what follows, we shall assume that each cavity within the sensor network has the same resonance frequency but differ in other aspects, e.g., by their intrinsic quality factors. Identical resonance frequency is not only optimal for enhancing the scan-rate, but also required so that cavity-induced phase shifts can be resolved by matching the center frequency of a JPA squeezed-source to the resonance frequencies (see Appendix~\ref{app:resonance_frequency_why}). We remark that the effects of resonance frequency fluctuations on network performance (assessed by the SNR) has been addressed in ref.~\cite{jeong2018concept} for an analogous setup of a multi-cell cavity.

\subsubsection{Near-optimal weights: theoretical analysis}
The expression to consider is the output amplitude $\hat{a}_1^{\rm(out)}$ of Eq.~\eqref{eq:a1_out}. From the discussions in the identical-sensors case, it should be apparent that, to maximize the signal power, we must maximize the signal-amplitude, which is a weighted sum of the amplitudes from all the cavities, with respect to the weights $\{w_{1k}\}$. To minimize the noise, we must make optimal use of the squeezing injected into the $\hat{a}_1^{\rm(in)}$ mode by optimizing with respect to the weights $\{w_{1k}^\prime\}$. We first maximize the signal along the direction of the homodyne measurement. Then, conditioned on the signal optimization, we minimize the noise. Detailed derivations are supplied in Appendix~\ref{app:opt_weights}. Results of the derivations and brief explanations are given below.

To maximize the signal-power in the primary mode $\hat{a}_1^{\rm(out)}$, we appropriately combine the signal-amplitudes from each cavity based on their relative contributions to the signal as well as resolve any potential phase differences between amplitudes which would otherwise lead to destructive interference. Such an optimization (explicitly provided in Appendix \ref{app:opt_weights}) leads to the following expression for the weights, 
\begin{equation}
    w_{1k}=\frac{\abs{\bm\chi_{m_ks_k}}}{\sqrt{\sum_{j=1}^M\abs{\bm\chi_{m_js_j}}^2}}\e^{-\im\theta_{m_ks_k}}.\label{eq:general_w}
\end{equation}
For identical sensors, $\abs{\bm\chi_{m_ks_k}}=\abs{\bm\chi_{ms}}$, implying that $w_{1k}=1/\sqrt{M}$, as shown in the previous section.

From here, we can calculate the signal power. First, define the quantity,
\begin{equation}
\begin{split}
   \ev{\!\!\ev{\abs{\bm{\chi}_{ms}}^2}\!\!}_{1/M}&\equiv \frac{\left(\sum_{k=1}^M\abs{w_{1k}}\abs{\bm\chi_{m_ks_k}}\right)^2}{M}\\
   &=\frac{\sum_{k=1}^M\abs{\bm\chi_{m_ks_k}}^2}{M},
  \end{split}
\end{equation}
which [after using Eq.~\eqref{eq:general_w}] is the uniform average of $\abs{\bm\chi_{ms}}^2$, due to this particular choice of weights. The signal power is then,
\begin{equation}
\begin{split}
    \mathbb{E}\left[\ev{\Re\left(\hat{a}_1^{\rm(out)}\right)}^2\right]&= \mathbb{E}\left[\left(\sum_{k=1}^M\abs{w_{1k}}\abs{\bm\chi_{m_ks_k}}\right)^2\mu_s^2\cos^2\phi_s\right]\\
    &=M\ev{\!\!\ev{\abs{\bm{\chi}_{ms}}^2}\!\!}_{1/M} n_s,\label{eq:signal_w}
\end{split}
\end{equation}
which is just $M$ times the average signal-power of a cavity in the network. Furthermore, this result is provably better than uniformly combining the signals. Indeed, if we chose uniform weights ($\abs{w_{1k}}=1/\sqrt{M}$), then the signal power would scale as $\ev{\!\ev{\abs{\bm{\chi}_{ms}}}\!}^2_{1/M}$, which is always less than or equal to $\ev{\!\!\ev{\abs{\bm{\chi}_{ms}}^2}\!\!}_{1/M}$.

To minimize the noise, we make optimal use of squeezing. Our attention is thus on the first term in Eq.~\eqref{eq:a1_out}---the squeezed noise. We must ensure that the squeezing is distributed to (1) modes with the highest cavity transmission, such that a maximal amount of squeezing is utilized, and (2) modes which contribute most to the output signal. This draws us to the following choice for the weights $w_{k1}^\prime$ (again, see Appendix \ref{app:opt_weights} for details),
\begin{equation}
    w_{k1}^\prime = \frac{\abs{\bm\chi_{m_ks_k}}\abs{\bm\chi_{m_km_k}}}{\sqrt{\sum_{j=1}^M\abs{\bm\chi_{m_js_j}}^2\abs{\bm\chi_{m_jm_j}}^2}}\e^{\im\theta_{m_ks_k}}.\label{eq:general_wprime}
\end{equation}

From the above equation, we can calculate an expression for the noise power. Before doing so, recall $\eta_l\equiv\abs{\bm\chi_{mm}}^2$ and define the quantity,
\begin{align}
    \ev{\!\!\ev{\abs{\bm{\chi}_{mm}}^2}\!\!}_{w^2}&\equiv \left(\sum_{l=1}^M\abs{w_{1l}}\abs{w_{l1}^\prime}\sqrt{\eta_l}\right)^2
    \nonumber
    \\
    &= \frac{\sum_{l=1}^M \abs{\bm\chi_{m_la_l}}^2\abs{\bm\chi_{m_lm_l}}^2}{\left(\sum_{j=1}^M\abs{\bm\chi_{m_js_j}}^2\right)},\label{eq:chi_w}
\end{align}
where the equality follows by substituting Eqs.~\eqref{eq:general_w} and \eqref{eq:general_wprime} into the above. Observe that $\ev{\!\!\ev{\abs{\bm{\chi}_{mm}}^2}\!\!}_{w^2}$ is the average of $\abs{\bm\chi_{mm}}^2$ with respect to the distribution $\abs{w_{1l}}^2$ (hence the subscript, $w^2$).
Using this definition along with Eq.~\eqref{eq:a1_out}, we find, after some algebra, the total noise power, 
\begin{align}
    &{\rm Var}\left(\Re\left(\hat{a}_1^{\rm(out)}\right)\right)
    \nonumber
    \\
    &\qq{} =N_T\Big( \ev{\!\!\ev{\abs{\bm{\chi}_{mm}}^2}\!\!}_{w^2}/G+1-\ev{\!\!\ev{\abs{\bm{\chi}_{mm}}^2}\!\!}_{w^2}\Big).\label{eq:noise_ww}
\end{align}
The first term is due to squeezing of the $\hat{a}_1^{\rm(in)}$ input mode, while the second and third terms follow from the fact that the other $M-1$ uncorrelated, quiet input-modes contribute a total amount of $N_T(1-\ev{\!\!\ev{\abs{\bm{\chi}_{mm}}^2}\!\!}_{w^2})$ thermal/vacuum fluctuations to the output noise-power. 

The SNR for the DQS setup is then formally given by the ratio of the signal power expressed in Eq.~\eqref{eq:signal_w} to the noise power expressed in Eq.~\eqref{eq:noise_ww},
\begin{align}
    &\overline{\rm SNR}_{M;\,\{w,w^\prime\}}=
    \nonumber
    \\
    &\qq{}
    \frac{M\ev{\!\!\ev{\abs{\bm{\chi}_{ms}}^2}\!\!}_{1/M} n_s}{N_T\Big( \ev{\!\!\ev{\abs{\bm{\chi}_{mm}}^2}\!\!}_{w^2}/G+1-\ev{\!\!\ev{\abs{\bm{\chi}_{mm}}^2}\!\!}_{w^2}\Big)}.\label{eq:theory_opt_snr}
\end{align}

\subsection{Scan-rate performance for multi-cavity network}
\begin{figure}[t]
    \centering
    \includegraphics[width=\linewidth]{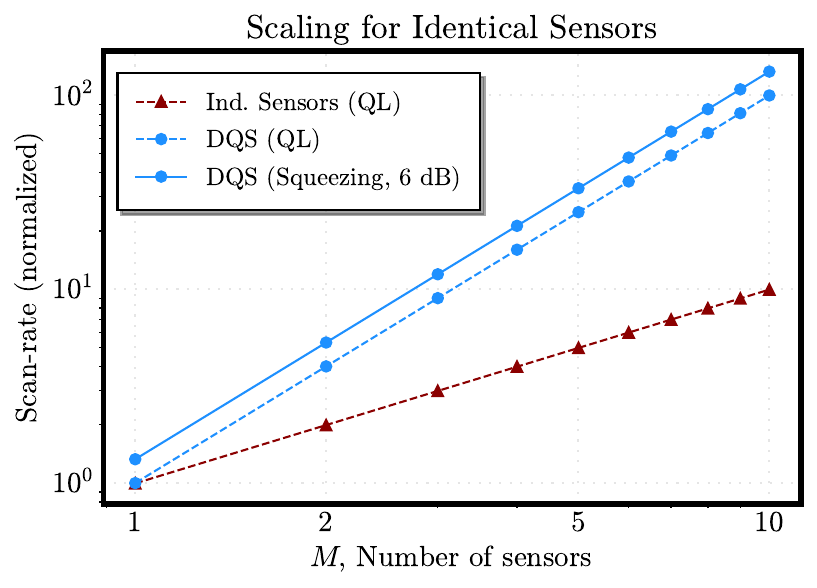}
    \caption{Scaling of the scan-rate with the number of sensors $M$ (log-log plot). Solid line corresponds to a squeezed input for a DQS setup, with gain $G\approx1.56$ (6dB of squeezing). Dashed lines correspond to QL setups with (QL-DCS; circles) and without (Ind. Sensors; triangles) joint post-processing. Observe quadratic scaling of the scan-rate for distributed sensing scenarios versus linear scaling for independent sensors (i.e., the slope of the former is twice that of the latter) as well as a constant factor improvement for all $M$ in the DQS setup due to squeezing with fixed gain. Normalization is with respect to the single cavity, quantum limited setup.}
    \label{fig:scanrate_dqs}
\end{figure}

\subsubsection{General observations and remarks}
We now assess the scan-rate performance of a general, non-identical $M$-sensor cavity network. As the theoretically derived weights from the two-step optimization, Eqs.~\eqref{eq:general_w} and \eqref{eq:general_wprime}, are near-optimal, we utilize these in the forthcoming analysis to benchmark performance. The main results are succinctly plotted in Fig.~\ref{fig:scanrate_dqs}. Data shows the scaling of the scan-rate with the number of sensors $M$ when the network is in coherent operation (blue lines; circles) versus a network of independent sensors (red lines; triangles). The plot also shows relative performance with (DQS, solid line) and without (QL-DCS, dashed line) squeezing. Even when the network is non-ideal, in the sense that individual cavities have different levels of performance, one can still achieve substantial performance enhancement from squeezing and a quadratic scaling in the number of sensors $M$ under coherent operation of the network (see below for an example of this with 5 cavity receivers).


Another aspect that we want to briefly consider is interference between the signal amplitudes when we combine them. Interference will occur, for instance, if the cavity-induced angles $\{\theta_{m_ks_k}\}$ of the signals are non-negligible or not locally resolved. As resolving such angles adds additional complexity to the linear networks used for distributing and combining microwave fields, we want to gain some intuition on how detrimental signal interference can be to the SNR if such phases are not taken care of. Consider a set of weights $\{w_{1k}\}$ that \textit{do not} resolve the angles $\theta_{m_ks_k}$. One can derive an expression for the signal power for these sets of weights, similar to relation~\eqref{eq:signal_w},
\begin{equation}
    \begin{split}
    &\mathbb{E}\left[\ev{\Re\left(\hat{a}_1^{\rm(out)}\right)}^2\right]
    \\
    &= \mathbb{E}\left[\left(\sum_{k=1}^M\abs{w_{1k}}\abs{\bm\chi_{m_ks_k}}\cos(\phi_s+\theta_{m_ks_k})\right)^2\mu_s^2\right]\\
    &= n_s\Bigg(\sum_{k=1}^M\abs{w_{1k}}^2\abs{\bm\chi_{m_ks_k}}^2
    \\
    &
    \ \ \ \  + \sum_{\substack{i,j \\ i\neq j}}^M\abs{w_{1i}}\abs{w_{1j}}\abs{\bm\chi_{m_ia_i}}\abs{\bm\chi_{m_js_j}}\cos(\theta_{m_ia_i}-\theta_{m_js_j})\Bigg)\label{eq:signal_interf}
\end{split}
\end{equation}
The first summation is just the average power of a single cavity in the network. The second summation contains interference terms between the signal amplitudes. 

\begin{figure}
    \centering
    \includegraphics[width=\linewidth]{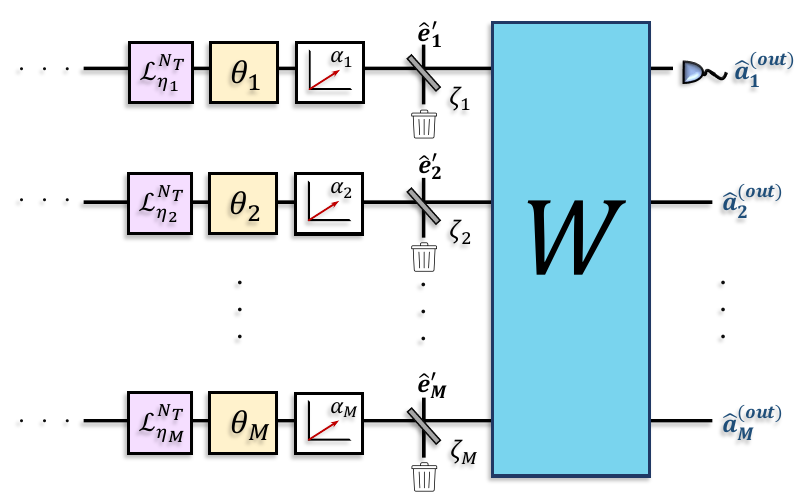}
    \caption{Including loss between the cavity--post-processing stage in a distributed sensing setup; $\zeta_k\in[0,1]$ is the transmission probability for the $k$th mode ($1-\zeta_k$ is the loss) and $\{\hat{e}_k^\prime\}$ are a set of uncorrelated environmental modes.}
    \label{fig:loss_circuit}
\end{figure}

To estimate the effects of interference on the scan-rate, we consider a superficial example. First, to single-out the interference effects, we operate at the quantum limit (i.e., zero squeezing, $G=1$), as squeezing does not effect the signal-power. Second, we assume each cavity to differ only in their intrinsic linewidths, which we choose to fall uniformly within the interval $\gamma_\ell\in[1, 3]$ (measured with respect to the smallest linewidth, taken arbitrarily as unity). We use this as an input into Eq.~\eqref{eq:signal_interf}, square the resulting expression, and integrate over frequencies to obtain an estimate for the scan-rate, assuming either uniform weights $\abs{w_{1k}}=1/\sqrt{M}$ or the near-optimal weights of Eq.~\eqref{eq:general_w} without the phase factor. To estimate relative performance, we normalize the results with respect to the near-optimal scan-rate when the phases $\{\theta_{m_ks_k}\}$ are corrected.

We find that, for a spread of linewidths $\gamma_\ell\in[1,3]$, interference causes a relative $\sim 3-5\%$ decrease in the scan-rate compared to when the phases are completely corrected. We furthermore find a notable differences in performance between choosing uniform weights versus choosing near-optimal weights when interference-effects are present, at about the $2\%$ level for this example. A more exhaustive and systematic study of signal-interference effects in the scan-rate performance can be accomplished using our formalism (e.g., relying on a more realistic model for inhomogeneities within the network), however we leave this problem to future, experimentally-driven research.

\subsubsection{Including loss}

\begin{figure}
    \centering
    \includegraphics[width=\linewidth]{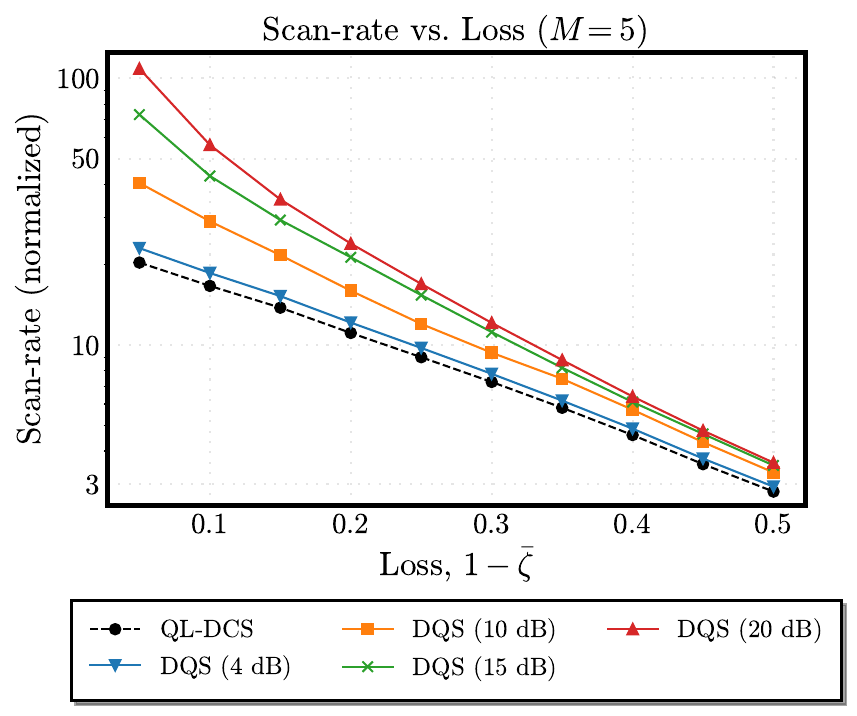}
    \caption{Scan-rate versus average loss for a distributed sensing setup with 5 cavities. For the QL-DCS setup, each cavity is quantum-limited and the signals are jointly post-processed. For the DQS setups, a squeezed vacuum (with some amount of squeezing, in dB) is distributed across the network and the signals are jointly post-processed. Normalization is with respect to the (zero loss) single cavity, quantum limited setup.}
    \label{fig:scanrate_loss}
\end{figure}

Any real experiment is plagued with inefficiency or loss. Our goal here is to weave loss into our network analysis and analyze its effects. We introduce loss between the cavity and detection/post-processing stage (Fig.~\ref{fig:loss_circuit}). Loss at this stage is most detrimental as the signal as well as the squeezing is hindered by such. 

Formally, including loss at the detection/post-processing stage amounts to making the substitution ${w_{0k}\rightarrow w_{0k}\sqrt{\zeta_k}}$ within our analysis, where $\zeta_k$ is the transmittance for the $k$th sensor (i.e., probability for a single photon to transmit from the output of the $k$th cavity to the post-processing stage), and including an additional noise term, $N_T(1-\sum_{k=1}^M \abs{w_{1k}}^2\zeta_k)$, due to vacuum/thermal fluctuations arising from the loss ports. The theoretically derived weights, $w_{1k}$ and $w^\prime_{k1}$ in Eqs.~\eqref{eq:general_w} and~\eqref{eq:general_wprime}, also pick up an extra factor $\sqrt{\zeta_k}$ during optimization when loss is present. We provide a concrete example highlighting the effect of loss on the scan-rate just below.

\subsubsection{Example: 5 sensor cavities}
Here, we examine the detriments of loss and inhomogeneity for an distributed sensing setup containing 5 cavities. We consider quantum limited configurations (QL-DCS) as well as DQS squeezed configurations. 

In our loss analysis, we assume that the transmittance at each mode varies (according to a random, uniform distribution) within the range $\zeta_k\in[\bar{\zeta}-\delta, \bar{\zeta}+\delta]$, where $\bar{\zeta}$ is the average transmittance (and thus $1-\bar{\zeta}$ is the average loss) and $\delta$ is the variance of the loss across the network, which we fix to $\delta=.05$. We run 100 simulations, choosing $\{\zeta_k\}$ randomly in each, and average the results, resulting in the data present in Fig.~\ref{fig:scanrate_loss}, where we plot the (average) scan-rate versus the (average) loss, $1-\bar{\zeta}$.\footnote{We remark on small subtlety. When loss is present, the optimal coupling regime is no longer given by $\gamma_m\approx 1.8G\gamma_l$ but instead depends subtly on the amount of loss as well as the amount of squeezing. We account for this in our numerical simulations.} For all values of the loss within the parameter regime considered ($1-\bar{\zeta}\in[.05, .5]$), the squeezed DQS setups always have a performance advantage over the QL-DCS setup, though the advantage is not as substantial when there is a significant amount of loss. Moreover, squeezing enhancement is ultimately ``loss-limited'', in the sense that: For a given amount of loss, $1-\bar{\zeta}$, there is a maximum amount of squeezing, above which there is no more benefit to be had by increasing the gain of the squeezer. We observe the loss-limited phenomenon in Fig.~\ref{fig:scanrate_loss} by the convergence of the various DQS squeezed curves as the loss increases. This was also observed in ref.~\cite{malnou2019} for the particular value $1-\bar{\zeta}\approx.3$, where the authors showed a loss-limited squeezing limit of $G\approx 20$ (i.e., $18-20$ dB of squeezing), which is comparable to our results near that regime. 

\begin{figure}[t]
    \centering
    \includegraphics[width=\linewidth]{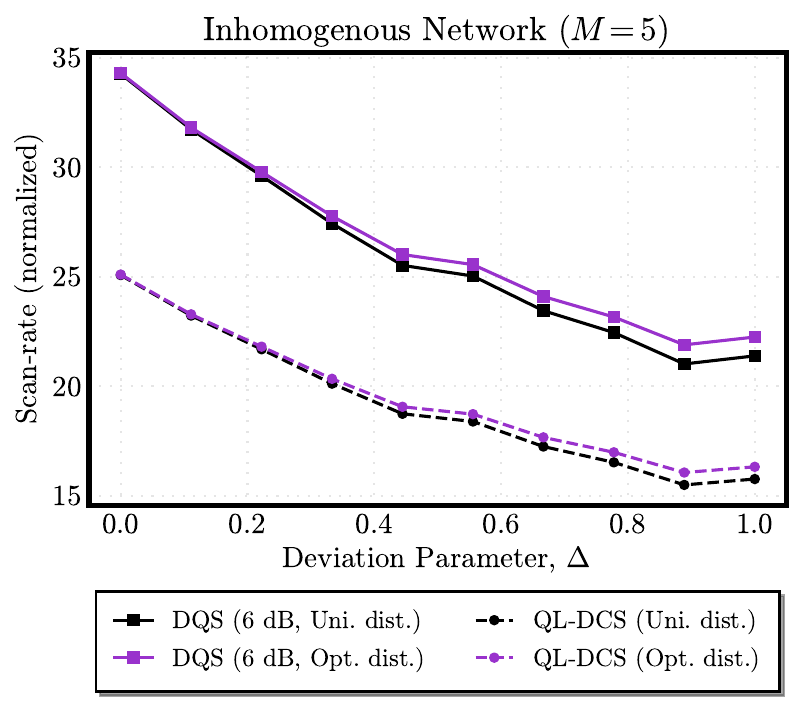}
    \caption{Scan-rate for a inhomogenous network of 5 cavities with varying linewidths. The deviation parameter, $\Delta$, measures how large the intrinsic linwidths of the cavities in the network differ from a (fixed) ``best'' cavity ($100\times\Delta$ is the average percent difference). QL-DCS and DQS setups are both shown for comparison. A squeezed vacuum with 6 dB of squeezing is distributed across the network in the DQS setup.}
    \label{fig:scanrate_Delta}
\end{figure}

We now consider the case where the 5 cavities are not identical (an inhomogenous network). To highlight the effects of inhomogeneity alone, we work in the zero loss regime. The signal/noise may vary from cavity-to-cavity in three ways: (1) different axion-photon coupling strengths, $\gamma_a$; (2) different resonance frequencies; and (3) different intrinsic linewidths (i.e., different intrinsic quality factors), $\gamma_l$. Regarding (1), an example of this would be that the strength of magnetic fields in each cavity (used to induce an axion$\rightarrow$photon conversion) vary by some degree. This would simply lead to some cavities having a larger signal than others. Of course, if the difference between the signal amplitudes is too large, then the ``bad'' cavities in the network do not provide much benefit. For only small differences here, we still expect a scaling and squeezing advantage (analogous to small differences in linewidths that we present below). Regarding (2), perhaps the largest detriment due to varying resonance frequencies is that the cavity response functions will not overlap, leading to a broadened signal with a lower peak. If the resonance frequencies differ by too large, then the coherent network effectively reduces to an independent set of sensors, with each sensor centered at different resonance frequencies. Since we have shown that it is advantageous to operate a coherent network versus a network of independent sensors, the resonance frequencies of the sensors should, therefore, be as closely matched as possible. We remark that small deviations of resonance frequencies was addressed in a similar network setting, for a multi-cell cavity~\cite{jeong2018concept} as well as in the ADMX collaboration's 4-cavity design~\cite{sikivie2020search}. Furthermore, in Appendix~\ref{app:squeezing}, we show how fluctuations in the resonance frequency of a single cavity can lead to, e.g., unwanted noise from the anti-squeezed quadrature. Similar effects will occur in a network setting as well, but as long as these fluctuations are not too large, anti-squeezing noise will be negligible.

For concreteness, we examine how the performance of the network diminishes when the cavities have different linewidths (intrinsic quality factors). We measure the variations in performance by choosing one cavity (the ``best'' cavity) with fixed linewidth, $\gamma_l$, and allowing all other cavities to have varying linewidths given by $\gamma_{l_j}=\gamma_l(1+\epsilon_j)$, where $\epsilon_j$ is a random, positive number drawn from an exponential distribution (to assure its positivity), $\Pr[\epsilon_j]=\exp(-\epsilon_j/\Delta)/\Delta$, where $\Delta$ is the ``deviation parameter'' that effectively measures how far the linewidth of the $j$th cavity differs from that of the best cavity, $\gamma_l$ (indeed, $100\times\Delta$ is the average percent difference). 

For simplicity, we assume the same deviation parameter, $\Delta$, for each cavity in the network. We consider QL-DCS setups as well as DQS setups (with 6 dB of squeezing distrubted across the network for the latter). The DQS setup always performs proportionally better here. We run 100 simulations, with each cavity taking an independent and random linewidth in each simulation, for a given $\Delta$, and average the results. We then vary $\Delta$ and observe how this affects the performance of the network, as measured by the scan-rate. The results are shown in Fig.~\ref{fig:scanrate_Delta}, where we plot the scan-rate for uniform weights, $\abs{w_{0k}^\prime}^2=1/5$ (black dashed/solid curves; Uni. dist.), and for the near-optimal weights of Eq.~\eqref{eq:general_wprime} (purple dashed/solid curves; Opt. dist.). As the deviation parameter becomes larger, the network performance degrades for both cases. For instance, at $\Delta\approx.2$, the percent difference from the identical sensors case ($\Delta=0$) is about $8\%$. For larger deviations in the network, our optimal weights become more favorable (for both QL-DCS setups and DQS setups), as seen by a splitting of the black and purple curves in Fig.~\ref{fig:scanrate_Delta} as $\Delta$ increases.

\section{Discussions and Conclusions}

In this work, we propose a compact, entangled sensor-network to accelerate the search for ultra-light, bosonic dark-matter particles. By coherently combining the signals from each sensor in the network, the sensor-network enjoys a $\sim M^2$ scaling of the scan-rate versus the number of sensors $M$, compared to $\sim M$ scaling for an independent set of sensors (imposed by the law of large numbers). By utilizing entanglement between the sensors, generated by splitting and distributing a single squeezed-vacuum, the sensors further enjoy a boosted scan-rate from squeezing---scaling as $\sim M^2G$ in the high-squeezing limit, where $G$ is the gain of the squeezed-vacuum. 

Our results may be immediately pertinent to multiple-cell haloscopes searching for DM in higher mass regions \cite{jeong2018concept,jeong2020prl}. Squeezing can be properly distributed to the multiple-cells in the cavity, and the output amplitudes can be coherently combined to surpass the quantum-limited search in such setups, as discussed in our paper. Another fascinating possibility is application of our techniques to the next generation of experiments with the ADMX collaboration's 4-cavity array~\cite{sikivie2020search}. Importantly, our DQS proposal will only add extra \emph{extrinsic} complexity to the setup, as a single squeezed vacuum can be externally prepared (along the same lines as in the recent HAYSTAC experiment~\cite{backes2021}) and then routed (via a passive linear network, $\bm W^\prime$) to the cavities in the array. The 4-cavity array can thus, in principle, obtain a quantum enhancement atop of the benefit from coherently combining the signals at the amplitude level, as we show in this work. 

We point out that one can also realize the same performance as our DQS scheme by, instead, injecting independent squeezed vacua into each cavity---thus requiring $M$ squeezers for the $M$ cavities. This further demands phase-locking of the individual squeezers to assure that the squeezed quadratures are aligned. On the other hand, our proposed DQS scheme achieves equivalent performance with only a \emph{single} squeezer, at the cost of an additional passive linear network (i.e., $\bm W^\prime$) which routes the ingoing microwave fields to the cavities. This resource reduction, from $M$ squeezers in the former to 1 squeezer in the latter, ultimately originates from the quantum correlations of the noises between microwave fields. Our proposal thus serves as an intriguing application of quantum entanglement towards probing fundamental physics (and towards quantum-enhanced broadband sensing with multiple cavity receivers, in general). 

Additionally, it is important to foresee potential technological impediments in the experimental implementation of the proposed protocol. Dark-matter search using a single squeezed microwave source has been demonstrated in Ref.~\cite{backes2021}, achieving a 1.9 enhancement in the scanning rate over a conventional classical dark-matter search scheme. Scaling up the system to $M$ sensors would pose a series of experimental challenges. First, the previous squeezing-enhanced experiment was carried out in a 1.7-liter microwave cavity at mK temperature in a dilution refrigerator. To accommodate multiple sensors subject to the limited volume of crystat chambers and a high magnetic field environment, one needs to design miniaturized cavities or use multiple cryostats connected with high-efficiency quantum transduction and interconnect~\cite{han2021microwave,wu2021} to accommodate the sensor cavities. Second, a multi-sensor system would need additional components such as circulators to prevent signal cross talk between different sensors. Such a requirement would increase transmission losses and thereby diminish the enhancement enabled by squeezing. To surmount such a hurdle one can resort to novel signal routing schemes such as a parametric swap interaction~\cite{burkhart2021} to eliminate the lossy non-reciprocal components. Third, the synchronization of the resonant frequencies and phases of distributed squeezed states at multiple cavities would call for advanced electronics for real-time data acquisition, processing, and feedforward. In this regard, field programmable gate arrays would offer the desired data-processing bandwidth and scalability.

We have focused on a sensor network distributed in a small volume, as we are mostly concerned with the initial search for dark matter particles. In the future, distributing the sensors at distance will enable extraction of more information about dark matter, as discussed in refs.~\cite{foster2021DM,derevianko2018network,chen2021vectorsensors}. Furthermore, assuming a functional (continuous-variable) quantum network~\cite{wehner2018quantum,wu2021continuous} with microwave-optical quantum transduction~\cite{han2021microwave,wu2021}, one can consider utilizing long-baseline interferometry~\cite{gottesman2012} on the microwave signals to boost the detection of finer characteristics of dark matter, after the existence of dark matter particles has been confirmed. We defer these fascinating analyses to future work. 

Before closing, we comment on other potential quantum resources for enhancing DM search. GKP states have already been explored for distributed displacement sensing~\cite{zhuang2020DQSgkp}. In Appendix~\ref{appendix:gkp_app}, we show that GKP states can indeed improve the scan-rate by a constant factor in the ideal case, however its advantage diminishes when a practical measurement scheme is adopted. It is still an open problem whether other exotic quantum resources can provide a boost under practical conditions.

\paragraph*{Note added.---} After the completion of our work, we became aware of a related but distinct quantum network design for a DM search with cavities~\cite{chen2021axion}. The authors of that work assume that each cavity mode is directly coupled to two other modes by a parametric process (squeezing) and a passive process (beam-splitter interaction), respectively. They then design a complex array of such interactions consisting of, e.g., many interleaved parametric processes to build a completely connected, active quantum network. The setup is quite different from ours in that we assume no direct coupling between cavity modes; we require only passive ``routing" of inputs and outputs to and from the cavities; and furthermore, we need only a single parametric process (e.g., to create a squeezed vacuum state), which is extrinsic to the internal evolution of the cavity modes.

\begin{acknowledgements} A.J.B. and Q.Z. thank M. Malnou and K. W. Lehnert for discussions on squeezing enhanced detection. The authors also thank L. Maccone for discussions. This material is based upon work supported by the U.S. Department of Energy, Office of Science, National Quantum Information Science Research Centers, Superconducting Quantum Materials and Systems Center (SQMS) under the contract No. DE-AC02-07CH11359. A.J.B. and Q.Z. also acknowledge support from the Defense Advanced Research Projects Agency (DARPA) under Young Faculty Award (YFA) Grant No. N660012014029. Z.Z. and Q.Z. also acknowledge support from NSF OIA-2134830 and NSF OIA-2040575. The work of CG and RH is also supported by the DOE QuantISED program through the theory  consortium ``Intersections of QIS and Theoretical Particle Physics'' at Fermilab.
\end{acknowledgements}


\appendix 

\section{Quick tutorial on Gaussian states and transformations}\label{appendix:gaussian}
In this paper, the relevant interactions between, say, $n$ modes of an electromagnetic field (axion-induced or otherwise) are quadratic in the annihilation and creation operators, $\hat{a}_j$ and $\hat{a}_j^\dagger$, where $j\in\{1,2,\dots,n\}$. Furthermore, the quantum states involved are mostly Gaussian (i.e., states generated by quadratic Hamiltonians); and even for non-Gaussian states such as the GKP state, the Gaussian approximation provides valuable insights. Such characteristics beg the use of the Gaussian formalism~\cite{weedbrook2012gaussian,serafini2017}, which is an efficient formalism that reduces the dynamics of an $n$-mode quantum state to matrix multiplication between a set of $2n\times2n$ matrices (encoding the dynamics) acting on the $2n\times1$ mean vector and $2n\times2n$ covariance matrix (to be described below) of the quantum state. We briefly review some relevant features of the Gaussian formalism below. The interested reader may consult refs.~\cite{weedbrook2012gaussian,serafini2017} for more details.

Given the annihilation and creation operators, $\hat{a}_j$ and $\hat{a}_j^\dagger$, define the quadrature operators
\begin{equation}
    \hat{Q}_j\equiv\frac{1}{\sqrt{2}}\left(\hat{a}_j+\hat{a}_j^\dagger\right)\qq{and} \hat{P}_j\equiv\frac{1}{\sqrt{2}}\left(\hat{a}_j^\dagger-\hat{a}_j\right),
\end{equation}
such that $[\hat{Q}_j,\hat{P}_k]=\im\delta_{jk}\,\forall\,j,k$, where we have let $\hbar=1$. Now define the $2n\times1$ vector of quadrature operators,
\begin{equation}
    \hat{\bm{R}}\equiv(\hat{Q}_1,\hat{P}_1, \hat{Q}_2, \hat{P}_2,\dots,\hat{Q}_n,\hat{P}_n)^\top,
\end{equation}
where the transpose is with respect to the vector space, not the operator space. The commutation relations can therefore be written as, 
\begin{equation}
    \left[\hat{\bm{R}}_j,\hat{\bm{R}}_k\right]=\im\bm{\Omega}_{jk} \qq{where}\bm{\Omega}=\mathbb{I}_n\otimes
    \begin{pmatrix}
    0 & 1\\
    -1 &0
    \end{pmatrix},
\end{equation}
where $\mathbb{I}_n$ is the $n\times n$ identity matrix. The matrix $\bm\Omega$ is known as the symplectic form and encodes the canonical commutation relations between the quadrature operators of all $n$ modes. 

We now define the mean vector, $\bm{\mu}$ (the vector of first moments), and covariance matrix, $\bm{\sigma}$ (the matrix of second moments), of an $n$-mode quantum state $\hat{\rho}$ as,
\begin{align}
    \bm{\mu}_j&\equiv\Tr\left(\hat{\bm{R}}_j\hat{\rho}\right)\\\qq{and}\bm{\sigma}_{jk}&\equiv\Tr\left(\left\{\hat{\bm{R}}_j-\bm{\mu}_j, \hat{\bm{R}}_k-\bm{\mu}_k\right\}\hat{\rho}\right),\label{eq:moments}
\end{align}
where $\{\cdot,\cdot\}$ denotes the symmetric anti-commutator. If $\hat{\rho}$ is a Gaussian state, then the mean vector and the covariance matrix completely determine all properties of the state. For example, a single-mode thermal state of mean quanta $\Bar{n}$ has zero first moments and covariance matrix $(1+2\Bar{n})\mathbb{I}_2$.

Any unitary operation $\hat{U}$ acting on the $n$-mode bosonic Hilbert space and generated from a Hamiltonian which is quadratic in the quadrature operators corresponds to $2n\times2n$ symplectic matrix, $\bm S$. The symplectic matrix acts linearly on the vector of quadrature operators such that: if $\hat{\bm{R}}^{(\text{in})}$ represents the quadrature operators of the input modes, then the quadrature operators of the output modes, $\hat{\bm{R}}^{(\text{out})}$, are found by matrix multiplication, i.e., $\hat{\bm{R}}^{(\text{out})}={\bm S}\hat{\bm{R}}^{(\text{in})}$. One calls the matrix $\bm S$ symplectic since it preserves the symplectic form, $\bm{S}\bm{\Omega S}^\top=\bm{\Omega}$. Taking the input/output transformation together with the definition of the first and second moments [Eq. \eqref{eq:moments}], one finds the output moments in terms of the input moments, 
\begin{equation}
    \bm{\mu}^{(\text{out})}={\bm S}\bm{\mu}^{(\text{in})}\qq{and}\bm{\mu}^{(\text{out})}=\bm{S}\bm{\sigma}^{(\text{in})}{\bm S}^\top.\label{eq:output_moments}
\end{equation}
Eq.~\eqref{eq:output_moments} holds for arbitrary input states, though such is sufficient to completely describe Gaussian states.

One can extend the discussion to include non-unitary dynamics in the Gaussian context, when the system of interest interacts (via quadratic interactions) with an inaccessible (Gaussian) environment. This leads to the general notion of a Gaussian completely-positive, trace-preserving (CPTP) map---better known as a (bosonic) Gaussian quantum channel. It can be shown that one can completely describe a Gaussian quantum channel, $\mathcal{G}$, by two $2n\times2n$ real matrices, $\bm X$ and $\bm Y$ (the scaling matrix and noise matrix, respectively) and a displacement vector $\bm{\nu}$, such that
\begin{equation}
    \bm{\mu}\overset{\mathcal{G}}{\rightarrow} {\bm X}\bm{\mu}+\bm{\nu}\qq{and}\bm{\sigma}\overset{\mathcal{G}}{\rightarrow} {\bm X}\bm{\sigma}\bm{X}^\top + \bm{Y},\label{eq:moment_channel_transform}
\end{equation}
where $\bm{Y}+\im\bm{\Omega}\geq\im \bm{X\Omega X}^\top$ is the only condition that the scaling matrix and noise matrix must satisfy in order for the above transformations to correspond to a proper Gaussian quantum channel. 

Interestingly, we can provide a unitary extension of the quantum channel $\mathcal{G}$ which corresponds to a symplectic matrix $\bm S$ acting jointly on the system $A$, with input mean and covariance $(\bm{\mu}_s, \bm{\sigma}_s)$, and an ``environment" $E$, with mean and covariance $(\bm{\mu}_E,\bm{\sigma}_E)$. To make correspondence with the scaling matrix and noise matrix from above, we write the symplectic matrix in block form, $\bm S=\big(\begin{smallmatrix}
  \bm A &\bm B\\
  \bm C & \bm D
\end{smallmatrix}\big)$
where $\bm A$ encodes the internal system dynamics and $\bm B$ encodes the coupling to the environment. It is then straightforward to show that one can associate the quantum channel $\mathcal{G}$ with the symplectic matrix $\bm S$, provided that
\begin{equation}
\bm X = \bm A ,\qq{} {\bm Y} = \bm B \bm{\sigma}_E {\bm B}^\top, \qq{and} \bm{\nu}=\bm{B}\bm{\mu}_E.\label{eq:x_y_nu}
\end{equation}


\section{Quantum model of the cavity}\label{appendix:cavity_model}
Here we provide details regarding our single-mode quantum-channel description for the input-output dynamics of an electromagnetic cavity. 

\subsection{Brief background}
We consider a damped cavity, defined by the mode $\hat{A}$ with (free) Hamiltonian $\hat{H}_c=\hbar\omega_c\hat{A}^\dagger\hat{A}$, where $\omega_c$ is the cavity resonance-frequency, linearly coupled to a set of memoryless (Markovian) ``bath" modes.
We model the full interaction between the cavity mode and the input/output modes of the bath with the Heisenberg-Langevin equations (see, for instance, Appendix E.2 from the arXiv version of ref. \cite{Clerk_2010}) in the rotating reference frame of the cavity,
\begin{equation}
    \dv{\hat{A}}{t}=-\frac{\gamma}{2}\hat{A}+\sum_{j\in\{m,s,\ell\}}\sqrt{\gamma_j}\hat{a}_j^{(\text{in})},\label{eq:cavity_dynamics}
\end{equation}
where $\gamma$ is the damping rate of the cavity which satisfies $\gamma=\sum_{j\in\{m,s,\ell\}}\gamma_j$, and $\hat{a}_j^{(\text{in})}$ represent the input modes that transfer energy within the cavity through their respective ports; see Fig. \ref{fig:cavity_channel} for an illustration. Here, the Markovian assumption is that $[\hat{a}_j^{(\text{in})}(t),\hat{a}_k^{(\text{in})\,\dagger}(t^\prime)]=\delta_{jk}\delta(t-t^\prime)$. Note that $\hat{a}_j^{(\text{in})}$ has units $\sqrt{\text{quanta/second}}$.

Though there is no explicit coupling between the ports, the cavity mode acts as an intermediary, allowing for an effective energy transfer from, e.g., the signal port, $\hat{a}_s$, to the measurement port, $\hat{a}_m$. After interaction with the cavity, the ports then exit the cavity as output modes---carrying the transferred energy either through inaccessible ports, such as the loss port $\hat{a}_\ell$ and signal-field port $\hat{a}_s$, or through the accessible measurement port $\hat{a}_m$. Each output mode satisfies the time-dependent relation \cite{Clerk_2010}
\begin{equation}
    \hat{a}_j^{(\text{out})}(t)=\hat{a}_j^{(\text{in})}(t)+\sqrt{\gamma}\hat{A}(t).\label{eq:in_out_cavity}
\end{equation}
After passing to the spectral domain by a Fourier transformation, one can solve Eq.~\eqref{eq:cavity_dynamics} for the spectral amplitude of the cavity-mode $\hat{A}$ and substitute that expression into Eq.~\eqref{eq:in_out_cavity}. The results are equations~\eqref{eq:in_out} and \eqref{eq:chi} of the main text.

\subsection{Channel reduction}

We now reduce the input-output relations of Eq.~\eqref{eq:in_out} for the 3 modes $(\hat{a}_m,\hat{a}_s,\hat{a}_\ell)$ to a single-mode input-output channel for the lone measurement port $\hat{a}_m$. First, define the vector of annihilation and creation operators for the input and output modes, 
\begin{equation}
    \hat{\bm{a}}\equiv(\hat{a}_m,\hat{a}_m^\dagger,\hat{a}_s,\hat{a}_s^\dagger,\hat{a}_\ell,\hat{a}_\ell^\dagger)^\top.\label{eq:avec}
\end{equation}
Now let $\bm{\chi}_{ij}=\abs{\bm{\chi}_{ij}}\e^{\im\theta_{ij}}$, where $\{\theta_{ij}\}$ are the angles of the complex susceptibility-matrix elements, which are only non-zero off-resonance; see e.g. Eqs.~\eqref{eq:theta_ma} and \eqref{eq:theta_mm}. An explicit expression for the susceptibility coefficient is in Eq.~\eqref{eq:chi}. From the definition~\eqref{eq:avec} and the input-output transformation~\eqref{eq:in_out}, it is easy to show that, 
\begin{equation}
    \hat{\bm{a}}^{(\rm{out})}=\bm{\Tilde{\chi}}\hat{\bm{a}}^{(\rm{in})},
\end{equation}
where
\begin{equation}
    \bm{\Tilde{\chi}}=
    \begin{pmatrix}
    \abs{\bm{\chi}_{mm}}\e^{\im\theta_{mm}\bm{\sigma_z}} &  \abs{\bm{\chi}_{ms}}\e^{\im\theta_{ms}\bm{\sigma_z}} & \abs{\bm{\chi}_{m\ell}}\e^{\im\theta_{m\ell}\bm{\sigma_z}} \\
    \abs{\bm{\chi}_{ms}}\e^{\im\theta_{ms}\bm{\sigma_z}} &  \abs{\bm{\chi}_{ss}}\e^{\im\theta_{ss}\bm{\sigma_z}} & \abs{\bm{\chi}_{s\ell}}\e^{\im\theta_{s\ell}\bm{\sigma_z}} \\
    \abs{\bm{\chi}_{m\ell}}\e^{\im\theta_{m\ell}\bm{\sigma_z}} &  \abs{\bm{\chi}_{s\ell}}\e^{\im\theta_{s\ell}\bm{\sigma_z}} & \abs{\bm{\chi}_{\ell\ell}}\e^{\im\theta_{\ell\ell}\bm{\sigma_z}}
    \end{pmatrix},
\end{equation}
with $\bm{\sigma_z}$ being the $2\times2$ Pauli-z matrix. We can go to the quadrature basis $\hat{\bm{R}}=(\hat{Q}_m,\hat{P}_m,\hat{Q}_s,\hat{P}_s,\hat{Q}_\ell,\hat{P}_\ell)^\top$ via a unitary transformation $\hat{\bm{R}}=\Bar{\bm{U}}\hat{\bm{a}}$, where 
\begin{equation}
\Bar{\bm{U}}={\rm diag}(\Bar{\bm{u}},\Bar{\bm{u}},\Bar{\bm{u}}) \qq{and} \Bar{\bm u}=\frac{1}{\sqrt{2}}
\begin{pmatrix}
1 & 1\\
  -\im & \im
\end{pmatrix}.
\end{equation}
From here, the input-output relations in the quadrature basis follows,
\begin{equation}
    \hat{\bm{R}}^{\rm (out)}=\left(\Bar{\bm{U}} \bm{\tilde{\chi}}\Bar{\bm{U}}^\dagger\right)\hat{\bm{R}}^{\rm (in)},
\end{equation}
with the symplectic orthogonal transformation $\Bar{\bm{U}} \bm{\tilde{\chi}}\Bar{\bm{U}}^\dagger$ given explicitly by,
\begin{equation}
    \Bar{\bm{U}} \bm{\tilde{\chi}}\Bar{\bm{U}}^\dagger = 
    \begin{pmatrix}
    \abs{\bm{\chi}_{mm}}\bm{O}(\theta_{mm}) &  \abs{\bm{\chi}_{ms}}\bm{O}(\theta_{ms}) & \abs{\bm{\chi}_{m\ell}}\bm{O}(\theta_{m\ell}) \\
    \abs{\bm{\chi}_{ms}}\bm{O}(\theta_{ms}) &  \abs{\bm{\chi}_{ss}}\bm{O}(\theta_{ss}) & \abs{\bm{\chi}_{s\ell}}\bm{O}(\theta_{m\ell}) \\
    \abs{\bm{\chi}_{m\ell}}\bm{O}(\theta_{m\ell}) &  \abs{\bm{\chi}_{s\ell}}\bm{O}(\theta_{s\ell}) & \abs{\bm{\chi}_{\ell\ell}}\bm{O}(\theta_{\ell\ell}),
    \end{pmatrix}
\end{equation}
where $\bm{O}(\theta_{ij})$ is a $2\times2$ symplectic orthogonal matrix corresponding to a rotation by an angle $\theta_{ij}$. To reduce the dynamics to a single-mode channel description, we first need to provide the scaling matrix, $\bm X$, and noise matrix, $\bm Y$, which one can derive by using Eq.~\eqref{eq:x_y_nu} and making the equivalences $\bm{A}=\abs{\bm{\chi}_{mm}}\bm{O}(\theta_{mm})$ and $\bm{B}=\big(\abs{\bm{\chi}_{ms}}\bm{O}(\theta_{ms}) \ \ \abs{\bm{\chi}_{m\ell}}\bm{O}(\theta_{m\ell})\big)$ ($\bm B$ is a $2\times4$ rectangular matrix). 

To make further progress, we make some simplifying (though physically reasonable) assumptions about the ``environmental" modes $\hat{a}_s$ and $\hat{a}_\ell$. First, we assume that the axion port and loss port ($\hat{a}_s$ and $\hat{a}_\ell$) are independent and that their input covariance matrices consist only of thermal fluctuations at temperature $T$, such that $\bm{\sigma}_E=N_T\mathbb{I}_4\delta(0)$, where $N_T=(1+2\Bar{n}_T)$~\footnote{This follows from the fact that $\ev{\hat{a}_j^{\rm (in)\,\dagger}(\omega)\hat{a}_j^{\rm (in)}(\omega^\prime)}\propto\delta(\omega-\omega^\prime)$. We shall drop the Dirac delta-distribution for brevity.} and $\bar{n}_T=1/(\e^{(\omega_c+\omega)/T}-1)$ in natural units ($\hbar=k_B=1$), where $\omega_c$ is the cavity resonance frequency. Second, we assume the signal field, referred to as mode $\hat{a}_s$, to be a classical coherent field with amplitude $\abs{\bm\mu_s}\sim\sqrt{\text{quanta/second/Hz}}$ and phase $\phi_s$, such that the input mean-vector for the signal port is $\bm{\mu}_s=\abs{\bm\mu_s}(\cos\phi_s,\sin\phi_s)^\top$. The mean vector for the loss port is taken to be the zero vector. From these assumptions, we have that
\begin{equation}
    \bm{\mu}_E= \bm{\mu}_s\oplus\bm{0}_\ell\qq{and}\bm{\sigma}_E=N_T\mathbb{I}_4.
\end{equation}

Using Eq.~\eqref{eq:x_y_nu}, we calculate an expression for the noise matrix,
\begin{align*}
    \bm{Y}&\propto\bm{B}\bm{B}^\top\\
    &=\begin{pmatrix}
    \abs{\bm{\chi}_{ms}}\bm{O}(\theta_{ms}) &  \abs{\bm{\chi}_{m\ell}}\bm{O}(\theta_{m\ell})
    \end{pmatrix}
    \begin{pmatrix}
    \abs{\bm{\chi}_{ms}}\bm{O}^\top(\theta_{ms}) \\
    \abs{\bm{\chi}_{m\ell}}\bm{O}^\top(\theta_{m\ell})
    \end{pmatrix} \\
    &=\sum_{k\neq m}\abs{\bm{\chi}_{mk}}^2\mathbb{I}_2\\
    &=\left(1-\abs{\bm{\chi}_{mm}}^2\right)\mathbb{I}_2.
\end{align*}
For the first equality, we used the correspondence $\bm{B}=\big(\abs{\bm{\chi}_{ms}}\bm{O}(\theta_{ms}) \ \ \abs{\bm{\chi}_{m\ell}}\bm{O}(\theta_{m\ell})\big)$; for the second equality, we used the orthogonality of the $\bm O$ matrices; and for the third equality, we used the unitarity relation $\sum_{j}\bm{\chi}_{ij}^*\bm{\chi}_{jk}=\delta_{ik}$. After performing similar calculations for the scaling matrix $\bm X$ and displacement $\bm\nu$, we thus find,
\begin{align}
    \bm{X}&=\abs{\bm{\chi}_{mm}}\bm{O}(\theta_{mm}),\label{eq:xy_decomp1}\\ \bm{Y}&=N_T\left(1-\abs{\bm{\chi}_{mm}}^2\right)\mathbb{I}_2,\\\qq{and}\bm{\nu}&=\abs{\bm{\chi}_{ms}}\bm{O}(\theta_{ms})\bm{\mu}_s,\label{eq:xy_decomp3}
\end{align}
where the angles, $\theta_{ms}$ and $\theta_{mm}$, are defined through the susceptibility coefficients [see Eq.~\eqref{eq:chi}] via
\begin{equation}
    \sin\theta_{ms}=\frac{\omega}{\sqrt{\gamma^2/4+\omega^2}}\label{eq:theta_ma}
\end{equation}
and
\begin{equation}
     \sin\theta_{mm} =\frac{\omega\gamma_m}{\sqrt{\left((\gamma_m-\gamma_\ell)^2/4+\omega^2\right)\left(\gamma^2/4+\omega^2\right)}}.\label{eq:theta_mm}
\end{equation}

\begin{figure*}[t]
    \centering
    \includegraphics[width=.49\linewidth]{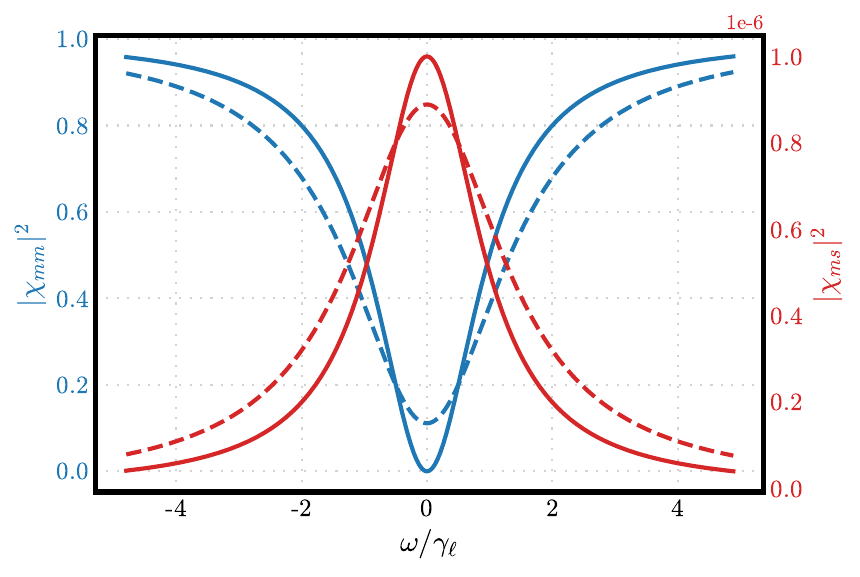}
    \includegraphics[width=.49\linewidth]{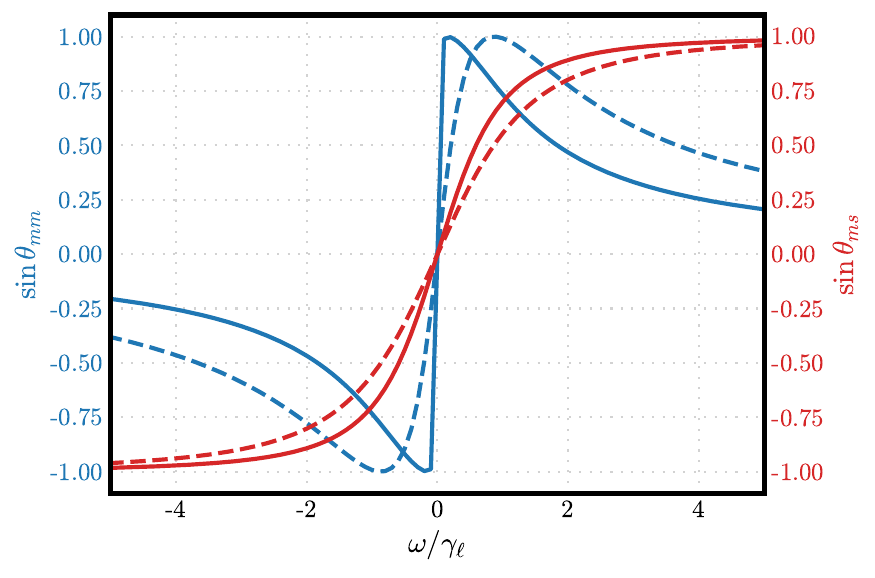}
    \caption{(Left) Plot of cavity the transmission $\abs{\bm{\chi}_{mm}}^2$ (blue curves) and the axion-photon transmission $\abs{\bm{\chi}_{ms}}^2$ (red curves); $\abs{\bm{\chi}_{ms}}^2$ governs the signal behavior.  (Right) Plot of mixing angles, $\theta_{mm}$ (blue curves) and $\theta_{ms}$ (red curves). Solid lines correspond to critical coupling $\gamma_m=\gamma_\ell$. Dashed lines correspond to the over-coupling, $\gamma_m=2\gamma_\ell$. We have arbitrarily chosen $\gamma_s=10^{-6}\gamma_\ell$ to obtain curves for $\abs{\bm{\chi}_{ms}}^2$.}
    \label{fig:susc_plot}
\end{figure*}

The expressions \eqref{eq:xy_decomp1}-\eqref{eq:xy_decomp3} correspond to a thermal-loss channel $\mathcal{L}_{\bm{\chi}_{mm}}^{N_T}$---with the (cavity) transmission parameter $\abs{\bm{\chi}_{mm}}^2$ and noise parameter $N_T=1+2\bar{n}_T$---followed by a (unitary) phase-shift channel $\Phi_{\theta_{mm}}$ and a (unitary) displacement channel $\mathcal{D}_{\bm{\nu}}$. See Fig.~\ref{fig:cavity_channel} for an illustration. The full quantum-channel describing the input-output relations for the measurement port therefore decomposes as,\footnote{The order of the loss channel $\mathcal{L}$ and the phase-shift $\Phi$ can be swapped, since these transformations commute.}
\begin{equation}
    \mathcal{G}=\mathcal{D}_{\bm{\nu}}\circ\Phi_{\theta_{mm}}\circ\mathcal{L}_{\bm{\chi}_{mm}}^{N_T},\label{eq:cavity_ch}
\end{equation}
as eluded to in the main text. Using the general moment transformations~\eqref{eq:moment_channel_transform} together with the channel decomposition of Eqs.~\eqref{eq:xy_decomp1}-\eqref{eq:xy_decomp3}, the input-output relations for the first and second moments of the measurement port are found to be
\begin{equation}
    \bm{\mu}_m^{\rm (out)}=\abs{\bm{\chi}_{mm}}\bm O(\theta_{mm})\bm{\mu}_m^{\rm (in)} + \abs{\bm{\chi}_{ms}}\bm{O}(\theta_{ms})\bm{\mu}_s,\label{eqapp:mu_out}
\end{equation}
and
\begin{multline}
    \bm{\sigma}_m^{\rm (out)}=\abs{\bm{\chi}_{mm}}^2\bm O(\theta_{mm})\bm{\sigma}_m^{\rm (in)}\bm O(\theta_{mm})^\top\\+N_T\left(1-\abs{\bm{\chi}_{mm}}^2\right)\mathbb{I}_2,
\end{multline}
which agree with Eqs.~\eqref{eq:cavity_out_mu} and~\eqref{eq:cavity_out_sigma} of the main text. We point out that we can cancel out the angle, $\theta_{mm}$, by applying a phase-rotation $\Phi_{-\theta_{mm}}$ on the measurement-port input fields prior to the cavity interaction. [In the quantum setting, by utilizing two-mode squeezing, this angle can automatically be taken care of without any extra phase rotation, as we explain in Appendix~\ref{app:resonance_frequency_why}.]

For reference, a plot of the cavity transmission, $\abs{\bm{\chi}_{mm}}^2$, and the signal-cavity coupling $\abs{\bm{\chi}_{ms}}^2$ is shown in Fig.~\ref{fig:susc_plot}, as well as a plot of the sine of the mixing-angles, $\sin\theta_{mm}$ and $\sin\theta_{ms}$. Peaks in $\sin\theta_{mm}$ appear at $\omega^\star=\pm\sqrt{(\gamma_m^2-\gamma_\ell^2)}/2$, at which point $\Im(\bm\chi_{mm})\rightarrow\pm1$. However, as we approach critical-coupling and cavity-resonance, $\gamma_m\rightarrow\gamma_\ell$  and $\omega\rightarrow0$, there is sharp transition in the phase due to the two competing limits. Observe that as $\omega\rightarrow\pm\infty$, $\theta_{mm}\rightarrow 0$ and $\theta_{ms}\rightarrow \pm\pi/2$.


\section{Homodyne vs. heterodyne detection}
\label{app:heterodyne}
We calculate the SNR inferred from heterodyne detection and show that the result is equivalent to a homodyne detection scheme when the signal is random. However, we show that this equivalence breaks whenever a phase-insensitive amplifier is introduced just prior to detection.

Heterodyne detection consists of (non-linearly) mixing the signal mode with a strong local-oscillator of a different frequency and measuring the corresponding output intensities. In this optical domain, this process is equivalent to first splitting the signal beam at a balanced beam-splitter into two (assuming vacuum-noise or low thermal-noise at the other input port of the beam-splitter), applying a relative phase-shift of $\pi/2$ to one output of the beam-splitter, and subsequently performing a homodyne detection on each signal---thus allowing one to measure both quadrature variables, $Q$ and $P$. Since one can simply sum the SNRs inferred from each detector (i.e., the SNRs of the independent homodyne measurements add in quadrature), it would appear that, in principle, a heterodyne detection scheme can have some benefit over the homodyne scheme. This intuition is not generically true however, since, in the example just described, the signal power arriving at each detector is half of the initial signal, due to the balanced beam-splitters. 

Quantitatively, assuming that each beam-splitter in the heterodyne detection scheme adds $N_T$ noise, we can find an expression for the SNR of the power for a heterodyne detection scheme directly from Eqs.~\eqref{eq:cavity_out_mu},
\begin{equation}
\begin{split}
    {\rm SNR}^{\rm (het)}&\equiv\frac{1}{2}\left(\frac{\ev*{\hat{Q}_m}^2}{{\rm Var}(\hat{Q}_m)}+\frac{\ev*{\hat{P}_m}^2}{{\rm Var}(\hat{P}_m)}\right)\\
    &=\frac{1}{2}\frac{\abs{\bm\chi_{ms}}^2\mu_s^2}{N_T}\\
    &=\frac{1}{2}\frac{\gamma_m\gamma_s\mu_s^2}{N_T\left((\gamma/2)^2+\omega^2\right)}.\label{eq:hetero_psd}
\end{split}
\end{equation}
Observe that the difference between homodyne and heterodyne detection schemes is the angle dependence in the $\cos^2\phi_s$ term for the former and the factor of 1/2 for the latter. Though, for a completely random signal, these detection schemes perform equally well after uniformly averaging over the phase $\phi_s$.

\subsection{Phase-insensitive amplification noise}
For a weak signal (as is the case for an axion signal), it is common to add a (high-gain) linear amplifier just prior to measurement in order to make the signal classically detectable \cite{caves1982quantum}. Here, we show that the addition of a phase-\textit{insensitive} amplifier to the detection-chain leads heterodyne detection to be the preferred detection method versus homodyne detection.  When performing homodyne measurements, it is thus pertinent to relegate most amplification to phase-\textit{sensitive} amplification, which does not contribute additional noise to the measured quadrature.

A (single-mode) phase-insensitive amplifier channel, $\mathcal{A}_g$, with gain $g\geq1$, is a Gaussian quantum channel which, given the first and second moments $\bm\mu$ and $\bm\sigma$,
\begin{align}
\mathcal{A}_g: \quad\bm\mu&\longrightarrow \sqrt{g}\bm\mu \\
\bm\sigma&\longrightarrow g\bm\sigma + N_{\mathcal{A}}(g-1)\mathbb{I}_2,
\end{align}
where $N_\mathcal{A}\geq1$ parameterizes the number of noisy quanta introduced into the signal ($N_\mathcal{A}=1$ for a vacuum- or ``quantum-limited" amplifier). In what follows, we shall simply assume that the amplifier adds the same amount of noisy quanta as all other non-unitary processes that we have considered so far and thus set $N_{\mathcal{A}}=N_T$. 

Assuming an initially quiet cavity mode (i.e., no driving or squeezing), the input moments to the amplifier channel, which originate from the output of the cavity, are ${\bm\mu=\abs{\bm\chi_{ms}}\bm O(\theta_{ms})\bm\mu_s}$ and $\bm\sigma=N_T\mathbb{I}_2$. The output moments of the amplifier are then,
\begin{align}
    \bm\mu^{(\rm out)} &= \sqrt{g}\abs{\bm\chi_{ms}}\bm O(\theta_{ms})\bm\mu_s,\\
    \bm\sigma^{(\rm out)} &= N_T(2g-1)\mathbb{I}_2.
\end{align}
The SNR for homodyne detection, after averaging over the coherence time of the axion field, is then,
\begin{equation}
    \overline{\rm SNR}_{g}^{(\rm hom)}= \left(\frac{g}{2g-1}\right)\frac{\abs{\bm\chi_{ms}}^2n_s}{N_T},
\end{equation}
where $n_s=\mathbb{E}[\mu_s^2/2]$, we have maintained the form of $\bm\chi_{ms}$, and omitted the sampling factor $\sqrt{\Delta_aT_O}$ for brevity. 

For heterodyne detection, it is straightforward to show that phase-insensitive amplification does not alter the SNR; in other words, $\overline{\rm SNR}_{g}^{(\rm het)}=\overline{\rm SNR}_{g=1}^{(\rm het)}=\abs{\bm\chi_{ms}}^2n_s/N_T$. Therefore, 
\begin{equation}
    \frac{1}{2}\leq\frac{\overline{\rm SNR}_{g}^{(\rm hom)}}{\overline{\rm SNR}_{g}^{(\rm het)}}=\frac{g}{2g-1}\leq 1,
\end{equation}
where the lower-bound is found in the high-gain limit and the upper-bound is achieved at $g=1$. Hence, heterodyne detection is the preferred detection method when high-gain phase-insensitive amplification is used. Though, this is not the case when squeezing is present in the homodyne setup.

\section{SNR for squeezing-enhanced search}\label{app:squeezing}

We explicitly calculate the SNR for the squeezing-enhanced protocol introduced in ref.~\cite{malnou2019} and implemented in ref.~\cite{dixit2021}. In what follows, to make optimal use of squeezing, we work in the natural reference frame of the measurement-port fields and thus set $\theta_{mm}=0$. An illustration of the setup, in the rotated frame, is shown in Fig.~\ref{fig:squeeze_cavity_channel}. 

The protocol starts by initially squeezing the input noise along one quadrature, say the $Q$-quadrature, by an amount $1/G$ where $G\geq1$. Assuming initial thermal equilibrium at the measurement port, this corresponds to taking the input moments as ${\bm\mu}_m^{\rm (in)}={\bm0}$ and ${\bm\sigma}_m^{\rm (in)}=N_T{\rm diag}(1/G,G)$. Substituting these expressions into Eqs.~\eqref{eq:cavity_out_mu} and~\eqref{eq:cavity_out_sigma}, one easily finds the output moments of the cavity (just before the anti-squeezer). We do not provide the explicit expression for this penultimate step, however after some thought, one deduces that the initial squeezer reduces the measurement-port input noise (relative to the signal) by a factor $G$ along the $Q$-quadrature. On the other hand, non-unity reflection from the cavity introduces (vacuum/thermal) noise to the measurement-port output, which we cannot reduce further with external operations (other than cooling the system and/or reducing the loss). 

Immediately proceeding the cavity interaction, we apply an anti-squeezer to the output of the cavity, which amplifies the $Q$-quadrature by a factor $G$, leading to the final output moments,
\begin{align}
    {\bm \mu}_m^{\rm(f)}&=\abs{\bm{\chi}_{ms}}{\rm diag}(G, 1/G)\bm{O}(\theta_{ms})\bm{\mu}_s\\
     \bm{\sigma}_m^{\rm (f)}&=N_T\abs{\bm{\chi}_{mm}}^2\mathbb{I}_2+N_T\left(1-\abs{\bm{\chi}_{mm}}^2\right){\rm diag}(G, 1/G).\label{eq:final_squeezed_sigma}
\end{align}
We see that the signal, as well as the cavity-noise term (i.e., the second term in $\bm{\sigma}_m^{\rm (f)}$), has been amplified along the $Q$-quadrature by a factor $G$ relative to the input noise (the first term in $\bm{\sigma}_m^{\rm (f)}$). These observations suggest some benefit to the squeezing protocol, as long as we restrict ourselves to measurements along the squeezed quadrature, $Q$.

We now calculate the SNR after a large observation time as a function of the squeezing $G$. After some algebra, we find the expression,
\begin{equation}
   \overline{{\rm SNR}}_{\rm sq}=\frac{\gamma_m\gamma_s n_s}{N_T\left(\frac{(\gamma/2)^2 + \omega^2-\gamma_m\gamma_\ell)}{G}+\gamma_m\gamma_\ell\right)}\sqrt{\Delta_aT_O},\label{eq:sq_snr}
\end{equation}
which is consistent with the results of ref.~\cite{malnou2019}. This reduces to the quantum-limited result of Eq.~\eqref{eq:alpha} at $G=1$ (no squeezing). Using this equation and the relationship between the SNR and the scan-rate (see Eq.~\eqref{eq:rate_vac}), one finds the squeezing-enhanced scan-rate quoted in the main text, Eq.~\eqref{eq:scan_rate_sq}. 

As pointed out in refs.~\cite{girvin2016axdm,malnou2019}, independently of squeezing, the SNR above has a maximum $\propto\gamma_s/\gamma_\ell$ at cavity resonance $\omega=0$ and at critical coupling $\gamma_m=\gamma_\ell$. Unfortunately, no amount of squeezing can push beyond this maximum, no matter the choice of the measurement-port coupling parameter $\gamma_m$. Physically, this is because the measurement-port susceptibility coefficient, $\bm{\chi}_{mm}$, which dictates the transfer of quanta from the measurement-port input to the measurement-port output, vanishes at these parameter settings. See Fig.~\ref{fig:susc_plot} for an illustration of this point. Therefore, when operating on-resonance and at critical coupling, no quanta incident on the measurement-port input---whether it be from squeezed-vacuum or initial thermal fluctuations---transfers to the measurement-port output. This establishes an intrinsic sensitivity limit on the peak SNR, which one can only increase by either decreasing the loss in the cavity or decreasing the operating temperature or both. On the other hand, squeezing can significantly increase the bandwidth over which the SNR is close to its peak value. To quickly see this, we can take the infinite squeezing limit of Eq.~\eqref{eq:sq_snr} and observe that $\lim_{G\rightarrow\infty}\overline{{\rm SNR}}_{\rm sq}\propto\gamma_s/\gamma_\ell$, which is independent of the cavity detuning, $\omega$. Increasing the bandwidth while maintaining peak-sensitivity is why squeezing can help DM-axion searches.

\subsection{Practical implementation to resolve cavity-induced phase shift}
\label{app:resonance_frequency_why}

\begin{figure}
    \centering
    \includegraphics[width=\linewidth]{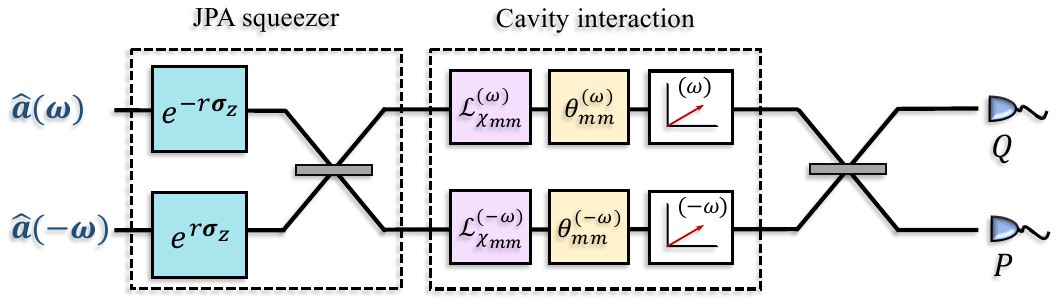}
    \caption{Squeezing-enhanced search with two-mode squeezing generated by a JPA~\cite{malnou2018jpa,backes2021}. To highlight the equivalence to the single-mode squeezing setup (see Fig.~\ref{fig:squeeze_cavity_channel}), we have decomposed the JPA squeezer into sub-elements consisting of single-mode squeezers on the individual modes $a(\omega)$ and $a(-\omega)$. Observe that we squeeze along $Q$ quadrature of the $a(\omega)$ mode and along the $P$ quadrature of the $a(-\omega)$ mode initially.}
    \label{fig:jpa_setup}
\end{figure}

Instead of single-mode squeezing, we can use two-mode squeezing generated by a Josephson Parametric Amplifier (JPA) \cite{malnou2018jpa}. The JPA is pumped with a strong field centered on twice the cavity resonance ($\omega_{\rm pump}=2\omega_c$) and generates photon-pairs in a two-mode squeezed vacuum state that are split in frequency across the cavity resonance---with one photon at frequency $\omega_c+\omega$ and its partner at $\omega_c-\omega$ ($\omega$ is the detuning). The two-mode squeezed state generated by this process is then injected into the cavity. As we shall show, a benefit of using two-mode squeezed vacuum, as opposed to single-mode squeezed vacuum, is that it naturally resolves the cavity-induced phase $\theta_{mm}$. Squeezing via a JPA was used in the recent axion-search demonstrations of \cite{malnou2019,backes2021}. 

A schematic of the setup is shown in Fig.~\ref{fig:jpa_setup}. For simplicity, we have not included an anti-squeezer (a ``JPA amp") after the cavity interaction, as including such in this setting would not change the forthcoming results for the SNR. Our goal is to show that the variance of the output (specifically, ${\rm Var}(Q)$ of the upper mode and ${\rm Var}(P)$ of the lower mode) is independent of $\theta_{mm}$---without introducing an external phase-shift---but otherwise equivalent to the single-mode squeezing case described previously, up to some small subtleties. To accomplish this goal, we track the covariance matrix through the circuit depicted in Fig.~\ref{fig:jpa_setup}. 

Assuming initial thermal equilibrium of both modes at temperature $T$, the covariance matrix just after the JPA squeezer but prior to the cavity interaction is,
\begin{align}
    \bm\sigma_{1}&\equiv \frac{N_T}{2} \begin{pmatrix}
    \mathbb{I}_2 & -\mathbb{I}_2 \\
     \mathbb{I}_2 & \mathbb{I}_2 
    \end{pmatrix}\begin{pmatrix}
    \e^{-2r\bm\sigma_z} & 0 \\
    0 & \e^{2r\bm\sigma_z}
    \end{pmatrix}\begin{pmatrix}
    \mathbb{I}_2 & \mathbb{I}_2 \\
     -\mathbb{I}_2 & \mathbb{I}_2 
    \end{pmatrix}\\
    &=N_T\begin{pmatrix}
    \cosh(2r)\mathbb{I} & \sinh(2r)\bm\sigma_{z} \\
    \sinh(2r)\bm\sigma_{z} &  \cosh(2r)\mathbb{I}
    \end{pmatrix},
\end{align}
where $\bm\sigma_z$ is the Pauli-z matrix. The above covariance matrix is that of a (noisy) two-mode squeezed vacuum. To find the covariance matrix after the cavity interaction, we make two observations: First, observe that $\abs{\bm\chi_{mm}(-\omega)}=\abs{\bm\chi_{mm}(\omega)}$---i.e. the cavity transmission is the same for each mode; see Fig.~\ref{fig:susc_plot} for an illustration. Second, $\theta_{mm}(-\omega)=-\theta_{mm}(\omega)$---i.e. the cavity-induced rotation is an odd function of frequency; see Eq.~\eqref{eq:theta_mm}. Now define the phase-rotation matrix, 
\begin{equation}
\begin{split}
    \bm V &\equiv \begin{pmatrix}
    \bm O[\theta_{mm}(\omega)] & 0 \\
    0 & \bm O[\theta_{mm}(-\omega)] 
    \end{pmatrix} \\
    &= \begin{pmatrix}
    \bm O[\theta_{mm}(\omega)] & 0 \\
    0 & \bm O^\top[\theta_{mm}(\omega)] 
    \end{pmatrix},
\end{split}
\end{equation}
where the second line follows since $\theta_{mm}$ is an odd function of frequency. From thus, the covariance matrix just after the cavity interaction is found,
\begin{equation}
\begin{split}
    \bm\sigma_{2}&=\abs{\bm\chi_{mm}(\omega)}^2 \bm V\bm\sigma_1\bm V^\top+ (1-\abs{\bm\chi_{mm}(\omega)}^2)\mathbb{I}_4\\
    &=\abs{\bm\chi_{mm}(\omega)}^2\bm\sigma_1+ (1-\abs{\bm\chi_{mm}(\omega)}^2)\mathbb{I}_4.
\end{split}
\end{equation}
The identity $\bm V\bm\sigma_1\bm V^\top=\bm\sigma_1$ can be derived from using the definitions of $\bm\sigma_1$ and $\bm V$ above, but this actually follows from the simple fact that $\bm O\bm\sigma_z\bm O=\bm\sigma_z$ for any $2\times2$ real, orthogonal matrix $\bm O$. We have thus shown that the output covariance matrix of the cavity is independent of the angle $\theta_{mm}$. 

Now we evaluate the variances of the quadrature measurements. Applying the second beam-splitter---situated just after the cavity interaction---undoes the beam-splitter interaction within the JPA squeezer, thereby reducing the output covariance matrix (prior to measurements) to a direct sum of independent single-mode squeezed states,
\begin{multline}
    \bm\sigma^{\rm(out)}=\abs{\bm\chi_{mm}(\omega)}^2\bm\left(\e^{-2r\bm\sigma_z}\oplus\e^{2r\bm\sigma_z}\right)\\+ (1-\abs{\bm\chi_{mm}(\omega)}^2)\mathbb{I}_4.
\end{multline}
We see that homodyne measurements along the $Q$ quadrature of mode $\hat{a}(\omega)$ and along the $P$ quadrature of mode $\hat{a}(-\omega)$ have equal variances, which are given by 
\begin{equation}
\begin{split}
    {\rm Var}\left[\hat{Q}(\omega)\right]&={\rm Var}\left[\hat{P}({-\omega})\right] \\ &= \abs{\bm\chi_{mm}(\omega)}^2\e^{-2r} + 1-\abs{\bm\chi_{mm}(\omega)}^2.
\end{split}
\end{equation}
Upon defining the gain, $G\equiv e^{2r}$, this result is equivalent to the single-mode squeezing case (without the anti-squeezer at the end).

Now to identify the signal. First, observe that $\abs{\bm\chi_{ms}(-\omega)}=\abs{\bm\chi_{ms}(\omega)}$; see Fig.~\ref{fig:susc_plot} for an illustration. Second, observe that $\theta_{ms}(-\omega)=-\theta_{ms}(\omega)$ (similar to $\theta_{mm}$), which follows from Eq.~\eqref{eq:theta_ma}. From these observations and upon inspecting Fig.~\ref{fig:jpa_setup}, we immediately write down the output signal,
\begin{equation}
\begin{split}
    \bm\mu^{\rm(out)}&\equiv \overbrace{\frac{1}{\sqrt{2}}\begin{pmatrix}
    \mathbb{I}_2 & \mathbb{I}_2 \\
     -\mathbb{I}_2 & \mathbb{I}_2 
    \end{pmatrix}}^{\text{Final beam-splitter}}
    \overbrace{\begin{pmatrix}
    \abs{\bm\chi_{ms}(\omega)}\bm O[\theta_{ms}(\omega)]\bm\mu_{s}\\
    \abs{\bm\chi_{ms}(-\omega)}\bm O[\theta_{ms}(-\omega)]\bm\mu_{s}
    \end{pmatrix}}^{\text{Axion-induced displacement}}\\
    &= \frac{\abs{\bm\chi_{ms}(\omega)}}{\sqrt{2}}\begin{pmatrix}
    \mathbb{I}_2 & \mathbb{I}_2 \\
     -\mathbb{I}_2 & \mathbb{I}_2 
    \end{pmatrix}
    \begin{pmatrix}
    \bm O[\theta_{ms}(\omega)]\bm\mu_{s}\\
    \bm O^\top[\theta_{ms}(\omega)]\bm\mu_{s}
    \end{pmatrix}\\
    &=\frac{\abs{\bm\chi_{ms}(\omega)}}{\sqrt{2}}\begin{pmatrix}
    \bm O[\theta_{ms}(\omega)]\bm\mu_{s}+\bm O^\top[\theta_{ms}(\omega)]\bm\mu_{s}\\
    \bm O^\top[\theta_{ms}(\omega)]\bm\mu_{s}-\bm O[\theta_{ms}(\omega)]\bm\mu_{s},
    \end{pmatrix}
\end{split}
\end{equation}
where $\bm\mu_s\equiv\mu_s(\cos\phi_s,\sin\phi_s)^\top$ is the axion field. 

From here, we find the signal amplitudes from the quadrature measurements, 
\begin{align}
    \ev{\hat{Q}(\omega)}&=\sqrt{2}\abs{\bm\chi_{ms}}\cos\left(\theta_{ms}(\omega)\right)\mu_s\cos\phi_s,\\
    \ev{\hat{P}(-\omega)}&=\sqrt{2}\abs{\bm\chi_{ms}}\sin\left(\theta_{ms}(\omega)\right)\mu_s\cos\phi_s.
\end{align}
Observe that, 
\begin{align}
    \ev{\hat{Q}(\omega)}^2+ \ev{\hat{P}(-\omega)}^2&=2\abs{\bm\chi_{ms}(\omega)}^2\mu_s^2\cos^2\phi_s,\\
    \implies \mathbb{E}\left[\ev{\hat{Q}(\omega)}^2+ \ev{\hat{P}(-\omega)}^2\right]&=2\abs{\bm\chi_{ms}(\omega)}^2n_s.
\end{align}
with the last equation being the average signal power. 

Using the fact that the SNRs for $\hat{Q}(\omega)$ and $\hat{P}(-\omega)$ add in quadrature for each detection step and integrating over many axion coherence times, we find the SNR of the PSD at the detuned frequency $\omega$,
\begin{equation}
\begin{split}
    \overline{\rm SNR}_{\rm JPA}&=\left(\frac{\mathbb{E}\left[\ev{\hat{Q}(\omega)}^2\right]}{{\rm Var}\left(\hat{Q}(\omega)\right)} + \frac{\mathbb{E}\left[\ev{\hat{P}(-\omega)}^2\right]}{{\rm Var}\left(\hat{P}(-\omega)\right)}\right)\sqrt{\Delta_aT_O}\\
    &=\frac{2\gamma_m\gamma_s n_s}{N_T\left(\frac{(\gamma/2)^2 + \omega^2-\gamma_m\gamma_\ell)}{G}+\gamma_m\gamma_\ell\right)}\sqrt{\Delta_aT_O},
\end{split}
\end{equation}
where Eqs.~\eqref{eq:chimm_expanded} and \eqref{eq:chima_expanded} have been used to expand $\bm\chi_{mm}$ and $\bm\chi_{ms}$ explicitly. This result is twice the SNR for the single-mode squeezing assisted search; see Eq.~\eqref{eq:sq_snr}. The factor of two comes from the fact that we are using two frequency modes. The performance of the JPA approach (per mode) is thus equivalent to that of the single-mode squeezing case. The real difference here is that we did not have to explicitly take care of the phase $\theta_{mm}$.

\subsubsection{Effect of resonance-frequency fluctuations}
We qualitatively investigate how a fluctuating cavity-resonance frequency affects the output noise power of the cavity when squeezing is present. Such fluctuations will generally introduce some anti-squeezing into the signal, due to the fluctuations induced in the angle $\theta_{mm}$. We can model this by assuming that the JPA pump-frequency is fluctuating. 

That is, consider the JPA pump to be at frequency $\Omega_p$. The JPA then generates photon-pairs with frequencies $(\Omega_p/2-\omega_c)\pm\omega$ in the cavity-rotating frame, where $\omega$ is the detuning from cavity resonance. Let us assume that $\Omega_p$ is a Gaussian random variable with mean $2\omega_c$ and standard deviation $\sigma_c$. Define $\varepsilon\equiv\Omega_p/2-\omega_c$, such that $\ev{\varepsilon}=0$ and $\ev{\varepsilon^2}=\sigma_c^2$ where $\varepsilon\sim\mathcal{N}(0,\sigma_c)$ is Gaussian distributed. The input modes to the cavity are then $a(\varepsilon\pm\omega)$. The relevant cavity parameters for each mode are $\abs{\bm\chi_{mm}(\varepsilon\pm\omega)}$ and $\theta_{mm}(\varepsilon\pm\omega)$. We calculate the variance in the $Q$ quadrature of the $\hat{a}(\varepsilon+\omega)$ output-mode (assuming this quadrature was squeezed initially), finding
\vspace{2em}
\begin{widetext}
\begin{multline}
    {\rm Var}\left[\hat{Q}(\varepsilon+\omega)\right]=N_T\abs{\bm\chi_{mm}(\varepsilon+\omega)}^2\Bigg(\e^{-2r}\left(\frac{1+\cos\left[\theta_{mm}(\varepsilon+\omega)+\theta_{mm}(\varepsilon-\omega)\right]}{2}\right) \\+\e^{2r}\left(\frac{1-\cos[\theta_{mm}(\varepsilon+\omega)+\theta_{mm}(\varepsilon-\omega)]}{2}\right)\Bigg)
    +N_T\left(1-\abs{\bm\chi_{mm}(\varepsilon+\omega)}^2\right).\label{eq:varq_fluctuations}
\end{multline}
\end{widetext}
A similar relation holds for ${\rm Var}[\hat{P}(\varepsilon-\omega)]$. 

We thus see that anti-squeezing appears in the noise power (the term proportional to $\e^{2r}$) when frequency-fluctuations are present, which can be detrimental to performance if the fluctuations are too large. We aim to establish a qualitative condition on the size of these fluctuations. Let us focus our attention on this anti-squeezing term in the above expression. 

Firstly, observe that $\theta_{mm}(\varepsilon-\omega)=-\theta_{mm}(\omega-\varepsilon)$ due to the oddness of $\theta_{mm}$. Using this fact, we can expand the argument of the cosine assuming the fluctuations are small, leading to the qualitative relation, 
\begin{equation}
\begin{split}
    \theta_{mm}(\varepsilon+\omega)+\theta_{mm}(\varepsilon-\omega)&=\theta_{mm}(\omega+\varepsilon)-\theta_{mm}(\omega-\varepsilon)\\ &\sim(\partial_\omega\theta_{mm})\varepsilon.
\end{split}
\end{equation}
Now the typical scale associated with changes in cavity quantities is the intrinsic linewidth, $\gamma_\ell$; hence, $\partial_\omega\theta_{mm}\sim1/\gamma_\ell$. If we substitute this qualitative expression into anti-squeezing term of Eq.~\eqref{eq:varq_fluctuations} and expand the cosine to first non-trivial order in its argument, we find that the anti-squeezed noise scales as $\e^{2r}\varepsilon^2/\gamma_\ell^2\sim\e^{2r}\sigma_c^2/\gamma_\ell^2$, where $\sigma_c$ is the typical size of the frequency fluctuations. For the anti-squeezed noise to contribute much less than thermal/vacuum fluctuations, we require that $\e^{2r}\sigma_c^2/\gamma_\ell^2\ll1\implies\sigma_c\ll \e^{-r}\gamma_\ell$. Recall that $\e^{r}=\sqrt{G}$, where $G$ is the gain of the squeezer. Once this constraint is no longer satisfied, the anti-squeezed noise roughly becomes the size of a thermal/vacuum fluctuation (the squeezing part can nonetheless tend to zero for large $r$), and the quadrature variance reduces approximately to the vacuum case without squeezing, at which point any performance enhancement gained by squeezing is completely lost.

A more stringent condition is to require the anti-squeezed noise to be much smaller than the squeezed noise, leading to the stronger constraint $\sigma_c\ll\e^{-2r}\gamma_\ell$. In this regime, anti-squeezing is completely negligible compared to all other noise terms in the quadrature variance.


\section{Derivation of near-optimal weights}\label{app:opt_weights}
We provide detailed derivations of the near-optimal weights---Eqs.~\eqref{eq:general_w} and \eqref{eq:general_wprime} in the main text---for a distributed network of quantum sensor-cavities.

Consider the signal amplitude along the real quadrature of $\hat{a}_1^{\rm(out)}$ from Eq.~\eqref{eq:a1_out}, $\sum_{k=1}^M\Re(w_{1k}\alpha_k)$.
Recall that, by definition, $\alpha_k=\abs{\bm\chi_{m_ks_k}}\mu_{s}\e^{\im(\phi_s+\theta_{m_ks_k})}$ and so
\begin{equation}
    \Re(w_{1k}\alpha_k)= \abs{w_{1k}}\abs{\bm\chi_{m_ks_k}}\mu_{s}\cos\left(\arg w_{1k}+\phi_s+\theta_{m_ks_k}\right).
\end{equation}
Assuming the axion-phase, $\phi_s$, is random and unknown, the amplitude is otherwise maximized in the phase variable for the choice $\arg w_{1k}=-\theta_{m_ks_k}$. This choice supports complete constructive interference between the output signal-amplitudes of the cavities when combining them, which, if left unaccounted for, would lead to a reduction in the total signal-power. 

For $\arg w_{1k}=-\theta_{m_ks_k}$, the total output amplitude $\sum_{k=1}^M\Re(w_{1k}\alpha_k)\propto\sum_{k=1}^M\abs{w_{1k}}\abs{\bm\chi_{m_ks_k}}$, up to a sensor-independent factor $\mu_s\cos\phi_s$. Thus, to further maximize the signal, we must maximize $\sum_{k=1}^M\abs{w_{1k}}\abs{\bm\chi_{m_ks_k}}$ with respect to the (magnitude of) the weights $\abs{w_{1k}}$, subject to the orthogonality constraint $\sum_{k=1}^M\abs{w_{1k}}^2=1$. Define the Lagrange function,
\begin{equation}
    \mathcal{L}=\sum_{k=1}^M\abs{w_{1k}}\abs{\bm\chi_{m_ks_k}}-\lambda\left(\sum_{k=1}^M\abs{w_{1k}}^2-1\right),
\end{equation}
where $\lambda$ is a Lagrange multiplier. Optimizing the Lagrangian with respect to the weights, $\partial_{\abs{w_{1k}}}\mathcal{L}=0$, suggests that $w_{1k}\propto\abs{\bm\chi_{m_ks_k}}$. Imposing the orthogonality condition on the weights supplies the pre-factor, from which one obtains, 
\begin{equation}
    w_{1k}=\frac{\abs{\bm\chi_{m_ks_k}}}{\sqrt{\sum_{j=1}^M\abs{\bm\chi_{m_js_j}}^2}}\e^{-\im\theta_{m_ks_k}},
\end{equation}
in accordance with Eq.~\eqref{eq:general_w} of the main text.

We now minimize the output noise power. For brevity, we temporarily define the real number $c_l\equiv \abs{w_{1l}}\abs{w^\prime_{l1}}\sqrt{\eta_l}$ and the phase $\varphi_l\equiv \arg w_{1l}+\arg w^\prime_{l1}$. A quick calculation shows that the $\hat{a}_1^{\rm(in)}$ mode contributes the following to the real quadrature of the $\hat{a}_1^{\rm(out)}$ mode,
\begin{equation}
\begin{split}
    \Re\left(\hat{a}_1^{\rm(out)}\right)&=\sum_{l=1}^Mc_l\Re\left(\e^{\im\varphi_l}\hat{a}_1^{\rm(in)}\right)+\dots\\
    &=\sum_{l=1}^Mc_l\left(\cos\varphi_l\Re\left(\hat{a}^{\rm(in)}_1\right)-\sin\varphi_l\Im\left(\hat{a}^{\rm(in)}_1\right)\right)\\&\quad\quad+\dots,
\end{split}
\end{equation}
where the dots represent contributions from thermal/vacuum noise. 

Assume that we squeeze along the real quadrature of the input, $\Re(\hat{a}^{\rm(in)}_1)$. The above relation tells us there there will be a contribution from anti-squeezing, as well as squeezing, when performing homodyne measurements along real quadrature of the output if $\varphi_l\neq0$. The anti-squeezed portion will cause an \textit{increase} in the noise. To neutralize this increase in noise, we must set $\varphi_l=0$. Using the result that $\arg w_{1l}=-\theta_{m_la_l}$ from Eq.~\eqref{eq:general_w}, the preceding argument implores use to choose $\arg w_{l1}^\prime=\theta_{m_la_l}$.

With the phases resolved, we minimize the noise with respect to the magnitude of the weights $\abs{w_{1l}^\prime}$. To accomplish this task, we maximize the function $\sum_{l=1}\abs{w_{1l}}\abs{w_{l1}^\prime}\sqrt{\eta_l}$ [coming from the first term in Eq.~\eqref{eq:a1_out}], subject to the orthogonality constraint $\sum_{l=1}^M\abs{w_{l1}^\prime}^2=1$. Define the Lagrange function,
\begin{equation}
    \mathcal{L}=\sum_{l=1}\abs{w_{1l}}\abs{w_{l1}^\prime}\sqrt{\eta_l}-\lambda\left(\sum_{l=1}^M\abs{w_{l1}^\prime}^2-1\right),
\end{equation}
where $\lambda$ is the Lagrange multiplier. Assuming the weights of $\bm W$ are set by Eq.~\eqref{eq:general_w} and optimizing the Lagrangian with respect to the weights, $\partial_{\abs{w_{k1}^\prime}}\mathcal{L}=0$, implies that $\abs{w_{k1}^\prime}\propto \abs{w_{1k}}\sqrt{\eta_k}$. Imposing the orthogonality condition on the weights supplies the pre-factor, from which we obtain
\begin{equation}
    w_{k1}^\prime = \frac{\abs{\bm\chi_{m_ks_k}}\abs{\bm\chi_{m_km_k}}}{\sqrt{\sum_{j=1}^M\abs{\bm\chi_{m_js_j}}^2\abs{\bm\chi_{m_jm_j}}^2}}\e^{\im\theta_{m_ks_k}},
\end{equation}
in accordance with Eq.~\eqref{eq:general_wprime} of the main text. 


\section{Making contact with classical cavity language}
\label{app:physical_property}

For a continuous spectrum, we define the axion signal field number spectral-density as
\begin{align}
    \expval{\hat{a}^\dagger_s(\omega) \hat{a}_s(\omega')}=2\pi n_s  \delta(\omega-\omega').
\end{align}
Under this convention, the commutation relation $[\hat{a}_s(\omega'),\hat{a}^\dagger_s(\omega)]=2\pi \delta(\omega'-\omega)$.
The time domain field operator is therefore
\begin{align}
    \hat{a}_s(t)=\int \frac{d\omega}{2\pi} \hat{a}_s(\omega) e^{i\omega t},
\end{align}
which satisfies the commutation relation $[\hat{a}_s(t'),\hat{a}^\dagger_s(t)]=\delta(t-t')$.
Here, $\hat{a}_s(t)$ has unit of $\sqrt{\mbox{quanta/second} }$ and $\hat{a}_s(\omega)$ has unit of $\sqrt{\mbox{quanta/HZ}}$. Hence, $n_s$ has units $\mbox{quanta/second/HZ}$. The axion-induced signal field flux is
\begin{equation}
\expval{\hat{a}^\dagger_s(t)\hat{a}_s(t)} =\int \frac{d\omega}{2\pi} n_s \approx  n_s\Delta_a.
\end{equation}
Therefore, the signal power coming out of the cavity from the measurement port, in units of $\text{energy/second}$, is
\begin{equation}\label{eq:psigqm}
\begin{split}
    \mathcal{P}_{\rm sig} &\approx \abs{\bm\chi_{ms}(0)}^2\hbar\omega_cn_s\Delta_a\\
    &=\frac{4\gamma_m/\gamma_\ell}{(\gamma_m/\gamma_\ell+1)^2}\hbar\omega_cn_s\Delta_a\gamma_s/\gamma_\ell,    
\end{split}
\end{equation}
where $\omega_c$ is the cavity resonance-frequency (equal to the axion mass, assuming the axion is resonant with the cavity). The signal power, $\mathcal{P}_{\rm sig}$, is related to the power inside the cavity, $\mathcal{P}_\mathrm{cav}$ of Eq.~\eqref{P_cav}, via $\mathcal{P}_{\rm sig}\approx\mathcal{P}_{\rm cav}\beta/(1+\beta)^2$, where $\beta\equiv\gamma_m/\gamma_\ell$ and the approximation holds for $Q_a\gg Q_c$.

To express the axion-conversion rate $\gamma_s$ in terms of physical parameters, we consider the classical expression for signal power,
\begin{equation}\label{eq:psigc}
    \mathcal{P}_{\rm sig}=\frac{\beta}{1+\beta}g_{a\gamma}^2\frac{\rho_a}{m_a}B^2V\eta Q_{\rm eff},
\end{equation}
where $Q_{\rm eff}^{-1}=(1+\beta)Q_c^{-1}+Q_a^{-1}$; $Q_c$ and $Q_a$ are the intrinsic cavity and axion quality factors, respectively. The readout/measurement process introduces additional loss, which is captured by $\beta=\gamma_m/\gamma_\ell$. The combination $Q_l^{-1}\equiv (1+\beta)Q_c^{-1}$ is often referred to as the loaded quality factor. Here, $m_a$ is the mass of the axion, which is equal to the resonant frequency of the cavity signal mode $\omega_c$; $\rho_a$ is the local axion dark matter density; $g_{a\gamma}$ is the coupling constant of mass dimension $-1$; $B$ is the magnetic field; $V$ is the volume of the cavity; and $\eta$ is the geometrical overlap between cavity mode and the axion dark matter field.

To relate \eqref{eq:psigc} to \eqref{eq:psigqm}, 
we define the coupling $\gamma_s$ between the axion and a cavity photon to be
\begin{equation}
    \gamma_s= (g_{a\gamma}B\sqrt{\eta})^2 2\pi \delta(\omega-m_a)\xrightarrow[]{\rm finite~\Delta_a} (g_{a\gamma}B\sqrt{\eta})^2\frac{1}{4\Delta_a}.
\end{equation}
We also identify $n_s =\rho_a V/m_a $ as the flux of axions per axion bandwidth to be found inside the cavity. We now rewrite \eqref{eq:psigqm} as
\begin{equation}
\begin{split}
    \mathcal{P}_{\rm sig}=\frac{4\gamma_m/\gamma_\ell}{(\gamma_m/\gamma_\ell+1)}\frac{\hbar\omega_c}{\gamma}  n_s\Delta_a\gamma_s
    \end{split}
\end{equation}
where $\gamma\approx\gamma_\ell+\gamma_m$ can be identified as the total width of the cavity given by $\omega_c/Q_{\rm eff}$, and $\frac{\gamma_m/\gamma_\ell}{(\gamma_m/\gamma_\ell+1)}$ is the same as $\beta/(1+\beta)$.

Equation \eqref{eq:psigqm} is completely general with respect to any signal field with occupation number $n_s$ and coupling rate $\gamma_s$. For instance, if the signal comes from dark-photon dark-matter that has a mass $m_{A'}$ and the kinetic mixing $\epsilon$ with photons, then one may take,
\begin{equation}
\begin{split}
    \gamma_s&\xrightarrow[]{\rm dark~photon}(\epsilon m_{A'}\sqrt{\eta})^2\frac 1{4\Delta_{A'}}\\
    n_s&\xrightarrow[]{\rm dark~photon} \rho_{A'} V/m_{A'},
    \end{split}
\end{equation}
where $\Delta_{A'}$ represents the bandwidth of the dark-photon.

\section{GKP-assisted DM search}\label{appendix:gkp_app}
We consider using a more exotic quantum resource---the GKP state~\cite{gottesman2012}---to assist in a DM search. We first provide some background and technical details regarding GKP states, the \textbf{SUM}-gate detection method, and then proceed with the GKP-assisted DM search protocol.
\subsection{Description of the GKP state}

The GKP state was originally developed for quantum error correction as a way to protect quantum information (hosted in qubits \cite{gkp2001} or other bosonic systems \cite{noh2020_o2o}) from noise.  Heuristically, the ideal, canonical GKP-state $\ket{\rm GKP}$ is an infinite lattice in the phase-space of a bosonic mode, which is translation invariant with respect to shifts in the $Q$ or $P$ quadrature by an amount $\sqrt{2\pi}$. Due to the translation-invariant property of the GKP state in phase-space, it is possible to simultaneously and precisely measure \textit{both} quadrature variables, $Q$ and $P$, modulo $\sqrt{2\pi}$. This is essentially due to the fact that each lattice point on the GKP-grid is infinitely squeezed along both the $Q$ and $P$ directions, which is most evident when we write out the canonical GKP state in the $Q$ and $P$ quadrature bases~\cite{gkp2001,duivenvoorden2017single,noh2020_o2o},
\begin{equation}
    \ket{\rm GKP}\propto\sum_{n\in\mathbb{Z}}\ket{n\sqrt{2\pi}}_q=\sum_{n\in\mathbb{Z}}\ket{n\sqrt{2\pi}}_p,
\end{equation}
where $\ket{\cdot}_q$ ($\ket{\cdot}_p$) represents a $Q$-quadrature ($P$-quadrature) basis state. This state however is unnormalizable, as infinite squeezing in each quadrature leads to an infinite number of quanta in the perfect GKP-state. We shall instead concern ourselves with finite GKP states constrained to a finite-region in phase space, defined as
\begin{equation}
    \ket{\rm GKP_{\Delta}}\propto\e^{-\Delta^2\hat{n}}\ket{\rm GKP},
\end{equation}
up to normalization, and $1/\Delta$ is the effective radius in phase-space (emanating from the origin) which supports the GKP grid. The above state is pure, however it is easier to deal with its noisy version, which is an incoherent mixture of GKP states~\cite{noh2020surface_code},
\begin{multline}
    \mathcal{N}_{{\bm\sigma}_{\rm GKP}}\left(\rm GKP\right)\propto\\
    \int_{\mathbb{R}^2}\dd{\bm\mu}\exp\left(-\bm\mu^\top{\bm\sigma}_{\rm GKP}^{-1}\bm\mu\right)\hat{D}_{\bm\mu}\dyad{\rm GKP}\hat{D}^\dagger_{\bm\mu},
\end{multline}
where $\bm\sigma_{\rm GKP}=2\left(\frac{1+\e^{-\Delta^2}}{1-\e^{-\Delta^2}}\right)\mathbb{I}_2\equiv\mathbb{I}_2/G$. One can multiply $\bm\sigma_{\rm GKP}$ by a factor of $N_T=1+2\Bar{n}_T$ to include initial thermal fluctuations in GKP state-preparation. Observe that the above description is a perfect GKP state going through an additive noise channel with equal noise added to each quadrature. We write the state in this seemingly complicated way as it is easier to generalize to arbitrary Gaussian processes acting on the GKP state that we consider later.

\subsection{\textbf{SUM}-gate}

In the GKP-assisted detection strategy, we couple the signal-mode, described by quadrature operators $\hat{Q}$ and $\hat{P}$, to an ancilla-mode in a GKP state, described by quadrature operators $\hat{Q}_{\rm anc}$ and $\hat{P}_{\rm anc}$, via the unitary operation, $\widehat{\textbf{SUM}}={\exp}(-\im\hat{Q}\hat{P}_{\rm anc})$, which acts on the quadrature operators as,
\begin{align}
\begin{aligned}
    \widehat{\textbf{SUM}}:\qq{} \hat{Q}&\rightarrow\hat{Q},& \qq{} \hat{P}&\rightarrow \hat{P}-\hat{P}_{\rm anc},&\\
    \hat{Q}_{\rm anc}&\rightarrow\hat{Q}_{\rm anc}+\hat{Q},&\qq{}\hat{P}_{\rm anc}&\rightarrow\hat{P}_{\rm anc}.&
\end{aligned}
\end{align}
It is easy to derive a symplectic-matrix representation of the \textbf{SUM}-gate and its inverse, which we immediately write in $2\times2$ blocks as
\begin{align}
    \textbf{SUM}&=
    \begin{pmatrix}
    \mathbb{I}_2 & -\Pi_P\\
    \Pi_Q & \mathbb{I}_2
    \end{pmatrix} \\ \qq{and}  \textbf{SUM}^{-1}&=
    \begin{pmatrix}
    \mathbb{I}_2 & \Pi_P\\
    -\Pi_Q & \mathbb{I}_2
    \end{pmatrix},\label{eq:symplectic_sum}
\end{align}
where $\Pi_Q={\rm diag}(1, 0)$ and $\Pi_P={\rm diag}(0, 1)$ represent projections along the $Q$ quadrature and $P$ quadrature of the respective modes.

\subsection{Joint distribution}
We formally derive the joint PDF for two imperfect GKP states which are coupled via a \textbf{SUM}-gate. This is precisely the situation for GKP-assisted axion-search, where the output of the cavity (in a noisy GKP state) couples to an imperfect GKP ancilla via the \textbf{SUM}-gate just before homodyne measurements are performed on each mode. Consider modes 1 and 2 in noisy GKP-states $\mathcal{N}_{\bm Y_k}(\rm GKP)$ with noise matrices $\bm{Y}_k=y_k\mathbb{I}_2$, where $y_k\geq0$ and $k\in\{1,2\}$. Defining $\bm Y\equiv \oplus_{k=1}^2\bm Y_k$, the joint state can then be written as,
\begin{multline}
    \mathcal{N}_{\bm Y}({\rm GKP}^{\otimes 2})\propto\\\int_{\mathbb{R}^4}\dd{\bm\mu}\exp\left(-\bm\mu^\top{\bm Y }^{-1}\bm\mu\right)\hat{D}_{\bm\mu}\left(\dyad{\rm GKP}^{\otimes 2}\right)\hat{D}^\dagger_{\bm\mu},
\end{multline}
where, for instance,
\begin{equation}
    {\bm Y}^{-1}=
    \begin{pmatrix}
    \frac{y_2}{y_1+y_2}\mathbb{I}_2 & 0\\
    0 & \frac{y_1}{y_1+y_2}\mathbb{I}_2
    \end{pmatrix}.
\end{equation}
We now apply a two-mode \textbf{SUM}-gate, formally resulting in the correlated state 
\begin{equation}
\mathcal{N}_{\bm Y}({\rm GKP}^{\otimes 2})\rightarrow\widehat{\textbf{SUM}}\Big(\mathcal{N}_{\bm Y}({\rm GKP}^{\otimes 2})\Big)\widehat{\textbf{SUM}}^\dagger.
\end{equation}
Two simplifying observations are in order. Firstly, the perfect GKP-states are invariant under the \textbf{SUM}-gate, i.e., 
\begin{equation}
\widehat{\textbf{SUM}}\left(\dyad{\rm GKP}^{\otimes 2}\right)\widehat{\textbf{SUM}}^\dagger=\dyad{\rm GKP}^{\otimes 2}.
\end{equation}
Secondly, since the \textbf{SUM}-gate is a symplectic transformation, the action of \textbf{SUM} on the Weyl operators $\hat{D}_{\bm\mu}$ can be taken care of by a redefinition of the integration variable,
\begin{equation}
\bm\mu^\prime\equiv(\textbf{SUM}^{-1})^\top\bm\mu
\end{equation}
where $\textbf{SUM}^{-1}$ (without a hat) is the inverse of the symplectic matrix for the \textbf{SUM}-gate from Eq.~\eqref{eq:symplectic_sum}. Note also that $\dd{\bm\mu}=\dd{\bm\mu^\prime}$ since \textbf{SUM} is a symplectic transformation. Upon defining a new noise matrix,
\begin{equation}
    {\bm Y}^\prime \equiv \left(\textbf{SUM}^{-1}\right)^\top{\bm Y}\left(\textbf{SUM}^{-1}\right),
\end{equation}
it follows that,
\begin{equation}
    \widehat{\textbf{SUM}}\Big(\mathcal{N}_{\bm Y}({\rm GKP}^{\otimes 2})\Big)\widehat{\textbf{SUM}}^\dagger=\mathcal{N}_{\bm Y^\prime}({\rm GKP}^{\otimes 2}),
\end{equation}
where the noise-matrix $\bm Y^\prime$ is the covariance matrix for the multi-variate Gaussian PDF of the two-mode state. 

In the GKP-assisted DM search protocol (see below), orthogonal homodyne measurements are performed on the signal and the ancilla, respectively; in the notation here, this corresponds to a quadrature measurement $P_1$ of mode 1 and $Q_2$ of mode 2, discarding the other quadratures $Q_1$ and $P_2$. The reduced PDF of the measurement outcomes, after discarding $Q_1$ and $P_2$, is uncorrelated in the remaining variables $P_1$ and $Q_2$. Thus, measurements along these quadrature directions are independent and described by uni-variate Gaussian PDFs.

\subsection{GKP search protocol}\label{sec:gkp}
Here, we consider using the GKP state~\cite{gkp2001} in a dark-matter search. We describe how, in the ideal case, the GKP search strategy can enhance the scan-rate by a constant factor relative to a squeezing-enhanced search with the same amount of squeezing, however, we also show a practical no-go for GKP-assisted scan when ancillary measurement noise is taken into consideration. 

The potential benefits of this new strategy derive from the non-Gaussian resource consumed in its implementation---the GKP state. Colloquially, the canonical GKP-state is a grid in phase-space, with unambiguously identifiable lattice-points separated by $\sqrt{2\pi}$. One can simultaneously measure the $Q$ and $P$ quadrature variables of the GKP state ${\rm mod}{\sqrt{2\pi}}$ (ignoring practical noise sources), due to the $\sqrt{2\pi}$ translation-invariance of grid~\cite{gkp2001,duivenvoorden2017single}. The Heisenberg uncertainty principle holds good due to the ${\rm mod}{\sqrt{2\pi}}$ structure~\cite{gkp2001}. 

Practically though, if a displacement of the grid occurs, one must infer in what direction (say, left or right for a 1D displacement) and with what magnitude the displacement occurs, however due to the $\sqrt{2\pi}$ translation-invariance, this inference is ambiguous if the displacement is close to (or known up to) half a lattice spacing, $\sqrt{\pi/2}$.~\footnote{For instance, if a displacement of $\sqrt{\pi}+\epsilon$ to the right occurs, say along the $Q$-quadrature, then one cannot distinguish this from a displacement of $\sqrt{\pi}-\epsilon$ to the left due to the $\sqrt{2\pi}$ translation-invariance of the GKP state. One can deal with this by making a biased decision based on which case is most likely to occur.} This ambiguity, however, is not a problem for very small displacements (and fairly low amount of noise), such as an axion-induced displacement, as we now discuss. For further details regarding the GKP state, see Appendix~\ref{appendix:gkp_app}; see also ref.~\cite{zhuang2020DQSgkp} for applications of GKP in general distributed-sensing scenarios.

\begin{figure}[t]
    \centering
    \includegraphics[width=\linewidth]{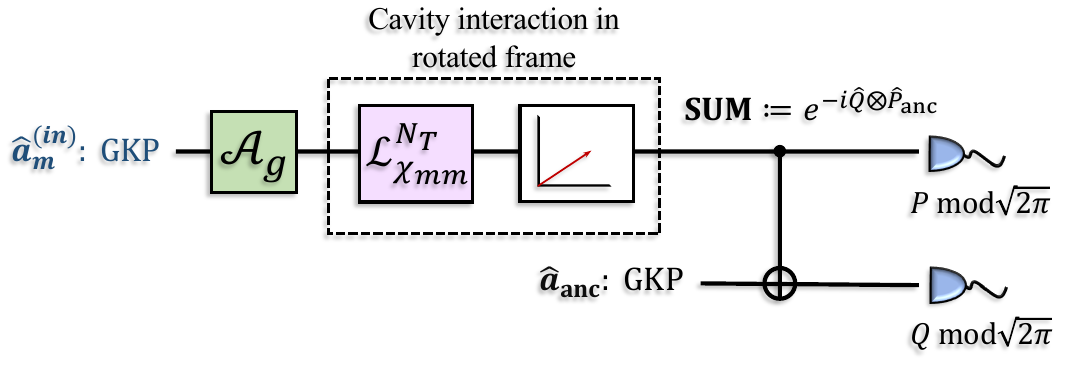}
    \caption{GKP-assisted search. A GKP state is prepared and injected into the microwave cavity. Just prior to injection, the input mode is amplified by a (quantum-limited) phase-insensitive amplifier $\mathcal{A}_g$ with gain chosen as $g=1/\abs{\bm\chi_{mm}}^2$, which converts the cavity-transmission loss to additive Gaussian noise---a necessary addition when using GKP states. The output mode of the cavity, which is slightly displaced due to the axion-field, then interacts via the \textbf{SUM}-gate with an ancillary mode prepared in a GKP state. Finally, orthogonal homodyne measurements are performed on each mode.}
    \label{fig:GKP_detection}
\end{figure}

The GKP-assisted protocol consists of the following steps (see Fig.~\ref{fig:GKP_detection}): 1) A GKP state is prepared and sent through a phase-insensitive amplifier $\mathcal{A}_g$, with gain $g=1/\abs{\bm\chi_{mm}}^2$ chosen in such a way to convert the cavity transmission-loss to additive Gaussian noise; 2) the amplified GKP-state is injected into a microwave cavity, where an axion-induced displacement of the cavity field occurs [Note here we have gone back to the rotated frame for a single cavity.]; 3) the output of the cavity (the signal, a displaced GKP state) is coupled via the \textbf{SUM}-gate to an ancilla-mode prepared in an ancillary GKP-state; 4) homodyne detection is performed along the $Q$-quadrature of the GKP-ancilla and along the $P$-quadrature of the signal-mode. The independent measurements are then combined in quadrature to infer the SNR.

For simplicity in this initial presentation, we shall assume added noise (vacuum-added noise from the cavity, intrinsic noise in the imperfect GKP resource-states etc.) is much smaller than half the lattice-spacing of the GKP grid, $\sqrt{\pi/2}$. In this regime, we can safely make a Gaussian approximation for the modes and focus our attention to a particular lattice point of the GKP grid (the origin in phase-space) about which all displacements are measured with respect to. We shall return to this approximation later. We can then describe a GKP state in terms of its Wigner function restricted to the origin, which is a Gaussian function with moments $(\bm\mu_{\rm GKP}=\bm0\mod\sqrt{2\pi},\bm\sigma_{\rm GKP})$, where the covariance matrix
\begin{equation}
    \bm\sigma_{\rm GKP}= \frac{N_T}{G}\mathbb{I}_2.\label{eq:gkp_noise}
\end{equation}
Observe that we have included initial thermal fluctuations that may be present during GKP-state preparation. Following convention, we define the squeezing of the GKP state in dB as $s_{\rm dB}\equiv10\log(G)$. Intuitively, the local-variance of the GKP-state, $\bm\sigma_{\rm GKP}$, represents the typical size of the fluctuations of a lattice-point on the GKP-grid, due to the finite squeezing used to prepare the GKP state (plus initial thermal fluctuations).

For a (quantum-limited) amplifier of gain $g$, the GKP state just before entering the cavity gets mapped to $\bm\sigma_{\rm GKP}\rightarrow g\bm\sigma_{\rm GKP} + (g-1)\mathbb{I}_2$. Choosing $g=1/\abs{\bm\chi_{mm}}^2$ and using the general input-output Eqs.~\eqref{eq:cavity_out_mu} and \eqref{eq:cavity_out_sigma}, the output of the cavity can be found,
\begin{align}
    {\bm \mu}_m^{\rm(out)}&=\abs{\bm{\chi}_{ms}}\bm{O}(\theta_{ms})\bm{\mu}_s\\
     \bm{\sigma}_m^{\rm (out)}&=\frac{N_T}{G}\mathbb{I}_2+2N_T\left(1-\abs{\bm{\chi}_{mm}}^2\right)\mathbb{I}_2,\label{eq:cavity_out_gkp}
\end{align}
where it is understood that ${\bm \mu}_m^{\rm(out)}$ is defined ${\rm mod}\sqrt{2\pi}$. We shall drop the modulo dependence for brevity. Observe that extra vacuum noise has been added to the output (seen as a factor of 2 in $\bm\sigma_m^{\rm(out)}$ above) from phase-insensitive amplification. The cavity output then couples via the \textbf{SUM}-gate to the GKP ancilla, which has local moments $(\bm\mu_{\rm anc}=\bm0, \bm\sigma_{\rm anc}=N_T/G_{\rm anc}\mathbb{I}_2)$, leading to a formal expression for the reduced moments,
\begin{align}
   \textbf{SUM}: 
   {\bm \mu}_m^{\rm(out)} &\rightarrow {\bm \mu}_m^{\rm(out)},
    {\bm \sigma}_m^{\rm(out)} \rightarrow {\bm \sigma}_m^{\rm(out)} + \Pi_P\bm\sigma_{\rm anc}\Pi_P,\\
    \bm\mu_{\rm anc} &\rightarrow\Pi_Q{\bm \mu}_m^{\rm(out)},
     \bm\sigma_{\rm anc} \rightarrow \bm\sigma_{\rm anc} + \Pi_Q{\bm \sigma}_m^{\rm(out)}\Pi_Q,
\end{align}
where $\Pi_Q={\rm diag}(1,0)$ and $\Pi_P={\rm diag}(0,1)$ are projections along the $Q$ quadrature and $P$ quadrature of the respective single-mode spaces. Since $\bm\mu_{\rm anc}=\bm0$, we see that the mean vector of the signal goes unchanged while the mean vector of the GKP-ancilla gets translated along the $Q$-quadrature by the $Q$-component of the signal, $\Pi_Q{\bm \mu}_m^{\rm(out)}$. We also observe a shuffling of $Q$- and $P$-quadrature noises between the signal and the ancilla. Written out explicitly, the noise in the $P$ ($Q$) quadrature of the signal (ancilla) is $N_T/G_{\rm eff}+2N_T(1-\abs{\bm\chi_{mm}}^2)$, where $1/G_{\rm eff}\equiv 1/G + 1/G_{\rm anc}$. 

After coupling the signal to the ancilla, the $Q$ quadrature of the ancilla and $P$ quadrature of the signal are then measured. The variances of these measurements add in quadrature, which---after integrating over a coherence time of the axion field---leads to an estimate for the SNR for GKP-assisted search,
\begin{equation}
    \overline{{\rm SNR}}_{\rm GKP}\approx\frac{2\gamma_m\gamma_s n_s}{N_T\left(\left(\frac{(\gamma/2)^2+\omega^2}{G_{\rm eff}}\right)+2\gamma_m\gamma_\ell\right)}\sqrt{\Delta_aT_O},\label{eq:gkp_snr}
\end{equation}
where the approximation indicates the Gaussian approximation for the GKP state, which is valid whenever the displacements and noise are much smaller than $\sqrt{\pi/2}$. Observe that the GKP-assisted detection cannot increase the peak SNR. This is easily seen in the infinite squeezing limit. In this limit, the first term in the denominator vanishes, and the factors of two in the numerator and denominator of the remaining terms cancel, leading to the same peak SNR as the critically-coupled, quantum-limited scenario. 

We now consider two limiting cases of the GKP-assisted search and analyze its performance relative to a squeezing-enhanced search, when the squeezing $G$ is taken as equal for each search method. In the first case, we relax the noise assumption about the ancillary measurement and assume that it is negligible. In this limit, it is shown that a GKP-assisted search has a scan-rate (almost) double that of a squeezing-enhanced search when the squeezing levels of each are comparable. In the second case, we include the effects of ancillary measurement noise and show a practical no-go in terms GKP-assisted performance. This results places strict constraints on any practical performance-enhancements to be gained from GKP states when searching for a random signal, at least within the framework of the cavity model presented here. 

\begin{figure}[t]
    \centering
    \includegraphics[width=\linewidth]{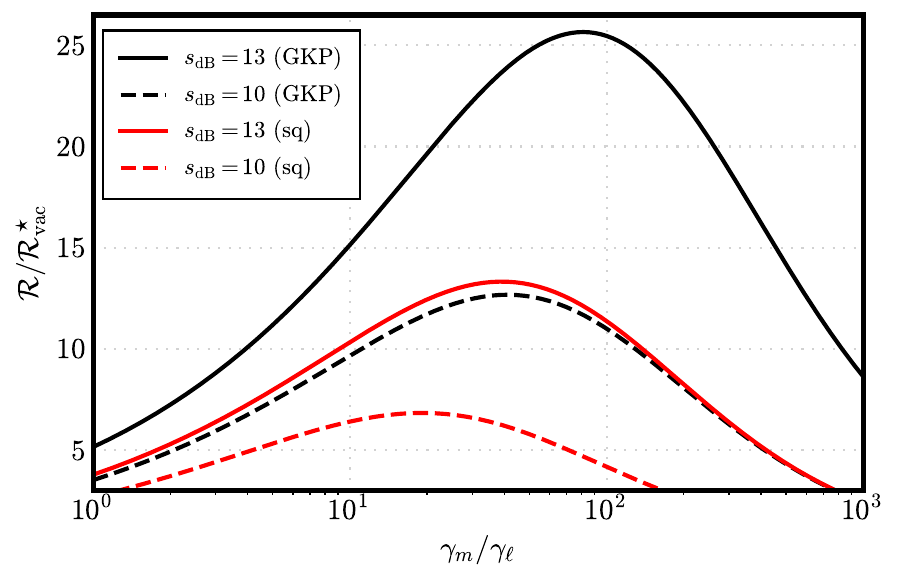}
    \caption{Comparison of scan-rate for squeezing-enhanced search [Eq.~\eqref{eq:scan_rate_sq}] and GKP-assisted search [Eq.\eqref{eq:scan_rate_gkp}] as a function of the ratio $\gamma_m/\gamma_\ell$ for several representative values of the squeezing in dB, $s_{\rm dB}=10\log_{10}(G)$. There exists optimal values for $\gamma_m$ depending on the squeezing level ($\gamma_m/\gamma_\ell\approx 2G$ for squeezing-enhanced and $\gamma_m/\gamma_\ell\approx 4G$ for GKP-assisted, respectively). We consider equal squeezing for each search method at $s_{\rm dB}= 10, 13$ dB (corresponding to $G= 10, 20$). The curves for the GKP-assisted search are for negligible ancillary-measurement noise and correspond to the Gaussian approximation for the GKP state.}
    \label{fig:rate_ratio}
\end{figure}

\subsubsection{Case 1: Negligible measurement-noise}
Assuming a negligible amount of ancillary measurement-noise, we show that the GKP-assisted scan performs about twice that of a squeezing-enhanced scan, in principle, when the squeezing levels for each method are comparable. If the GKP ancillary noise is negligible, then $G_{\rm anc}\gg G$ and therefore $G_{\rm eff}=G$. Substituting this value into the SNR of Eq.~\eqref{eq:gkp_snr}, we find the scan-rate of the GKP-assisted search relative to the optimal quantum-limited search,
\begin{equation}
    \frac{\mathcal{R}_{\rm GKP}}{\mathcal{R}^\star_{\rm QL}}= \frac{27\sqrt{G}\left(\frac{\gamma_m}{\gamma_\ell}\right)^2}{8\left(\frac{\left(\frac{\gamma_m}{\gamma_\ell}+1\right)^2}{4G}+2\frac{\gamma_m}{\gamma_\ell}\right)^{3/2}}.\label{eq:scan_rate_gkp}
\end{equation}

We plot this ratio in Fig.~\ref{fig:rate_ratio} and compare with the squeezing-enhanced search. The optimal coupling value for the GKP case is $\gamma_m/\gamma_\ell\approx4G$. At this optimal setting and in the large squeezing limit, the above ratio reduces to $\mathcal{R}^\star_{\rm GKP}/\mathcal{R}^\star_{\rm QL}\approx 1.4 G$ and thus $\mathcal{R}^\star_{\rm GKP}/\mathcal{R}^\star_{\rm sq}\approx 2 G_{\rm GKP}/G_{\rm sq}$, which comes from the analysis just below Eq.~\eqref{eq:scan_rate_sq}. Here we have explicitly notated the squeezing levels in the GKP-assisted scan and squeezing-enhanced scan by $G_{\rm GKP}$ and $G_{\rm sq}$, respectively. Therefore, when the squeezing levels for each scan method are comparable---i.e., $G_{\rm GKP}\approx G_{\rm sq}$---the GKP-assisted scan asymptotically outperforms the squeezing-enhanced scan by about a factor of 2.

\begin{figure}
    \centering
    \includegraphics[width=\linewidth]{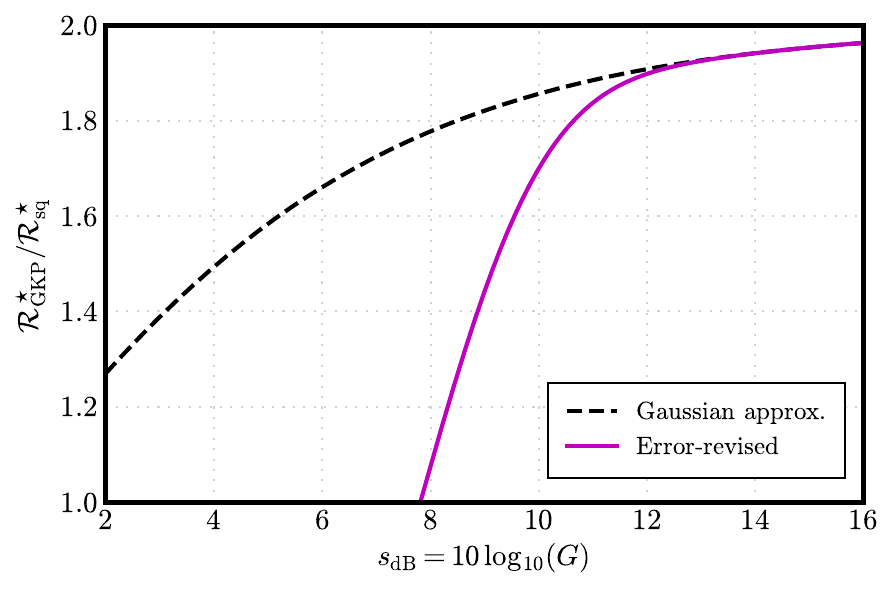}
    \caption{GKP-assisted scan-rate enhancement relative to squeezing-enhanced scan, for negligible GKP-ancillary measurement-noise. Vertical axis is the ratio between the GKP-assisted scan-rate, $\mathcal{R}_{\rm GKP}^\star$, and squeezing-enhanced scan-rate, $\mathcal{R}_{\rm sq}^\star$, evaluated at their respective optimal-coupling values, $\gamma_m/\gamma_\ell\approx 4G$ and $\gamma_m/\gamma_\ell\approx 2G$, respectively, assuming equal squeezing level $G$ for each.}
    \label{fig:rate_gkp_v_sq}
\end{figure}

In Fig.~\ref{fig:rate_gkp_v_sq}, we show a more detailed comparison between the two search methods by considering the ratio of the scan-rates evaluated at their respective optimal-coupling values, $\mathcal{R}_{\rm GKP}^\star/\mathcal{R}_{\rm sq}^\star$. The dashed-line is the theoretical prediction for GKP-assisted scan-rate from the Gaussian analysis just presented, while the solid error-revised curve takes into consideration the ${\rm mod}\sqrt{2\pi}$-structure of the GKP grid (and thus deviations from Gaussianity). See Appendix~\ref{appendix:gkp_app} for further details of the latter. Observe that the asymptotic enhancement-value of 2 is reached in the large-squeezing limit. Two further observations are noteworthy: (1) When the squeezing is greater than 10dB, the discrepancy between the Gaussian approximation and the actual estimate is quite small ($\lesssim10\%$); (2) There is a ``break-even" point at $\approx$8dB of squeezing when the GKP-assisted scan begins to outperform the squeezing-enhanced scan. 

Before moving to the next section, we point out that we are comparing the two schemes with the same level of squeezing, while the GKP state has more energy than the squeezed vacuum state. We have chosen such a comparison as the level of squeezing represents the capability of state-engineering and an energy constraint is irrelevant in our sensing scenario.


\subsubsection{Case 2: Non-negligible measurement-noise}
In the case where the GKP-ancilla has the same noise as the GKP-state injected into the cavity, $G_{\rm anc}=G$, which is perhaps the most practical case, we find that there is a practical no-go when it comes to performance enhancement of a GKP-assisted search versus a single-mode squeezing-enhanced search. The reason is that, in GKP-assisted detection, although one can measure both quadrature variables of the signal and thus gain a factor of 2, the noise also gets increased by (more than) a factor of 2, due to the noise in the ancilla and noise from the amplifier that is required to convert the cavity-transmission loss to additive Gaussian noise. In particular, from the previous analysis, the noise in the P (Q) quadrature of the signal (ancilla) just after the \textbf{SUM}-gate is $1/G_{\rm eff}+2(1-\abs{\bm\chi_{mm}}^2)$ where, without loss of generality, we have assumed vacuum-dominated noise and thus set $N_T=1$. Assuming an ancilla GKP-state which is identical to the GKP-state that is injected into the cavity, we have $G_{\rm eff}=G/2$, and thus the quadrature noise is $2\left(1/G+1-\abs{\bm\chi_{mm}}^2\right)$. Observe that the noise in the single-mode squeezing case is $\abs{\bm\chi_{mm}}^2/G+1-\abs{\bm\chi_{mm}}^2$. Since $0\leq\abs{\bm\chi_{mm}}^2\leq1$, there is (more than) twice as much quadrature noise in the GKP-assisted detection than the squeezing assisted detection. In the GKP-assisted detection, the factor of two from the noise cancels out the factor of two in the signal which was acquired from measuring both quadrature variables. One can also see this by comparing Eqs.~\eqref{eq:sq_snr} and~\eqref{eq:gkp_snr} when $G_{\rm eff}=G/2$. Therefore, the GKP-assisted scan does no better (if not worse) than the single-mode squeezing-enhanced scan in this practical limit.

\subsection{Detailed error analysis}
We consider the error in our ideal Gaussian approximation for the GKP-state as a general function of additive Gaussian noises, which leads to the error-revised curve in Fig.~\ref{fig:rate_gkp_v_sq}. All the practically relevant noise sources---cavity-added vacuum-noise, thermal noise, imperfect GKP resource-states---can be converted to additive Gaussian noises; see, e.g., ref.~\cite{Noh2019_capacity} for details of such a conversion. Further, since our detection strategy consists of independent homodyne measurements on the signal and ancilla and since the noises added to the $Q$ and $P$ quadrature variables are equivalent, it is sufficient to restrict ourselves to one quadrature, say the $Q$ quadrature, in the current analysis. The PDF describing fluctuations of the GKP grid (of the signal or the ancilla) due to noise is given by a Gaussian distribution,
\begin{equation}
    p_{y}(q)=\frac{1}{\sqrt{\pi y}}\e^{-\frac{q^2}{y}},
\end{equation}
where $\ev*{q^2}=y/2$ is the variance ($\sqrt{y/2}$ is the standard deviation). For instance, cavity-transmission losses correspond to $y/2=(1-\abs{\bm\chi_{mm}}^2)$. 
With the modulo structure of the GKP grid in mind, we provide a strategy to estimate an unknown displacement, which has been previously analyzed in the applications of continuous-variable quantum error correction~\cite{noh2020_o2o} and distributed quantum sensing (DQS)~\cite{zhuang2020DQSgkp}. The estimator $\tilde{q}$ for a given displacement $q$ is chosen as,
\begin{multline}
    \tilde{q}=R_{\sqrt{2\pi}}(q)\equiv\sum_{n\in\mathbb{Z}}(q-n\sqrt{2\pi})\\\times{\rm I}\left(q\in[(n-1/2)\sqrt{2\pi}, (n+1/2)\sqrt{2\pi}]\right),
\end{multline}
where
\begin{multline}
  {\rm I}\left(q\in[(n-1/2)\sqrt{\pi}, (n+1/2)\sqrt{\pi}]\right)=\\
    {\begin{cases}
    1 &  q\in[(n-1/2)\sqrt{2\pi}, (n+1/2)\sqrt{2\pi}] \\
    0 & \text{else}
    \end{cases}}, 
\end{multline}
is an indicator function~\cite{noh2020_o2o}. The value $n\sqrt{2\pi}$ quantifies how many lattice-spacings the displacement $q$ is from the origin at $n=0$. The relative displacement from the $n$th lattice-point, $q-n\sqrt{2\pi}$, then lies within half a lattice spacing $\sqrt{\pi/2}$ of this point. Assuming an unknown axion-induced displacement of $\epsilon_s\equiv\abs{\bm\chi_{ms}}\mu_s\sin\phi_s$ along the $Q$-quadrature, an estimate for the $k$th moment is then, 
\begin{equation}
    \ev*{\tilde{q}^k}\equiv\sum_{n\in\mathbb{Z}}\int_{(n-1/2)\sqrt{2\pi}}^{(n+1/2)\sqrt{2\pi}}\dd{q}p_y(q-\epsilon_s)(q-n\sqrt{2\pi})^k,\label{eq:gkp_moments}
\end{equation}
where the axion-induced displacement has been absorbed into the mean for the PDF. When $\sqrt{y},\epsilon_s\ll\sqrt{\pi}$, $\ev*{\tilde{q}}\approx\epsilon_s$ and ${\rm Var}(\tilde{q})\approx y/2$, and the Gaussian approximation for the GKP state holds good.

\begin{figure}
    \centering
    \includegraphics[width=\linewidth]{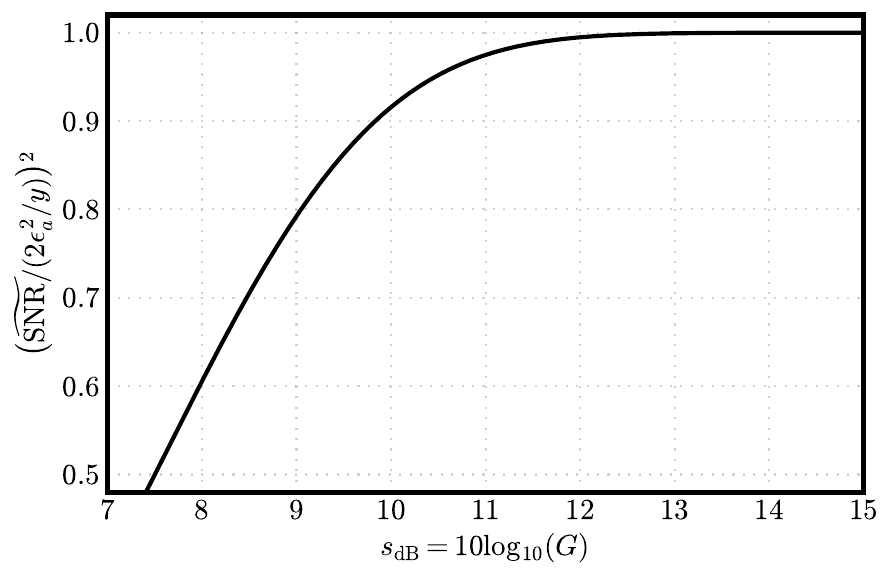}
    \caption{Plot of the (square of the) estimated SNR, $\widetilde{\rm SNR}$, to its Gaussian counterpart $2\epsilon_s^2/y$ as a function of the squeezing $G$ when the additive noise $y$ obeys Eq.~\eqref{eq:y_G}. An arbitrarily low value of $\epsilon_s=.001$ was chosen for the axion-induced displacement to generate the plot. This choice does not appreciably change the analysis for all $\epsilon_s\lesssim.1$.}
    \label{fig:snr_gkp}
\end{figure}

From this estimation strategy, we can estimate the SNR for the power, which we formally define as $\widetilde{{\rm SNR}}\equiv\ev{\tilde{q}}^2/{\rm Var}(\tilde{q})$. We wish to compare this estimation to its Gaussian counterpart of $2\epsilon_s^2/y$, within the relevant scenarios/parameter-regimes discussed in the main text, in which case $2\epsilon_s^2/y$ reduces to Eq.~\eqref{fig:snr_gkp}. To do so, we make the correspondence $y=1/G+2(1-\abs{\bm\chi_{mm}}^2)$, which is the additive noise in the GKP-assisted search scenario. See Fig.~\ref{fig:GKP_detection} and accompanying analysis. Upon optimization of the scan-rate, in general, the squeezing $G$ and the cavity-transmission $\abs{\bm\chi_{mm}}^2$ are related, however from the Gaussian analysis, we find an optimal-coupling of $\gamma_m/\gamma_\ell\approx 4G$, from which the following relation may be derived,
\begin{equation}
    1-\abs{\bm\chi_{mm}}^2= \frac{4G}{\left(\frac{(4G+1)^2}{4}+\omega^2\right)} \leq \frac{16G}{(4G+1)^2},
\end{equation}
where we have taken the on-resonance ($\omega=0$) value as a worst-case approximation. We shall assume this value from hereon to simplify the analyses. The additive noise then reduces to a simple function of the squeezing,
\begin{equation}
    y= 1/G + \frac{32G}{(4G+1)^2}.\label{eq:y_G}
\end{equation}
Since the scan-rate scales as the square of the SNR, the relevant quantity to consider is $\widetilde{\rm SNR}^2$. Figure~\ref{fig:snr_gkp} shows the deviation of the estimated SNR, $\widetilde{\rm SNR}$, relative to its Gaussian counterpart, $2\epsilon_s^2/y$, as a function of the squeezing in dB. For squeezing levels above 10dB, there is less than a $10\%$ discrepancy between $\widetilde{\rm SNR}$ [evaluated via Eq.~\eqref{eq:gkp_moments}] and the Gaussian approximation for the GKP state [i.e. Eq.~\eqref{eq:gkp_snr}]. We note that, at 10dB of squeezing, $y\approx.3$. 

Finally, considering the (optimal) GKP-assisted scan-rate, $\mathcal{R}_{\rm GKP}^\star$, from the Gaussian analysis, Eq.~\eqref{eq:scan_rate_gkp}, we can estimate the optimal error-revised GKP-assisted scan-rate $\widetilde{\mathcal{R}}_{\rm GKP}^\star$ by the following. Let $\widetilde{\rm SNR}_{\rm GKP}$ be the SNR evaluated from the estimation procedure just presented and $\overline{\rm SNR}_{\rm GKP}$ be the SNR taken from the Gaussian analysis [Eq.~\eqref{eq:gkp_snr}]. Then,
\begin{align*}
    \widetilde{\mathcal{R}}_{\rm GKP}&\propto\int\dd{\omega}\widetilde{\rm SNR}_{\rm GKP}^2\\
    &=\int\dd{\omega}\overline{\rm SNR}_{\rm GKP}^2 \left(\frac{\widetilde{\rm SNR}_{\rm GKP}}{\overline{\rm SNR}_{\rm GKP}}\right)^2\\
    &\geq \left(\frac{\widetilde{\rm SNR}_{\rm GKP}}{\overline{\rm SNR}_{\rm GKP}}\right)^2_{\omega=0}\int\dd{\omega}\overline{\rm SNR}_{\rm GKP}^2,
\end{align*}
where we use the fact that there is a maximal amount of vacuum noise on resonance, $\omega=0$, at which point the ratio $\widetilde{\rm SNR}_{\rm GKP}/\overline{\rm SNR}_{\rm GKP}$ is the smallest. Now observe that,
\begin{equation*}
\mathcal{R}_{\rm GKP}\propto\int\dd{\omega}\overline{\rm SNR}_{\rm GKP}^2,
\end{equation*}
where $\mathcal{R}_{\rm GKP}$ is the scan-rate from the Gaussian analysis. It thus follows that,
\begin{equation}
    \widetilde{\mathcal{R}}_{\rm GKP}^\star\geq \left(\frac{\widetilde{\rm SNR}_{\rm GKP}}{\overline{\rm SNR}_{\rm GKP}}\right)^2 \mathcal{R}_{\rm GKP}^\star\label{eq:gkp_scan_rate_bound}
\end{equation}
where it is understood that the pre-factor is evaluated at $\omega=0$ and $\gamma_m/\gamma_\ell\approx4G$ (the optimal point inferred from the Gaussian analysis). The error-revised (solid) curve referred to in Fig.~\ref{fig:rate_gkp_v_sq} corresponds to the estimate on the right-hand side of Eq.~\eqref{eq:gkp_scan_rate_bound}.

%

\end{document}